\documentclass[onecolumn,draftcls]{IEEEtran}
\usepackage[utf8]{inputenc}
\usepackage{epstopdf}
\usepackage[skip=2pt]{caption}
\usepackage[mathscr]{eucal}
\usepackage[cmex10]{amsmath}
\usepackage{amssymb,amsmath,amsfonts,latexsym}
\usepackage{graphicx,bm,xcolor,url}
\usepackage[caption=false]{subfig} 
\usepackage{array}%array and tabular environments
\usepackage{verbatim}
\usepackage{bm}
\usepackage{cite}
\usepackage{algpseudocode}
\usepackage{algorithm}
\usepackage{verbatim}
\usepackage{textcomp}
\usepackage{mathrsfs}
\usepackage{multirow}
\usepackage{hyperref}
\usepackage{epstopdf}
\usepackage{centernot}
\usepackage{dsfont}
 \usepackage{color,pgfplots,tikz}
\usepackage[nolist]{acronym}

\usepackage{thmtools}

\usepackage{xpatch}

\makeatletter
\newif\ifthmt@listswap
\def\thmt@TRUE{true}
\def\thmt@FALSE{false}
\define@key{thmt-listof}{swapnumber}[true]{%
  \def\thmt@tmp{#1}%
  \ifx\thmt@tmp\thmt@TRUE
    \thmt@listswaptrue
  \else\ifx\thmt@tmp\thmt@FALSE
    \thmt@listswapfalse
  \else
    \PackageError{thmtools}{Unknown value `#1' to key swapnumber}{}%
  \fi\fi
}

\xpatchcmd\thmt@mklistcmd
  {\protect\numberline{\protect\let\protect\autodot\protect\@empty}}
  {%
    \protect\ifthmt@listswap
    \protect\else
      \protect\numberline{\protect\let\protect\autodot\protect\@empty}%
    \protect\fi
  }
  {}{\fail}

\xpatchcmd\thmt@mklistcmd
  {%
    \protect\numberline{\csname the\thmt@envname\endcsname}%
    \thmt@thmname
  }
  {%
    \protect\ifthmt@listswap
      \thmt@thmname~\csname the\thmt@envname\endcsname
    \protect\else
      \protect\numberline{\csname the\thmt@envname\endcsname}%
      \thmt@thmname
    \protect\fi
  }
  {}{\fail}
\makeatother

\catcode`~=11 \def\UrlSpecials{\do\~{\kern -.15em\lower .7ex\hbox{~}\kern .04em}} \catcode`~=13 

\allowdisplaybreaks[1]
 
\newcommand{\nn}{\nonumber}

\begin{acronym}
\acro{DMC}{discrete memoryless channel}
\acro{BSC}{binary symmetric channel}
\acro{CC}{constant composition}
\acro{FSC}{finite-state channels}
\acro{i.i.d.}{independently and identically distributed}
\acro{JSCC}{joint source-channel coding}
\acro{ML}{maximum likelihood}
\acro{RCU}{Random Coding Union}
\acro{RGV}{Random Gilbert-Varshamov}
\acro{TRC}{typical random coding}
\acro{RCE}{random coding exponent}
\end{acronym}

\newcommand{\calB}{\mathcal{B}}
\newcommand{\calD}{\mathcal{D}}

\newcommand{\calI}{\mathcal{I}}

\newcommand{\calK}{\mathcal{K}}
\newcommand{\calL}{\mathcal{L}}
\newcommand{\calN}{\mathcal{N}}

\newcommand{\calP}{\mathcal{P}}
\newcommand{\calT}{\mathcal{T}}
\newcommand{\calU}{\mathcal{U}}
\newcommand{\calY}{\mathcal{Y}}

\newcommand{\rme}{\mathrm{e}}

\DeclareMathAlphabet{\mathbsf}{OT1}{cmss}{bx}{n}
\DeclareMathAlphabet{\mathssf}{OT1}{cmss}{m}{sl}% slanted sans serif

\DeclareSymbolFont{bsfletters}{OT1}{cmss}{bx}{n}  
\DeclareSymbolFont{ssfletters}{OT1}{cmss}{m}{n}
\DeclareMathSymbol{\bsfGamma}{0}{bsfletters}{'000}
\DeclareMathSymbol{\ssfGamma}{0}{ssfletters}{'000}
\DeclareMathSymbol{\bsfDelta}{0}{bsfletters}{'001}
\DeclareMathSymbol{\ssfDelta}{0}{ssfletters}{'001}
\DeclareMathSymbol{\bsfTheta}{0}{bsfletters}{'002}
\DeclareMathSymbol{\ssfTheta}{0}{ssfletters}{'002}
\DeclareMathSymbol{\bsfLambda}{0}{bsfletters}{'003}
\DeclareMathSymbol{\ssfLambda}{0}{ssfletters}{'003}
\DeclareMathSymbol{\bsfXi}{0}{bsfletters}{'004}
\DeclareMathSymbol{\ssfXi}{0}{ssfletters}{'004}
\DeclareMathSymbol{\bsfPi}{0}{bsfletters}{'005}
\DeclareMathSymbol{\ssfPi}{0}{ssfletters}{'005}
\DeclareMathSymbol{\bsfSigma}{0}{bsfletters}{'006}
\DeclareMathSymbol{\ssfSigma}{0}{ssfletters}{'006}
\DeclareMathSymbol{\bsfUpsilon}{0}{bsfletters}{'007}
\DeclareMathSymbol{\ssfUpsilon}{0}{ssfletters}{'007}
\DeclareMathSymbol{\bsfPhi}{0}{bsfletters}{'010}
\DeclareMathSymbol{\ssfPhi}{0}{ssfletters}{'010}
\DeclareMathSymbol{\bsfPsi}{0}{bsfletters}{'011}
\DeclareMathSymbol{\ssfPsi}{0}{ssfletters}{'011}
\DeclareMathSymbol{\bsfOmega}{0}{bsfletters}{'012}
\DeclareMathSymbol{\ssfOmega}{0}{ssfletters}{'012}

\newcommand{\tilE}{\tilde{E}}

\newcommand{\hatm}{\hat{m}}

\newcommand{\tilm}{\tilde{m}}

\newcommand{\hatP}{\hat{P}}
\newcommand{\tilp}{\tilde{p}}
\newcommand{\tilP}{\tilde{P}}

\newcommand{\tilX}{\tilde{X}}

\newcommand{\tilbx}{\tilde{\bx}}
\newcommand{\tilbX}{\tilde{\bX}}

\newcommand{\barx}{\bar{x}}

\newcommand{\eps}{\varepsilon}
\newcommand{\floor}[1]{\lfloor{#1}\rfloor}

\newcommand{\dotleq}{\stackrel{.}{\leq}}

\newcommand{\dotgeq}{\stackrel{.}{\geq}}

\DeclareMathOperator*{\argmax}{arg\,max}
\DeclareMathOperator*{\argmin}{arg\,min}

\DeclareMathOperator{\var}{\mathrm{Var}}

\newtheorem{theorem}{Theorem} 
\newtheorem{lemma}{Lemma}
\newtheorem{proposition}{Proposition}
\newtheorem{corollary}{Corollary}
\newtheorem{definition}{Definition}

\newcommand{\qednew}{\nobreak \ifvmode \relax \else
      \ifdim\lastskip<1.5em \hskip-\lastskip
      \hskip1.5em plus0em minus0.5em \fi \nobreak
      \vrule height0.75em width0.5em depth0.25em\fi}

\usepackage{ifthen}
\usepackage{xargs}
\newcommandx{\yaHelper}[2][1=\empty]{%
\ifthenelse{\equal{#1}{\empty}}%
  { \ensuremath{ \scriptstyle{ #2 } } } % no offset
  { \raisebox{ #1 }[0pt][0pt]{ \ensuremath{ \scriptstyle{ #2 } } } }  % with offset
}
\newcommandx{\yrightarrow}[4][1=\empty, 2=\empty, 4=\empty, usedefault=@]{%
  \ifthenelse{\equal{#2}{\empty}}
  { \xrightarrow{ \protect{ \yaHelper[ #4 ]{ #3 } } } } % there's no text below
  { \xrightarrow[ \protect{ \yaHelper[ #2 ]{ #1 } } ]{ \protect{ \yaHelper[ #4 ]{ #3 } } } } % there's text below
}
\newcommand{\pto}{\smash{\stackrel{({\rm p})}{ \,\longrightarrow\,}}}

\usepackage{xspace}
\DeclareFontFamily{U}{mathc}{}
\DeclareFontShape{U}{mathc}{m}{it}%
{<->s*[1.03] mathc10}{}

\DeclareMathAlphabet{\mathscr}{U}{mathc}{m}{it}

\newcommand{\bx}{\boldsymbol{x}}
\newcommand{\bX}{\boldsymbol{X}}
\newcommand{\by}{\boldsymbol{y}}
\newcommand{\bY}{\boldsymbol{Y}}

\newcommand{\calA}{\mathcal{A}}
\newcommand{\calC}{\mathcal{C}}
\newcommand{\calE}{\mathcal{E}}
\newcommand{\calX}{\mathcal{X}}
\newcommand{\calS}{\mathcal{S}}
\newcommand{\calF}{\mathcal{F}}
\newcommand{\calQ}{\mathcal{Q}}
\newcommand{\calM}{\mathcal{M}}
\newcommand{\calV}{\mathcal{V}}

\newcommand{\cn}{\mathscr{c}_n}

\newcommand{\indicator}{\mathds{1}}

\newcommand{\Ecn}{E_n(\cn)}

\newcommand{\calG}{\mathcal{G}} % Free energy function (Markov source)
\newcommand{\bbR}{\mathbb{R}} % Real Numbers
\newcommand{\bbZ}{\mathbb{Z}} % Integer
\newcommand{\bbE}{\mathbb{E}} % Expectation
\newcommand{\bbP}{\mathbb{P}} % Probability of a set

\newcommand{\PP}{\mathbb{P}} % Expectation

\IEEEoverridecommandlockouts

\newcommand{\ddleq}{\enspace \mathring{\leq} \enspace}
\newcommand{\ddgeq}{\enspace \mathring{\geq} \enspace}
\newcommand{\ddeq}{\enspace \mathring{=} \enspace}
\newcommand{\bbarx}{\boldsymbol{\bar{x}}}

\begin{document}

\title{Generalized Random Gilbert-Varshamov Codes: Typical Error Exponent and Concentration Properties}
	
\author{Lan V.~Truong and Albert Guill\'en i F\`abregas
		\thanks{L. V. Truong is with the Department of Engineering, 
			University
			of Cambridge, Cambridge CB2 1PZ, U.K. (e-mail: lt407@cam.ac.uk).
			A.~Guill\'en i F\`abregas is with the  Department of Engineering, 
			University
			of Cambridge, Cambridge CB2 1PZ, U.K. and the Department of Information and 
			Communication
			Technologies, Universitat Pompeu Fabra, Barcelona 08018, Spain (e-mail: guillen@ieee.org). %He was with the Instituci\'o Catalana de Recerca i Estudis Avan\c{c}ats (ICREA),Barcelona 08010, Spain .
			
			}
		%\thanks{This work has been presented in part at the 2021 IEEE Information Theory Workshop, Kanazawa, Japan.}
		\thanks{This work has been funded in part by the European Research Council under ERC grant agreement 725411 and by the Spanish Ministry of Economy and Competitiveness under grant PID2020-116683GB-C22.
}
	}

\maketitle

\begin{abstract}
We find the exact typical error exponent of constant composition generalized random Gilbert-Varshamov (RGV) codes over DMCs channels with generalized likelihood decoding. We show that the typical error exponent of the RGV ensemble is equal to the expurgated error exponent, provided that the RGV codebook parameters are chosen appropriately. We also prove that the random coding exponent converges in probability to the typical error exponent, and the corresponding non-asymptotic concentration rates are derived. Our results show that the decay rate of the lower tail is exponential while that of the upper tail is double exponential above the expurgated error exponent. The explicit dependence of the decay rates on the RGV distance functions is characterized.
\end{abstract}

%%%%%%%%%%%%%%%%%%%%%%%%%%%%%%%%%%%%%%%%%%
%\section{Introduction}

%Random coding is the key technique employed in information theory in order to show that a code with low error probability exists without explicitly constructing it. Codes are constructed at random, and the average error probability over all randomly generated codes is bounded. Then, it follows that there must exist a code with error probability at least as low as the ensemble average error probability over the codes. In most proofs of coding theorems for discrete memoryless channels, codewords of random codes are typically generated independently.

%Error exponents

%RGV

%TRC and tails

%This is the introduction. Define RGV and GLD.

\section{Introduction}

Introduced by Shannon \cite{Shannon48}, random coding is the key technique employed in information theory in order to show that a code with low error probability exists without explicitly constructing it. Codes are constructed at random, and the average error probability over all randomly generated codes is bounded. Then, it follows that there must exist a code with error probability at least as low as the ensemble average error probability over the codes. In particular, for \ac{DMC}, Shannon showed that there exists a code of rate smaller than the channel capacity with vanishing probability of error as the codeword length increases.

Since Shannon's work, random coding has not only been applied extensively, but has been refined in a number of ways. For rates below capacity, Fano \cite{Fano} characterized the exponential decay of the error probability defining the \ac{RCE} as the negative normalized logarithm of the ensemble-average error probability. In \cite{Gallager1965a}, Gallager derived the \ac{RCE} in a simpler way and introduced the idea of expurgation in order to show the existence of a code with an improved exponent the at low rates. An upper bound to the error exponent for the \ac{DMC}, called sphere-packing bound, was first introduced in \cite{sgb} and it was shown to coincide with the \ac{RCE} for rates higher than a certain critical rate. Nakibo{\u{g}}lu in~\cite{Nakiboglu2020} recently derived sphere-packing bounds for some stationary memoryless channels using Augustin's method~\cite{nakibouglu2019augustin}.

Most proofs invoking random coding arguments, assume that codewords are independent. Random Gilbert-Varshamov (RGV) codes were first introduced in \cite{somekh_2019}, and are a family of random codes inspired by the basic construction attaining the Gilbert-Varshamov bound for codes in Hamming spaces. The code construction is based on drawing codewords recursively from a fixed type class, in such a way that a newly generated codeword must be at a certain minimum distance from all previously chosen codewords, according to some generic distance function. For suitably optimized distance functions, RGV codes attain Csiszár and K{ö}rner's exponent \cite{CK81}, which is known to be at least as high as both the random-coding and expurgated exponents. 

Most works on random coding and error exponents study the error exponent of the ensemble-average error probability. In \cite{Barg2002a}, Barg and Forney  studied the i.i.d. random coding over the \ac{BSC} with maximum likelihood decoding and showed that the error exponent of most random codes is close to the so-called \ac{TRC} exponent, strictly larger than the \ac{RCE} at low rates. Upper and lower bounds on the \ac{TRC} for constant-composition codes and general \ac{DMC}s were provided in \cite{Nazari}. For the same type of codes and channels, Merhav \cite{Merhav2018a} determined the exact \ac{TRC} error exponent for a wide class of stochastic decoders called generalized likelihood decoders (GLD), of which maximum-likelihood is a special case. Merhav derived the \ac{TRC} exponent for spherical codes over coloured Gaussian channels~\cite{merhav2019error} and for random convolutional code ensembles \cite{merhav2019error2}. Merhav provided a dual expression of the \ac{TRC} for i.i.d. codes in \cite{merhav2019lagrange}. Tamir {\em et al.} \cite{Tamir2020a} studied the upper and lower tails of the error exponent around the \ac{TRC} exponent for random pairwise-independent constant-composition codes with GLD. It was shown that the tails behave in a non-symmetric way: the lower tail decays exponentially while the upper tail decays doubly-exponentially; the latter was first established for a limited range of rates in \cite{Ahlswede1982}. By studying the behavior of both tails,  work in \cite{Tamir2020a} proves concentration in probability. 
The TRC was shown to be universally achievable with a likelihood mutual-information decoder in \cite{tamir2021universal}. For pairwise-independent ensembles and arbitrary channels, Cocco~\emph{et al.} showed in~\cite{cocco2022} that the probability that a code in the ensemble has an exponent smaller than a lower bound on the \ac{TRC} exponent is vanishingly small. Recently, Truong~\emph{et al.} showed that, for \ac{DMC}s, the error exponent of a randomly generated code with pairwise-independent codewords converges in probability to its expectation -- the typical error exponent \cite{Truong2022PO}. For high rates, the result is a consequence of the fact that the random-coding error exponent and the sphere-packing error exponent coincide. For low rates, instead, the convergence is based on the fact that the union bound accurately characterizes the probability of error. The paper also zooms into the behavior at asymptotically low rates and shows that the error exponent converges in distribution to  Gaussian-like distributions.

\subsection{Contributions}
This work focusses on the RGV code ensemble and discusses concentration properties of error exponents around its \ac{TRC}.   Compared with constant-composition codes, the dependence among RGV codewords causes standard concentration inequalities such as Hoeffding's inequality not  to hold. In this work, we develop new techniques to overcome the challenges presented by RGV codeword dependence. Our main contributions include:    
\begin{itemize}
	\item We find the exact typical error exponent \ac{TRC} for the RGV ensemble by proving matched upper and lower bounds on the TRC and show it is equal to its \ac{RCE}, i.e., to the maximum of the expurgated and random-coding exponent.
	\item We show that the random error exponent converges in probability to the TRC.
	\item We characterize the convergence rates of the above convergence and show that it is exponential for the lower tail and double-exponential for the upper tail under some technical conditions.
\end{itemize}

\subsection{Notation}
Random variables will be denoted by capital letters,  and their realizations will be denoted by the corresponding lower case letters. Random vectors and their realizations will be denoted, respectively, by boldfaced capital and lower case letters. Their alphabets will be superscripted by their dimensions. For a generic joint distribution $P_{XY}=\{P_{XY}(x,y), x \in \calX, y \in \calY\}$, which will often be abbreviated by $P$, information measures will be denoted in the conventional manner, but with a subscript $P$, that is $I_P(X;Y)$ is the mutual information between $X$ and $Y$, and similarly for other quantities. Natural logarithms are assumed in the derivations; examples will employ base $2$. The probability of an event $\calE$ will be denoted by $\bbP(\calE)$, the indicator function of event $\calE$ will be denoted by $\indicator\{\calE\}$, and the expectation operator will be denoted by $\bbE[\cdot]$. The notation $[t]_+$ will stand for $\max\{t,0\}$. 

For two positive sequences, $\{a_n\}$ and $\{b_n\}$, the notation $a_n\doteq b_n$ will stand for exponential equality, that is $\lim_{n\to \infty} \frac{1}{n}\log(\frac{a_n}{b_n})=0$. Exponential inequalities $a_n \dotleq b_n$ and $a_n \dotgeq b_n$ are defined as $\lim_{n\to \infty} \frac{1}{n}\log(\frac{a_n}{b_n})\leq0$ and $\lim_{n\to \infty} \frac{1}{n}\log(\frac{a_n}{b_n})\geq0$, respectively. Accordingly, the notation $a_n\doteq e^{-n\infty}$ means that $a_n$ decays super-exponentialy.
For two positive sequences, $\{a_n\}$ and $\{b_n\}$, whose elements are both smaller than one for all large enough $n$, the notation $a_n \ddeq b_n$ will stand for double-exponential equality, that is
\begin{align}
\lim_{n\to \infty} \frac{1}{n}\log \bigg(\frac{\log b_n}{\log a_n}\bigg)= 0.
\end{align}
Similarly, $a_n \ddleq b_n$ means that
\begin{align}
\limsup_{n\to \infty} \frac{1}{n}\log \bigg(\frac{\log b_n}{\log a_n}\bigg)\leq 0,
\end{align} and $a_n \ddgeq b_n$ stands for
\begin{align}
\liminf_{n\to \infty} \frac{1}{n}\log \bigg(\frac{\log b_n}{\log a_n}\bigg)\geq 0.
\end{align}

A sequence of random variables $\{A_n\}_{n=1}^{\infty}$ converges to $A$ in probability, denoted as $A_n \pto A$ if for all $\delta>0$ \cite[Sec.~2.2]{Durrett},
\begin{align}
\lim_{n\to\infty}\bbP[|A_n-A|>\delta]= 0.
\label{eq:convp}
\end{align}

The empirical distribution, or type, of a sequence $\bx \in \calX^n$, which will be denoted by $\hatP_{\bx}$, is the vector of relative frequencies, $\hatP_{\bx}(x)$, of each symbol $x \in \calX$ in $\bx$. The set of all possible empirical distributions of sequences of length $n$ on alphabet $\calX$ is denoted by $\calP_n(\calX)$. The joint empirical distribution of a pair of sequences, denoted by $\hatP_{\bx \by}$, is similarly defined. The set of all possible joint empirical distributions of sequences of length $n$ on alphabets $\calX$ and $\calY$ is denoted by $\calP_n(\calX\times\calY)$.
The type class of $Q_X$, denoted by $\calT(Q_X)$, is the set of all vectors $\bx \in \calX^n$ with $\hatP_{\bx}=Q_X$. The joint type class of $P_{XY}$, denoted by $\calT(P_{XY})$, is the set of pairs of sequences $(\bx,\by) \in \calX^n \times \calY^n$ with $\hatP_{\bx \by}=Q_{XY}$. In addition, we also define 
$
\calQ(Q_X)\triangleq \big\{P_{XX'} \in \calP_n(\calX \times \calX): P_X=P_{X'}=Q_X \big\}.
$
Finally, $[M]$ denotes the set $\{1,2,\cdots,M\}$, and $[M]_*^2\triangleq \{(m,m') \in [M]^2: m\neq m'\}$.

\subsection{Structure of the Paper}
In Section \ref{sec:pre}, we introduce error probability and error exponent of a RGV code. In Section \ref{RGV:intro}, we introduce the generation of RGV random codebook ensembles. We also mention about properties of RGV codes and type-numerators in this section.  We derive the typical error exponent for the RGV in Section \ref{sec:dmc}. Finally, we study concentration properties of this ensemble in Section \ref{sec:concen}. Proofs of the main results can be found in the corresponding sections while the proofs of auxiliary results can be found in the Appendices. 
%%%%%%%%%%%%%%%%%%%%%%%%%%%%
\section{Preliminaries} \label{sec:pre}
We assume that the RGV code $\calC_n=\{\bx_1,\bx_2,\dotsc,\bx_M\} \in \calX^n, M=e^{nR}$ is employed for transmission over a DMC channel with channel law $W(y|x)$ for $x \in \calX, y \in \calY$. More specifically, when the transmitter wishes to convey a message $m \in \{1,2,\cdots,M\}$, it sends codeword $\bx_m=(x_{m,1},\dotsc,x_{m,n})\in\calX^n$ over the channel. The channel produces an output vector $\by=(y_1,y_2,\dotsc,y_n) \in \calY^n$, according to 
\begin{align}
W(\by|\bx_m)=\prod_{i=1}^n W(y_i|x_{m,i}).
\end{align}

At the decoder side, we assume that a GLD  \cite{Merhav2017a} is used to infer what the transmitted message was. The GLD \cite{Merhav2017a} extends the likelihood decoder in \cite{yassaee} and \cite{Scarlett2015e}, and is a stochastic decoder that randomly selects the message estimate $\hatm$ according to the posterior probability distribution given the channel output $\by$ as follows
\begin{align}
\Pr(\hatm=m|\by)=\frac{\exp\{ng(\hatP_{\bx_m,\by})\}}{\sum_{m=1}^M \exp\big\{ng(\hatP_{\bx_{m'},\by})\big\}},
\end{align} where $g(\cdot)$, henceforth referred to as the \emph{decoding metric}, is an arbitrary continuous function of a joint distribution $P_{XY}$ on $\calX \times \calY$. For
\begin{align}
g(P_{XY})=\sum_{x \in \calX} \sum_{y \in \calY} P_{XY}(x,y)\log  W(y|x),
\end{align} we recover the ordinary likelihood decoder \cite{Scarlett2015e}. For
\begin{align}
g(P_{XY})=\beta \sum_{x \in \calX} \sum_{y \in \calY} P_{XY}(x,y)\log  W(y|x),
\end{align} $\beta\geq 0$ being a free parameter, we extend this to a parametric family of decoders, where $\beta$ controls the skewness of the posterior \cite{Merhav2018a}. In particular, $\beta \to \infty$ leads to the (deterministic) ML decoder. Other interesting choices are associated with mismatched metrics,
\begin{align}
g(P_{XY})=\beta \sum_{x\in \calX} \sum_{y \in \calY} P_{XY}(x,y)\log  W'(y|x),
\end{align} $W'$ being different from $W$, and 
\begin{align}
g(P_{XY})=\beta I_P(X;Y),
\end{align}
which is the stochastic version of the well-known universal maximum mutual information (MMI) decoder \cite{Csi97}, which has been recently proven to be universal in a typical error exponent sense \cite{Tamir2022}. The MMI decoder is approached for $\beta \to \infty$. 

The average probability of error, associated with a given RGV code $\mathscr{c}_n$ and the GLD, is given by
\begin{align}
P_{\rme}(\mathscr{c}_n)=\frac{1}{M}\sum_{m=1}^M \sum_{m'\neq m} \sum_{\by \in \calY^n} W(\by|\bx_m)\cdot\frac{\exp\{ng(\hatP_{\bx_{m'},\by})\}}{\sum_{\tilm=1}^M \exp\{ng(\hatP_{\bx_{\tilm},\by})\}} \label{defPeCn}.
\end{align}
The error exponent of code $\mathscr{c}_n$ is defined  as 
\begin{align}
\Ecn = -\frac{1}{n}\log P_{\rme}(\mathscr{c}_n).
\label{eq:en0}
\end{align}
Let $R=\lim_{n\to \infty} \frac{1}{n}\log M_n$ be the rate of the code in bits per channel use. An error exponent $E(R)$ is said to be achievable when there exists a sequence of codes $\{\mathscr{c}_n\}_{n=1}^{\infty}$ such that $\liminf_{n\to \infty} \Ecn\geq E(R)$. The channel capacity $C$ is the supremum of the code rates $R$ such that $E(R)>0$. 

In the next sections, we introduce RGV codebook ensemble and derive concentration properties of the error exponent~\eqref{eq:en0} of sequences of RGV codes ${\cal C}_n$ in the asymptotic regime.

%%%%%%%%%%%%%%%%%%%%%%%%%%%%%%%%%%%%%%%%%%%
\section{RGV Random Codebook Ensembles and Properties}
\subsection{RGV Random Codebook Ensembles} \label{RGV:intro}
In this section, we describe basic RGV codebook construction, channel model and GLD. The RGV codebook was first introduced in \cite{somekh_2019}, which extended code constructions that attain the Gilbert-Varshamov bound on the Hamming space \cite{Gilbert1952ACO,Varshamov1957a}. The RGV construction is a randomized constant composition counterpart of such codes for arbitrary DMCs and arbitrary distance functions. 

\begin{definition} Let $\Omega$ be the set of bounded, continuous, symmetric, and type-dependent functions $d(\cdot,\cdot): \calX^n \times \calX^n \to \bbR$, i.e., bounded functions that satisfy $d(\bx,\bx')=d(\bx',\bx)$ for all $\bx,\bx' \in \bbR^n$, that depend on $(\bx,\bx')$ only through the joint distribution $\hatP_{\bx \bx'}$, and that are continuous on the probability simplex. 
\end{definition}

We refer to $d \in \Omega$ as a distance function, although it need not to be a distance in the topological space (e.g., it may be negative). Some examples of such distance function include Hamming distance, Bhattacharyya distance, and equivocation distance \cite{somekh_2019}.

The RGV code $\calC_n=\{\bx_1,\bx_2,\dotsc,\bx_M\} \in \calX^n$ with $M$ codewords of length $n$ is constructed such that any two distinct codewords $\bx,\bx' \in \calC_n$ satisfy $d(\bx,\bx')>\Delta$ for a given distance function $d(\cdot,\cdot) \in \Omega$ and $\Delta \in \bbR$. This guarantees that the minimum distance of the codebook exceeds the minimum distance $\Delta$. The construction depends on the input distribution $Q_X \in \calP_n(\calX)$ and is described by the following steps: 
\begin{enumerate}
	\item The first codeword, $\bx_1$, is drawn equiprobably from $\calT(Q_X)$;
	\item The second codeword, $\bx_2$, is drawn equiprobably from
	\begin{align}
	\calT(Q_X,\bx_1)&\triangleq \big\{\bbarx \in \calT(Q_X): d(\bbarx,\bx_1)>\Delta \big\}\\
	&=\calT(Q_X)\setminus \big\{\bbarx \in \calT(Q_X): d(\bbarx,\bx_1)\leq \Delta \big\},
	\end{align} i.e., the set of sequences with composition $Q_X$ whose distance to $\bx_1$ exceeds $\Delta$;
	\item Continuing recursively, the $i$-th codeword $\bx_i$ is drawn equiprobably from
	\begin{align}
	\calT(Q_X,\bx_1^{i-1})&\triangleq \big\{\bbarx \in \calT(Q_X): d(\bbarx, \bx_j)>\Delta, j=1,2,\dotsc,i-1\big\}\\
	&=\calT(Q_X,\bx_1^{i-2})\setminus  \big\{\barx \in \calT(Q_X,\bx_1^{i-2}): d(\bbarx,\bx_{i-1})\leq \Delta\big\}
	\end{align}
	where for $j<k$, $\bx_j^{k} = (\bx_j,\dotsc,\bx_k)$ is a shorthand notation to denote previously drawn codewords.
\end{enumerate}

For a given RGV code with rate $R$, type $Q_X$, distance function $d$, and minimum distance $\Delta$, we define the
random coding exponent (RCE) associated with the decoding metric $g$ as
\begin{align}
E_{\rm{rce}}^{\rm rgv}(R,Q_X,d,\Delta)\triangleq \lim_{n\to \infty} -\frac{1}{n} \log  \bbE[P_{\rme}(\calC_n)] \label{eq:Pe}
\end{align}
and the typical random coding (TRC) error exponent associated with the decoding metric $g$ as
\begin{align}
E_{\rm{trc}}^{\rm rgv}(R,Q_X,d,\Delta)\triangleq \lim_{n\to \infty} -\frac{1}{n}\bbE[\log  P_{\rme}(\calC_n)] \label{eq:en},
\end{align} provided that these limits exist\footnote{For the ensembles and channels considered in this paper, it will be shown in Section \ref{sec:dmc} that these limits exist.}, where the expectation is with respect to the randomness of the code $\calC_n$. 

Let $Q_X \in \calP(\calX),\Delta \in  \bbR, d \in \Omega$, and define the following quantity
\begin{align}
\Gamma(P_{XX'},R)\triangleq \min_{P_{Y|XX'}}\Big\{D(P_{Y|X}\|W|Q_X)+I_P(X';Y|X)+[\max\{g(P_{XY}),\alpha(R,P_Y)\}-g(P_{X'Y})]_+\Big\},
\label{eq:def_gammao}
\end{align}	
where
\begin{align}
\alpha(R,P_Y)\triangleq \max_{P_{X'|Y}: P_{X'}=Q_X,\atop I_P(X'; Y)\leq R}\big(g(P_{X'Y})-I_P(X';Y)\big)+R \label{defalphathmo}. 
\end{align}

The expurgated error exponent for RGV ensemble is defined as 
\begin{align}
E_{\rm{ex}}^{\rm{rgv}}(R,Q_X,d,\Delta)(R,Q_X,d,\Delta)&\triangleq  \min_{P_{XX'}\in \calQ(Q_X): d(P_{XX'})>\Delta, I_P(X;X')\leq R}  \bigg\{\Gamma(P_{XX'},R)+ I_P(X;X')-R\bigg\} \label{defexRGV}.
\end{align}

The main result of \cite{somekh_2019} is that for ML decoding, and suitably optimized distance function and minimum distance, the constant composition RGV ensemble attains a random coding exponent equal to the expurgated exponent, i.e., 
\begin{align}
E_{\rm{rce}}^{\rm rgv}(R,Q_X,d,\Delta) = E_{\rm{ex}}(R,Q_X)
\end{align}
where 
\begin{align}
E_{\rm{ex}}(R,Q_X)=\min_{P_{X'|X}: I_P(X;X')\leq R, P_{X'}=Q_X} \{\Gamma(P_{XX'},R)+I_P(X;X')-R\} 
\label{Eex}.
\end{align}
is the expurgated exponent of the independent constant composition ensemble with composition $Q_X$ and GLD. In this paper, we study the TRC of the RGV ensemble $E_{\rm{trc}}^{\rm rgv}(R,Q_X,d,\Delta)$ as well as the concentration of the exponent around the TRC. Specifically, we give a generic expression of $E_{\rm{trc}}^{\rm rgv}(R,Q_X,d,\Delta)$ and show that $E_{\rm{trc}}^{\rm rgv}(R,Q_X,d,\Delta)=E_{\rm{ex}}(R,Q_X)$ for suitably optimized minimum distance and distance functions. In addition, we provide bounds on the exponential and double-exponential concentration rates of the lower and upper tails of the error exponent of RGV codes.
%%%%%%%%%%%%%%%%%%%%%%%%%%%
\subsection{Properties of RGV Codebooks} 
In this section, we introduce technical several results characterizing the key properties of the generalized RGV construction. We begin by restating some known properties from \cite{somekh_2019}; we will then introduce a number of other properties that will be helpful in the derivation of our main results.

\begin{lemma} \cite[Lemma 1]{somekh_2019} \label{lem1some} Under the condition:
\begin{align}
R\leq \min_{P_{XX'} \in \calQ(Q_X): d(P_{XX'})\leq \Delta} I(X;X')-2 \delta 
\label{keycond}
\end{align} for some $\delta>0$ and $\bx_1^{i-1}$ occurring with non-zero probability, we have that
\begin{align}
(1-e^{-n\delta}) |\calT(Q_X)|\leq |\calT(Q_X,\bx_1^{i-1})|\leq |\calT(Q_X)|, \quad \forall i \in [M].
\end{align}
\end{lemma}
\begin{lemma} \cite[Lemma 2]{somekh_2019} \label{lem2some} Under the condition \eqref{keycond}, for any $k,m \in [M], k\neq m$ and $\bx_k, \bx_m \in \calT(Q_X)$ such that $d(\bx_k,\bx_m)>\Delta$, then we have
\begin{align}
\frac{1-4 \delta_n^2}{|\calT(Q_X)|^2} e^{-2 \delta_n}\leq \bbP[\bX_k=\bx_k,\bX_m=\bx_m] \leq \frac{1}{(1-e^{-n\delta})^2 |\calT(Q_X)|^2}
\end{align} 
while $\bbP[\bX_k=\bx_k,\bX_m=\bx_m] =0$ whenever $d(\bx_k,\bx_m)\leq \Delta$, where,
\begin{align}
\delta_n\triangleq \frac{e^{-n\delta}}{1-e^{-n\delta}}.
\end{align}
\end{lemma}
\begin{lemma} \cite[Lemma 4]{somekh_2019} \label{lem4some} For any message index $m$, the marginal distribution of codeword $\bX_m$ is $\bbP(\bx_m)=\frac{1}{|\calT(Q_X)|}$ for $\bx_m \in \calT(Q_X)$. 
\end{lemma}

In order to derive the TRC and convergence properties of the RGV code ensemble, we need to derive new properties of this random codebook. Some properties of the pairwise independent fixed-composition code ensemble \cite{Merhav2018a, Tamir2020a} are proven to hold for the RGV codebook under some extra conditions by other proof techniques. First, the following lemma can be easily proved using the same arguments as \cite{somekh_2019}.
\begin{lemma} \label{lem:aux0} Consider the generalized RGV construction with the rate $R$ satisfying \eqref{keycond}. Then, for any $\calA \subset [M]$ and $\delta>0$ as chosen in \cite{somekh_2019}, under the condition that $\min_{k,l \in \calA:k \neq l} d(\bx_k,\bx_l)>\Delta$, it holds that
	\begin{align}
	\bbP\bigg[\bigcap_{k \in \calA} \{\bX_k=\bx_k\}\bigg]\leq \frac{1}{(1-e^{-n\delta})^{|\calA|} |\calT(Q_X)|^{|\calA|}} \label{mot}.
	\end{align}
In addition, if $\min_{k,l \in \calA:k \neq l} d(\bx_k,\bx_l)\leq \Delta$, it holds that
\begin{align}
\bbP\bigg[\bigcap_{k \in\cal A} \{\bX_k=\bx_k\}\bigg]=0 \label{amote}.
\end{align}
Furthermore, if $\min_{k,l \in [M']:k\neq l}d(\bx_k,\bx_l)>\Delta$ for any $M'\leq M$, it holds that 
	\begin{align}
	\bbP\bigg[\bigcap_{m \in [M']} \{\bX_m=\bx_m\}\bigg]\geq \frac{1}{|\calT(Q_X)|^{M'}} \label{atigt}. 
	\end{align}
\end{lemma} 
In general, \eqref{atigt} does not hold for any $\calA \subset [M]$ as \eqref{mot}, but it holds for the class of subsets $\{[M']\}_{M'\leq M}$. Compared with Lemma \ref{lem2some}, \eqref{atigt} is tighter at $M=2$ if $\{k,m\}=\{1,2\}$. However, Lemma \ref{lem2some} is more general, i.e., it holds for any subset $\{k, m\}: (k,m)\in [M]\times [M], \enspace k\neq m\}$. 
\begin{IEEEproof}
	See Appendix \ref{lem:aux0proof}.
\end{IEEEproof}

Denote by
\begin{align}
\calI(m,m')\triangleq \indicator \{(\bx_m,\bx_{m'}) \in \calT(P_{XX'})\} 
\label{defImmprime}.
\end{align}
Then, the following result, whose proof can be found in Appendix \ref{aux1lemproof}, holds. 
\begin{lemma} \label{aux1:lem} Let $P_{XX'}$ be a joint-type in $\calQ(Q_X)$ such that $d(P_{XX'})>\Delta$. Define
	\begin{align}
	L(P_{XX'})\triangleq \frac{|\calT(P_{XX'})|}{|\calT(Q_X)|^2} \label{ta1}.
	\end{align}
	Then, under the condition \eqref{keycond} and $d(P_{XX'})>\Delta$, for any two pairs $(i,j), (k,l) \in [M]_*^2$ such that $(i,j)\neq (k,l)$, it holds that
	\begin{align}
	(1-4\delta_n^2) e^{-2\delta_n} L(P_{XX'}) \leq \bbE[\calI(i,j)]\leq \frac{1}{(1-e^{-n\delta})^2}  L(P_{XX'}) \label{aAp0},
	\end{align} and
	\begin{align}
	\bbE[\calI(i,j)\calI(k,l)]\leq \frac{1}{(1-e^{-n\delta})^4}  L^2(P_{XX'}). \label{aAp1}
	\end{align} 
	This implies that
	\begin{align}
	\bbE[\calI(i,j)]&\doteq \exp\{-nI_P(X;X')\} \label{Ap0},\\
	\bbE[\calI(i,j)\calI(k,l)] &\dotleq \exp\{-2n I_P(X;X')\} \label{Ap1}.
	\end{align}
\end{lemma}

\subsection{Useful Properties of Type Enumerators} 
In this section, we state some important properties of the type enumerator of RGV codebooks. For a given joint-type $P_{XX'} \in \calQ(Q_X)$, the type enumerator $N(P_{XX'})$ is defined as the number of codeword pairs with joint type $P_{XX'}$, i.e., 
\begin{align}
N(P_{XX'})&\triangleq \sum_m \sum_{m'\neq m} \indicator \{(\bx_m,\bx_{m'}) \in \calT(P_{XX'})\}\\
&=\sum_{(m,m') \in [M]_*^2} \calI(m,m'),
\label{eq:type_enum}
\end{align}  where $\calI(m,m')$ is defined in \eqref{defImmprime}.
 %We start with the following lemma. 
\begin{lemma} \label{lem:aux1}  Fix arbitrary small positive numbers $\delta>0$ and $\eps>0$. 
%Define
%	\begin{align}
%	\delta_n\triangleq \frac{e^{-n\delta}}{1-e^{-n\delta}} \label{defdeltan}.
%	\end{align} 
	Let $P_{XX'} \in \calQ(Q_X)$ be a joint distribution that satisfies $I_P(X;X')<2R-\eps$ and $d(P_{XX'})>\Delta$. Define 
	\begin{align}
	\calE(P_{XX'})=\bigg\{\calC_n: N(P_{XX'})< (1-4 \delta_n^2) e^{-2\delta_n}  \exp\{n[2R-I_P(X;X')-\eps]\}\bigg\}.
	\label{eq:defE}
	\end{align}
Then, for any rate $R$ satisfying \eqref{keycond}, it holds (as $n$ sufficiently large) that
	\begin{align}
	\bbP\big[ \calE(P_{XX'})\big]\leq \frac{1}{(1-e^{-n\eps/2})^2}\bigg[\frac{e^{4\delta_n}}{\big(1-4\delta_n^2\big)^2\big(1-e^{-n\delta}\big)^2}e^{-n\eps/2} + \frac{e^{4\delta_n}}{(1-4\delta_n^2)^2(1-e^{-n\delta})^4}-1\bigg] \to 0
	\end{align} as $n\to \infty$ for any fixed $\delta>0$.
\end{lemma}
\begin{IEEEproof}
See Appendix \ref{lem:aux1proof}.	
\end{IEEEproof}
%Next, we prove some other results. 
\begin{lemma} \label{lem:ba1} Let $\eps>0$ be given and assume that the condition \eqref{keycond} holds. Then, for any $P_{XX'} \in \calQ(Q_X)$ such that $I_P(X;X')\leq 2R$ and $d(P_{XX'})>\Delta$,
	\begin{align}
	\PP\big[N(P_{XX'})\geq e^{n(2R-I_P(X;X')+\eps)}\big]&\ddleq \exp\big\{-e^{n(2R-I_P(X;X')+\eps)}\big\} \label{afact}\\
	&\dotleq e^{-n \infty} \label{bfact}. 
	\end{align}
\end{lemma}
\begin{IEEEproof}
See Appendix \ref{lem:ba1proof}.	
\end{IEEEproof}
%Similarly, the following lemma can be proved.
\begin{lemma}\label{lem:ba2}  Let $\eps>0$ be given. Then, for any $P_{XX'} \in \calQ(Q_X)$ such that $I_P(X;X')\geq 2R-\eps$ and $d(P_{XX'})>\Delta$ such that the condition \eqref{keycond} holds,
	\begin{align}
	\bbP\big[N(P_{XX'})\geq e^{n\eps}\big]&\ddleq \exp\big\{-e^{n\eps}\big\}\label{uv}\\
	&\dotleq e^{-n\infty} \label{uv2}.
	\end{align}
\end{lemma}
\begin{IEEEproof}
See Appendix \ref{lem:ba2proof}.	
\end{IEEEproof}
\begin{lemma}\label{lem:ab2} For any $P_{XX'} \in \calQ(Q_X)$ such that $I_P(X;X')\geq 2R$ and $d(P_{XX'})>\Delta$ such that the condition \eqref{keycond} holds, we have
	\begin{align}
	\bbP\big[N(P_{XX'})\geq 1\big]\doteq \exp\{n(2R-I_P(X;X'))\} \label{eq190}.
	\end{align}
\end{lemma}
\begin{IEEEproof}
See Appendix \ref{lem:ab2proof}.
\end{IEEEproof}
The following lemma is a key result for showing the exponentially-decay of the lower tail decay.
\begin{lemma} \label{prop3} Let $P_{XX'} \in \calQ(Q_X)$ such that $d(P_{XX'})>\Delta$. Then, under the condition \eqref{keycond}, we have
	\begin{align}
	\bbP\big[N(P_{XX'})\geq e^{ns}\big]\doteq e^{-n E(R,P_{XX'},s)} \qquad \forall s\in \bbR,
	\end{align}
	where
	\begin{align}
	E(R,P,s)=\begin{cases} [I_P(X;X')-2R]_+,&\qquad [2R-I_P(X;X')]_+> s\\ +\infty,&\qquad [2R-I_P(X;X')]_+<s \end{cases}.
	\end{align}
\end{lemma}
\begin{IEEEproof}
	See a detailed proof in Appendix \ref{proofprop3}.	
\end{IEEEproof}
The following lemma is a key enabling result to attain the double-exponential bound for the concentration properties of the random coding exponent in the RGV codebook. As opposed to the independent fixed-composition ensemble \cite{Tamir2020a}, a direct application of Suen's correlation inequality as ~\cite[Proof of Lemma 2]{Tamir2020a} does not give the double-exponential bound. More specifically, since all RGV codewords are correlated, the number of adjacent pairs of a fixed pair $(m,m')$ is now $e^{2nR}$ which causes the term in \cite[Eq.~(B.18)]{Tamir2020a} to be equal to $1$. For the independent fixed-composition code ensemble, this term is $e^{nR}$.

To overcome this difficulty, we develop another proof for this lemma which is not based on the Suen's correlation inequality. See Appendix \ref{lemSuenproof} for a detailed proof.
\begin{lemma} \label{lemSuen} Let $\eps>0$ and $\calD \subset \{P_{XX'} \in \calQ(Q_X): d(P_{XX'})>\Delta\}$ be given. Then, under the condition 
	\begin{align}
	\min_{P_{XX'} \in \calD} I_P(X;X')-2\delta \leq R\leq  \min_{P_{XX'} \in \calQ(Q_X): d(P_{XX'})\leq \Delta} I_P(X;X')- 2 \delta \label{condkeyb},
	\end{align}
	or
	\begin{align}
	R\leq \min\bigg\{\min_{P_{XX'}\in \calD}I_P(X;X')- \min_{P_{XX'}\in \calQ(Q_X): d(P_{XX'})\leq \Delta}I_P(X;X') ,\min_{P_{XX'} \in \calQ(Q_X): d(P_{XX'})\leq \Delta}I_P(X;X') \bigg\}-2\delta \label{condkeyc}
	\end{align}
	for some $\delta>0$, we have
	\begin{align}
	\min_{P_{XX'} \in \calD} \bbP\bigg\{N(P_{XX'})\leq e^{-n\eps} \bbE[N(P_{XX'})]\bigg\} \ddleq \exp\big\{-\min\big(e^{n(R-2\delta)},e^{n(2R-\min_{P_{XX'}\in D} I_P(X;X'))}\big)\big\} \label{eq220}.
	\end{align} 
\end{lemma}
%\begin{remark} \label{rmk1}

Observe that for $d(P_{XX'})=-I_P(X;X')$ and $\Delta=-(R+2\delta)$, the condition \eqref{condkeyb} holds since 
\begin{align}
\min_{P_{XX'} \in \calD} I_P(X;X')-2\delta &\leq \max_{P_{XX'}: d(P_{XX'})>\Delta} I_P(X;X')-2\delta\\
&=\max_{P_{XX'}: I_P(X;X')<-\Delta} I_P(X;X')-2\delta\\
&<-(\Delta+2\delta),
\end{align}
and
\begin{align}
\min_{P_{XX'}: d(P_{XX'})\leq \Delta} I_P(X;X')-2\delta &=\min_{P_{XX'}: I_P(X;X')\geq -\Delta} I_P(X;X')-2\delta\\
&=-(\Delta+2\delta). 
\end{align}
Hence, the double-exponential expression in \eqref{eq220} holds for this special distance $d$ and $\Delta$. The condition \eqref{condkeyc} also holds for many other classes of distances $d$ and different values of $\Delta$.
%\end{remark}

Finally, we state the following key lemma, whose proof can be found in Appendix \ref{uplem1proof}.
\begin{lemma} \label{uplem1} Recall the definition of $E_{\rm{ex}}^{\rm{rgv}}(R,Q_X,d,\Delta)$\footnote{$E_{\rm{ex}}^{\rm{rgv}}(R,Q_X,d,\Delta)$ is the expurgated error exponent of the RGV code.} in \eqref{defexRGV}. Let 
	\begin{align}
	\calA_1&=\bigg\{P_{XX'} \in \calQ(Q_X): d(P_{XX'})>\Delta, I_P(X;X')>2R  \bigg\} \label{defcalA1},\\
	\calA_2&=\bigg\{P_{XX'} \in \calQ(Q_X): d(P_{XX'})>\Delta, I_P(X;X')\leq 2R,  \Gamma(P_{XX'},R-\eps) +I_P(X;X')-R\leq E_0+\eps  \bigg\} \label{defcalA2},
	\end{align}
and define
	\begin{align}
	\calF_0\triangleq \bigcap_{P_{XX'} \in \calA_1 \cup \calA_2} \big\{ N(P_{XX'})=0 \big\} \label{defF0}.
	\end{align}
	Under the conditions that $R< E_{\rm{ex}}^{\rm{rgv}}(R,Q_X,d,\Delta)$ and
	\begin{align}
	\min_{P_{XX'} \in \calQ(Q_X): d(P_{XX'})\leq \Delta} I_P(X;X') &\geq  \max_{P_{XX'} \in \calQ(Q_X): d(P_{XX'})> \Delta} I_P(X;X') \label{ek1cond},\\
	R &\leq \min_{P_{XX'} \in \calQ(Q_X): d(P_{XX'})\leq \Delta} I_P(X;X')-2 \delta \label{condkeymu}
	\end{align} for some $\delta>0$,  it holds that
	\begin{align}
	\bbP(\calF_0) \ddgeq \exp\big\{-e^{n \max_{P_{XX'} \in \calA_2} (2R-I_P(X;X')\delta)}\big\} \label{ex}.
	\end{align}
\end{lemma}

%\begin{remark} \label{rmk2}
	Similarly to the preceeding discussion, setting $d(P_{XX'})\triangleq -I_P(X;X')$, we obtain that
	\begin{align}
	\min_{P_{XX'} \in \calQ(Q_X): d(P_{XX'})\leq \Delta} I_P(X;X')&\geq -\Delta,\\
	\max_{P_{XX'} \in \calQ(Q_X): d(P_{XX'})> \Delta} I_P(X;X')&<-\Delta,
	\end{align} 
	so \eqref{ek1cond} holds. For \eqref{condkeymu} being hold, it is required that $R\leq -(\Delta+2\delta)$.
%\end{remark}

In connection to Lemma \ref{lemSuen}, the proof of the related result in \cite[Prep.~6]{Tamir2020a} cannot be applied here since it uses the Suen's correlation inequality, i.e. \cite[Fact 3]{Tamir2020a}. Since all codewords in RGV ensemble are dependent, the number of adjacent nodes in the corresponding adjacency graph is too big which makes this type of arguments invalid. To overcome this difficulty, in Appendix \ref{uplem1proof}, we develop a new technique. However, the double-exponential constant in \eqref{ex} is smaller than the one in \cite[Prep.~6]{Tamir2020a} for the fixed-composition code ensemble.

%%%%%%%%%%%%%%%%%%%%%%%%%%%%%%%%%%%%%%%
\section{Typical Random Coding Exponent of Gilbert-Varshamov Codes}
\label{sec:dmc}
In this section, we show an expression for the TRC of the RGV code ensemble. The expression, when optimized over the distance function $d(\cdot,\cdot)$ and minimum distance $\Delta$, recovers the expurgated exponent for the GLD proposed in \cite{Merhav2017a}. The main result, proven in Section \ref{sec:proofth1}, is stated in the following.

\begin{theorem} \label{mainthm1}
Let $Q_X \in \calP(\calX),\Delta \in  \bbR, d \in \Omega$, and define the following quantity
\begin{align}
\Gamma(P_{XX'},R)\triangleq \min_{P_{Y|XX'}}\Big\{D(P_{Y|X}\|W|Q_X)+I_P(X';Y|X)+[\max\{g(P_{XY}),\alpha(R,P_Y)\}-g(P_{X'Y})]_+\Big\},
\label{eq:def_gamma}
\end{align}	
where
\begin{align}
\alpha(R,P_Y)\triangleq \max_{P_{X'|Y}: P_{X'}=Q_X,\atop I_P(X'; Y)\leq R}\big(g(P_{X'Y})-I_P(X';Y)\big)+R \label{defalphathm}. 
\end{align}
Then, for any $R$ satisfying the condition in \eqref{keycond}, the typical random coding exponent of the RGV code ensemble with the GLD is given by
\begin{align}
E_{\rm{trc}}^{\rm rgv}(R,Q_X,d,\Delta)= \min_{P_{X'|X}:P_{X'}=Q_X ,\atop I_P(X;X')\leq 2R, d(P_{XX'})>\Delta} \big\{\Gamma(P_{XX'},R)+I_P(X;X')-R\big\} \label{akathm}.
\end{align}
\end{theorem}

Before proceeding with the proof of the result, some discussion is in order. Observe that if we remove the constraint $d(P_{XX'})>\Delta$ (i.e., no constraint on the distance between each codeword pair), the expression of the TRC for the RGV ensemble code in \eqref{akathm} becomes the TRC of the constant composition code ensemble with composition $Q_X$ under GLD decoding in \cite[Eq.~(18)]{Merhav2018a}. 
In addition, as shown below, when the distance function $d(\cdot,\cdot)$ is optimized, and $\Delta$ is chosen appropriately, the TRC expression \eqref{akathm} recovers the expurgated bound in \cite[Theorem 1]{Merhav2018a} $E_{\rm ex}(R,Q_X)$.

The following results are similar to ones in \cite[Section IV]{somekh_2019}.
\begin{corollary} Let $\eps>0$ be given, and let $R,P$, and $d \in \Omega$ be given. The TRC of the generalized RGV construction with sufficiently small $\delta$, $d(P_{XX'})=-I_P(X;X'), \Delta=-(R+2\delta)$, sufficiently large $n$, and GLD rule is at least as high as $E_{\rm{ex}}(R,Q_X)-\eps$.
\end{corollary}
\begin{IEEEproof} First, it is easy to see that the choices $d(P_{XX'})=-I_P(X;X')$ and $\Delta=-(R+2\delta)$ are valid for all $R$ in the sense of satisfying the rate condition in \eqref{keycond} (see proof of \cite[Cor. 2]{somekh_2019}). Now, under the same choices, we have
	\begin{align}
	&E_{\rm{trc}}^{\rm rgv}(R,Q_X,d,\Delta)\bigg|_{d(P_{XX'})=-I_P(X;X'), \Delta=-(R+2\delta)}\\
	&\qquad =\min_{P_{X'|X}:P_{X'}=Q_X ,\atop I_P(X;X')\leq 2R, I_P(X;X')\leq R+2\delta} \big\{\Gamma(P_{XX'},R)+I_P(X;X')-R\big\}\\
	&\qquad= \min_{P_{X'|X}:P_{X'}=Q_X ,\atop I_P(X;X')\leq R+2\delta} \big\{\Gamma(P_{XX'},R)+I_P(X;X')-R\big\} \label{x0}.
	\end{align}	The result follows by taking $\delta \to 0$ and using the continuity of $E_{\rm{trc}}^{\rm rgv}(R,Q_X,d,\Delta)$ in $R$.
\end{IEEEproof}

The following proposition reveals that the above choice of $(d,\Delta)$ is a choice that maximizes the TRC given in Theorem \ref{mainthm1}.
\begin{lemma} Under the setup of Theorem \ref{mainthm1} with 
	\begin{align}
	R\leq \min_{P_{XX'} \in \calQ(Q_X): d(P_{XX'})\leq \Delta} I_P(X;X')-2\delta  \label{cond108}
	\end{align} for some $\delta>0$, we have
	\begin{align}
	E_{\rm{trc}}^{\rm rgv}(R,Q_X,d,\Delta)\leq E_{\rm{trc}}^{\rm rgv}(R,Q_X,d,\Delta)\bigg|_{d(P_{XX'})=-I_P(X;X'), \Delta=-(R+2\delta)}.
	\end{align}
\end{lemma}
\begin{IEEEproof}
From \eqref{cond108}, for all joint type $P_{XX'} \in \calQ(Q_X)$ such that $d(P_{XX'})\leq \Delta$, we have $R+2\delta \leq I_P(X;X')$. Hence, if $R+2\delta>I_P(X;X')$, it holds that $d(P_{XX'})>\Delta$. This means that
\begin{align}
\bigg\{P_{XX'}\in \calQ(Q_X): I_P(X;X')<R+2\delta\bigg\} \subset \bigg\{P_{XX'}\in \calQ(Q_X): d(P_{XX'})>\Delta \bigg\} \label{AQ0}.
\end{align}
It follows from \eqref{AQ0} that for $\delta$ sufficiently small, 
\begin{align}
E_{\rm{trc}}^{\rm rgv}(R,Q_X,d,\Delta)&= \min_{P_{X'|X}:P_{X'}=Q_X ,\atop I_P(X;X')\leq 2R, d(P_{XX'})>\Delta} \bigg\{\Gamma(P_{XX'},R)+I_P(X;X')-R\bigg\}\\
&\leq  \min_{P_{X'|X}:P_{X'}=Q_X ,\atop I_P(X;X')\leq 2R, I_P(X;X')<R+2\delta} \bigg\{\Gamma(P_{XX'},R)+I_P(X;X')-R\bigg\}\\
&= \min_{P_{X'|X}:P_{X'}=Q_X ,\atop I_P(X;X')<R+2\delta} \bigg\{\Gamma(P_{XX'},R)+I_P(X;X')-R\bigg\}\\
&=E_{\rm{trc}}^{\rm rgv}(R,Q_X,d,\Delta)\bigg|_{d(P_{XX'})=-I_P(X;X'), \Delta=-(R+2\delta)} \label{las},
\end{align} where \eqref{las} follows from the continuity of $E_{\rm{trc}}^{\rm rgv}(R,Q_X,d,\Delta)$ in $R$ and \eqref{x0}.
\end{IEEEproof}

As in \cite{somekh_2019}, the choice $d(P_{XX'})=-I_P(X;X')$ is universally optimal in maximizing the TRC in Theorem \ref{mainthm1} (subject to \eqref{keycond}), in the sense that it does not depend on the channel or input distribution. 

\begin{figure}[htp]
	\centering
\definecolor{mycolor1}{rgb}{0.00000,1.00000,1.00000}%
\definecolor{mycolor2}{rgb}{1.00000,0.00000,1.00000}%
\begin{tikzpicture}

\begin{axis}[%
width=4.521in,
height=3.566in,
at={(0.758in,0.481in)},
scale only axis,
xmin=0,
xmax=0.8,
xlabel style={font=\color{white!15!black}},
xlabel={$R$},
ymin=0,
ymax=1.8,
ylabel={Error Exponents},
axis background/.style={fill=white},
xmajorgrids,
ymajorgrids,
legend style={legend cell align=left, align=left}
]

\addplot [color=mycolor1, dashdotted, line width=1.0pt]
  table[row sep=crcr]{%
0	1.72718894482893\\
0.05	1.23344509608977\\
0.1	1.01879634546913\\
0.15	0.82137807607759\\
0.2	0.683920485053404\\
0.25	0.593861873995012\\
0.3	0.472156780168197\\
0.35	0.391127726172941\\
0.4	0.313902196267716\\
0.45	0.241148250104092\\
0.5	0.173882551585589\\
0.55	0.113838378565535\\
0.6	0.0638383785655351\\
0.65	0.0207482546488783\\
0.7	0.000500250166791782\\
0.75	0.000500250166791782\\
};
\addlegendentry{$E_{\rm ex}(R)$}

\addplot [color=red, line width=1.0pt]
  table[row sep=crcr]{%
0	1.72718894482893\\
0.05	1.06879634546913\\
0.1	0.783920485053404\\
0.15	0.622156780168197\\
0.2	0.513902196267716\\
0.25	0.423882551585589\\
0.3	0.363838378565535\\
0.35	0.313838378565535\\
0.4	0.263838378565535\\
0.45	0.213838378565535\\
0.5	0.163838378565535\\
0.55	0.113838378565535\\
0.6	0.0638383785655351\\
0.65	0.0207482546488783\\
0.7	0.000500250166791782\\
0.75	0.000500250166791782\\
};
\addlegendentry{$E_{\rm trc}(R)$}

\addplot [color=mycolor2, dotted, line width=1.0pt]
  table[row sep=crcr]{%
0	0.000500250166791767\\
0.05	0.0505002501667918\\
0.1	0.100500250166792\\
0.15	0.150500250166792\\
0.2	0.200500250166792\\
0.25	0.250500250166792\\
0.3	0.300500250166792\\
0.35	0.313838378565535\\
0.4	0.263838378565535\\
0.45	0.213838378565535\\
0.5	0.163838378565535\\
0.55	0.113838378565535\\
0.6	0.0638383785655351\\
0.65	0.0207482546488783\\
0.7	0.000500250166791782\\
0.75	0.000500250166791782\\
};
\addlegendentry{$E_0^{\rm min}(R)$}

\addplot [color=black, dashed, line width=1.0pt]
  table[row sep=crcr]{%
0	0.662056516702387\\
0.05	0.612056516702387\\
0.1	0.562056516702387\\
0.15	0.512056516702387\\
0.2	0.462056516702387\\
0.25	0.412056516702387\\
0.3	0.362056516702387\\
0.35	0.312056516702387\\
0.4	0.262056516702387\\
0.45	0.212056516702387\\
0.5	0.162056516702387\\
0.55	0.112056516702387\\
0.6	0.0620565167023868\\
0.65	0.0205482413134113\\
0.7	0.00703325688261087\\
0.75	0.00703325688261087\\
};
\addlegendentry{$E_{\rm r}(R)$}

\addplot [color=green, dashed, line width=1.0pt, mark=x, mark options={solid, green}]
  table[row sep=crcr]{%
0	1.72718894482893\\
0.05	1.23344509608977\\
0.1	1.01879634546913\\
0.15	0.82137807607759\\
0.2	0.683920485053404\\
0.25	0.593861873995012\\
0.3	0.472156780168197\\
0.35	0.391127726172941\\
0.4	0.313902196267716\\
0.45	0.241148250104092\\
0.5	0.173882551585589\\
0.55	0.113838378565535\\
0.6	0.0638383785655351\\
0.65	0.0207482546488783\\
0.7	0\\
0.75	0\\
};
\addlegendentry{$E_{\rm trc}^{\rm rgv}(R)$}

\addplot [color=blue, dotted, line width=1.0pt, mark=+, mark options={solid, blue}]
  table[row sep=crcr]{%
0	1.72718894482893\\
0.05	1.23344509608977\\
0.1	1.01879634546913\\
0.15	0.821378076077591\\
0.2	0.683920485053404\\
0.25	0.593861873995012\\
0.3	0.472156780168197\\
0.35	0.391127726172941\\
0.4	0.313902196267716\\
0.45	0.241148250104092\\
0.5	0.173882551585589\\
0.55	0.113838378565535\\
0.6	0.063838378565535\\
0.65	0.0205482413134113\\
0.7	0\\
0.75	0\\
};
\addlegendentry{$E_{\rm rce}^{\rm rgv}(R)$}

\end{axis}

\end{tikzpicture}%
	\caption{Error Exponents for the $Z$-channels with crossover probability $0.001$.}
	\label{fig:fig11}
\end{figure}
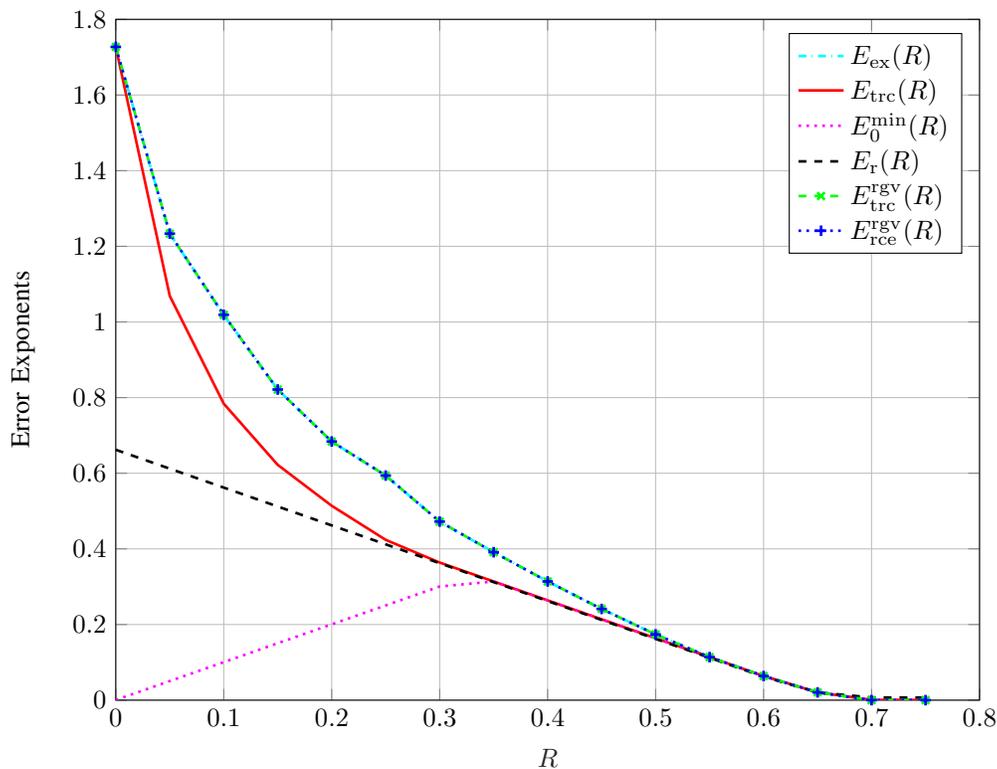

In Fig.~\ref{fig:fig11}, we plot various error exponents for the $Z$-channel with crossover probability $0.001$ and let $Q_X(0)=Q_X(1)=1/2$. This example was considered in \cite{Merhav2017a,Tamir2020a}. Specifically, for reference we plot the random coding exponent $E_{\rm r}(R)$, the expurgated exponent $E_{\rm ex} (R)$, and the TRC $E_{\rm trc}(T)$ for constant composition codes. For the RGV ensemble exponents, we choose $d(P_{XX'})=-I_P(X;X')$ and $\Delta=-R$ so as to achieve the largest possible exponents. We plot the corresponding random coding exponent $E_{\rm rce}^{\rm rgv}(R)$ and its corresponding TRC $E_{\rm trc}^{\rm rgv}(R)$ and illustrate that they both coincide with the expurgated exponent $E_{\rm ex} (R)$.

%%%%%%%%%%%%%%%%%%%%%%%%%%
\subsection{Proof of Theorem \ref{mainthm1}}
\label{sec:proofth1}

The proofs for both upper and lower bounds follow similar lines to those in \cite{Merhav2018a}. The main difference is the dependence among codeword induced by the RGV ensemble. In order to analyze this dependence, we developed new concentration inequalities and applied generalized versions of Hoeffding's inequality. 

%%%%%%%%%%%%%%%%%%%%%%%%%%%%%%%%%%
\subsubsection{Lower bound on TRC}
First, we prove the following result.
\begin{lemma} \label{lem:aut1} Recall the definition of $\alpha(R,P_Y)$ in \eqref{defalphathm}. Fix an $\eps>0$. For any $m \in [M]$, let
\begin{align}
Z_m(\by)\triangleq \sum_{\tilm \neq m} e^{n g(\hatP_{\bX_{\tilm},\by})} \label{defZmy}.
\end{align}
and
\begin{align}
\calA_m\triangleq \{ Z_m(\by)\leq \exp\big\{n\alpha(R-\eps, \hatP_{\by} ) \}\}. 
\end{align}
Then, under the condition \eqref{keycond}, it holds that
\begin{align}
\bbP\big[\calA_m \big] \dotleq \exp\bigg\{- e^{n\eps}\bigg[1- \frac{e^{-n (\eps+\delta)}}{1-e^{-n\delta}}-e^{-n\eps}(1+n\eps)\bigg]  \bigg\}, \qquad \forall m \in [M] \label{pus0}.
\end{align}
\end{lemma}
\begin{IEEEproof}
See Appendix \ref{lem:aut1proof}.
\end{IEEEproof}
\begin{proposition} \label{lem:trclow} Under the same assumptions as Theorem \ref{mainthm1}, the RGV code ensemble satisfies
\begin{align}
E_{\rm{trc}}^{\rm rgv}(R,Q_X,d,\Delta) \geq \min_{P_{XX'}\in \calQ(Q_X),\atop I_P(X;X')\leq 2R, d(P_{XX'})>\Delta} \bigg\{\Gamma(P_{XX'},R)+I_P(X;X')-R\bigg\} \label{aka}.
\end{align}
\end{proposition}
\begin{IEEEproof}
Using the GLD, the error probability is 
\begin{align}
P_{\rme}(\calC_n)=\frac{1}{M}\sum_{m=1}^M \sum_{m'\neq m} \sum_{\by \in \calY^n} W(\by|\bx_m)\frac{\exp\{ng(\hatP_{\bx_{m'},\by})\}}{\exp\{ng(\hatP_{\bx_m,\bY})\}+\sum_{\tilm\neq m}\exp\{{ng(\hatP_{\bx_{\tilm},\by})}\}}.
 \label{eq3b}
\end{align}
From \eqref{eq3b}, we obtain
\begin{align}
\bbE\big[P_{\rme}(\calC_n)\big]  &\leq \bbE \bigg[\frac{1}{M}\sum_{m=1}^M \sum_{\by} W(\by|\bX_m)\sum_{m'\neq m}\min\bigg\{1,\frac{e^{n g(\hatP_{\bX_{m'},\by})}  } {e^{n g(\hatP_{\bX_{m,\by}})}+\sum_{\tilm \neq m} e^{n g(\hatP_{\bX_{\tilm},\by})}}\bigg\}\bigg]\\
& = \bbE\bigg[\frac{1}{M}\sum_{m=1}^M \sum_{\by} W(\by|\bX_m)\sum_{m'\neq m\atop d(\bX_m,\bX_{m'})>\Delta}\min\bigg\{1,\frac{e^{n g(\hatP_{\bX_{m'},\by})}  } {e^{n g(\hatP_{\bX_{m,\by}})}+\sum_{\tilm \neq m} e^{n g(\hatP_{\bX_{\tilm},\by})}}\bigg\}\bigg] \label{eq20},
\end{align} where \eqref{eq20} follows from the fact that $\min_{m'\neq m} d(\bx_m,\bx_{m'})>\Delta$ for any code $\calC_n=(\bx_1,\bx_2,\cdots,\bx_M)$ in the RGV codebook ensemble.

Now, we use similar arguments as \cite{Merhav2018a} with some changes to cooperate the condition $d(x_m,x_{\tilm})>\Delta$ in the sum in \eqref{eq20}.
From \eqref{eq20} and Lemma \ref{lem:aut1}, for any $\eps>0$, we obtain
\begin{align}
\bbE[P_{\rme}(\calC_n)] \leq \bbE \bigg[\frac{1}{M}\sum_{m=1}^M \sum_{\by} W(\by|\bX_m)\sum_{m'\neq m\atop d(\bX_m,\bX_{m'})>\Delta}\min\bigg\{1,\frac{e^{n g(\hatP_{\bX_{m'},\by})}  } {e^{n g(\hatP_{\bX_{m,\by}})}+e^{n \alpha(R-\eps, \hatP_{\by} )}}\bigg\}\bigg] \label{eq21}.
\end{align} 
From the method of types \cite{Csis00} we have that
\begin{align}
W(\by|\bx_{\tilm})=e^{-n\big[H(\hatP_{\bx_{\tilm},y})-H(Q_X)+D\big(\hatP_{\bx_{\tilm},y}\|Q_X \times W \big) \big]} \label{eq5b}.
\end{align} 
Thus, it follows from \eqref{eq5b} that
\begin{align}
&\sum_{m=1}^M \sum_{\by} W(\by|\bx_m)\sum_{m'\neq m\atop d(\bx_m,\bx_{m'})>\Delta}\min\bigg\{1,\frac{e^{n g(\hatP_{\bx_{m'},\by})}  } {e^{n g(\hatP_{\bx_{m,\by}})}+e^{n \alpha(R-\eps, \hatP_{\by} )}}\bigg\}\\
&\qquad \doteq  \sum_{m=1}^M\sum_{\by} W(\by|\bx_m)\sum_{m'\neq m\atop d(\bx_m,\bx_{m'})>\Delta} \exp\Big\{-n\big[\max\{g(\hatP_{\bx_m,\by}),\alpha(R-\eps,\hatP_{\by})\}-g(\hatP_{\bx_{m'},\by})\big]_+\Big\}\\
%&\qquad \doteq   \sum_{m=1}^M\sum_{\by} \sum_{m'\neq m\atop d(\bx_m,\bx_{m'})>\Delta} W(\by|\bx_m) \exp\Big\{-n\big[\max\{g(\hatP_{\bx_m,\by}),\alpha(R-\eps,\hatP_{\by})\}-g(\hatP_{\bx_{m'},\by})\big]_+\Big\}\\
&\qquad =  \sum_{m=1}^M\sum_{\by}\sum_{m'\neq m\atop d(\bx_m,\bx_{m'})>\Delta}\exp\big\{\big(-n\big[H(\hatP_{\bx_{m,y}})-H(Q_X)+D\big(\hatP_{\bx_m,y}\|Q_X \times W \big) \big]\big)\big\}\nn\\
&\qquad \qquad \qquad\qquad\qquad\qquad\qquad \times  \exp\Big\{-n\big[\max\{g(\hatP_{\bx_m,\by}),\alpha(R-\eps,\hatP_{\by})\}-g(\hatP_{\bx_{m'},\by})\big]_+\Big\} \label{A1}\\
&\qquad \doteq \sum_{P_{XX'}\in \calQ(Q_X): d(P_{XX'})>\Delta} N(P_{XX'})\sum_{P_{Y|XX'}}\exp\big\{n H_P(Y|XX')\big\}\nn\\
&\qquad \qquad \qquad\qquad\qquad\qquad\qquad \times  \exp\Big\{\big(-n\big[H(P_{XY})-H(Q_X)+D\big(P_{XY}\|Q_X \times W \big) \big]\big)\Big\}\nn\\ 
&\qquad \qquad \qquad\qquad\qquad\qquad\qquad \times \exp\Big\{-n[\max\{g(P_{XY}),\alpha(R-\eps,P_Y)\}-g(P_{X'Y})]_+\Big\}\\
&\qquad\doteq \sum_{P_{XX'}\in \calQ(Q_X): d(P_{XX'})>\Delta} N(P_{XX'})\exp\Big\{-n \min_{P_{Y|XX'}}\Big(-H_P(Y|XX')+H(P_{XY})-H(Q_X)\nn\\
&\qquad \qquad \qquad \qquad\qquad\qquad+D\big(P_{XY}\|Q_X \times W \big) +\big[\max\{g(P_{XY}),\alpha(R-\eps,P_Y)\}-g(P_{X'Y})\big]_+\Big)\Big\}\\
&\qquad\doteq \sum_{P_{XX'}\in \calQ(Q_X): d(P_{XX'})>\Delta} N(P_{XX'})\exp\Big\{-n \min_{P_{Y|XX'}}\big(D(P_{Y|X}\|W|Q_X)+I_P(X';Y|X)\nn\\
&\qquad \qquad \qquad \qquad\qquad\qquad +\big[\max\{g(P_{XY}),\alpha(R-\eps,P_Y)\}-g(P_{X'Y})\big]_+\big) \Big\} \label{A2}\\
&\qquad=\sum_{P_{XX'}\in \calQ(Q_X): d(P_{XX'})>\Delta} N(P_{XX'})\exp\{-n \Gamma(P_{XX'},R-\eps)\} \label{A3},
\end{align} where \eqref{A1} follows from \eqref{eq5b}, and \eqref{A3} follows from \eqref{eq:def_gamma},
where the joint type enumerator $N(P_{XX'})$ has been defined in \eqref{eq:type_enum}.
From \eqref{eq21}, \eqref{A3}, and \eqref{eq:type_enum}, we obtain
\begin{align}
\bbE[\log  P_{\rme}(\calC_n)] &\leq \log  \Big(\bbE\big[P_{\rme}(\calC_n)\big]\Big)\label{AF3}\\
& \leq \log  \bigg(\sum_{P_{XX'} \in \calQ(Q_X): d(P_{XX'})>\Delta} \bbE\big[N(P_{XX'})\big] \exp\big\{-n \Gamma(P_{XX'},R)\big\}
\bigg)-nR \label{A5},
\end{align} where \eqref{AF3} follows from the concavity of $\log  x$ in $(0,\infty)$ and Jensen's inequality.

Now, for any $P_{XX'} \in \calQ(Q_X)$ such that $d(P_{XX'})>\Delta$, from Lemma \ref{aux1:lem},we obtain
\begin{align}
\bbE\big[N(P_{XX'})\big]&=\sum_{m=1}^M \sum_{m'\neq m}\bbP\big[(\bX_m,\bX_{m'}) \in \calT(P_{XX'})\big]\\
&\doteq e^{n(2R-I_P(X;X'))} \label{basa}.
\end{align}
Hence, from \eqref{A5} and \eqref{basa}, we obtain 
\begin{align}
&\bbE[\log  P_{\rme}(\calC_n)]\dotleq \log  \bigg(\sum_{P_{XX'} \in \calQ(Q_X): d(P_{XX'})>\Delta}  e^{n(2R-I_P(X;X'))} \exp\big\{-n \Gamma(P_{XX'},R)\big\} \bigg)-nR\label{mage}.
\end{align} 
From \eqref{mage}, we finally have
\begin{align}
E_{\rm{trc}}^{\rm rgv}(R,Q_X,d,\Delta) \geq \min_{P_{XX'}:P_{X'}=P_X ,\atop I_P(X;X')\leq 2R, d(P_{XX'})>\Delta} \bigg\{\Gamma(P_{XX'},R)+I_P(X;X')-R\bigg\}\label{C1}.
\end{align} This concludes the proof of Proposition \ref{lem:trclow}.
\end{IEEEproof}

%%%%%%%%%%%%%%%%%%%%%%%%%%%%%%
\subsubsection{Upper bound on TRC}
\begin{proposition} \label{lem:upTRC} Under the same assumptions as Theorem \ref{mainthm1}, the RGV code ensemble satisfies
	\begin{align}
	E_{\rm{trc}}^{\rm{rgv}}(R,Q_X,d,\Delta)\leq \min_{P_{XX'}\in \calQ(Q_X):\atop I_P(X;X')\leq 2R, d(P_{XX'})>\Delta} \bigg\{\Gamma(P_{XX'},R)+I_P(X;X')-R\bigg\} \label{akab}.
	\end{align}
\end{proposition}
\begin{IEEEproof}
The following proof follows similar lines to the proof in \cite[Sect.~5.2]{Merhav2018a}. However, the same proof cannot be used for the RGV ensemble. In addition to the difference in proofs of Lemmas \ref{lem:aux1} and \eqref{eqv}, we also need to make additional changes in since the decay rate of $\bbP[\calE(P_{XX'})]$ in Lemma \ref{lem:aux1} is not exponential as \cite[Eq.~(48)]{Merhav2018a}.
 
Given a joint-type $P_{XX'} \in \calQ(Q_X)$ such that  $I_P(X;X')<2R-\eps$ and $d(P_{XX'})>\Delta$, let us define
\begin{align}
Z_{mm'}(\by)=\sum_{\tilm \neq m, m'} \exp\{ng(\hatP_{\bX_{\tilm},\by})\},
\end{align}
and
\begin{align}
&\calG_n(P_{Y|XX'}) =\bigg\{\calC_n: \sum_m \sum_{m'\neq m} \calI(m,m')\sum_{\by \in \calT(P_{Y|XX'})}\indicator \{ Z_{mm'}(\by)\leq \exp\{n[\alpha(R+2\eps,P_Y)+\eps] \}\}\nn\\
&\qquad \qquad \qquad \qquad \qquad  \geq (1-4 \delta_n^2) e^{-2\delta_n}\exp\{n[2R-I_P(X;X')-3\eps/2]\}.|\calT(P_{Y|XX'})|\bigg\} \label{defGPYX}.
\end{align} 
where $\calI(m,m')$ is defined in \eqref{defImmprime}.
Recall the definition of $\calE(P_{XX'})$ in Eq. \eqref{eq:defE} Lemma \ref{lem:aux1}. Then, similarly to \cite[Sect.~5.2]{Merhav2018a} we have
\begin{align}
&\bbP\big[\calG_n^c(P_{Y|XX'})\cap \calE^c(P_{XX'})\big]\nn\\
&\leq \bbP\bigg[ \sum_m \sum_{m'\neq m} \calI(m,m')\sum_{\by \in \calT(P_{Y|XX'})}\indicator \{ Z_{mm'}(\by)\leq (1-4 \delta_n^2) e^{-2\delta_n} \exp\{n[\alpha(R+2\eps,P_Y)+\eps] \}\}\nn\\
&\qquad \qquad \qquad  \leq \exp\{n[2R-I_P(X;X')-3\eps/2]\}.|\calT(P_{Y|XX'})|,\nn\\
&\qquad \qquad \qquad  N(P_{XX'})\geq  (1-4 \delta_n^2) e^{-2\delta_n}  \exp\{n[2R-I_P(X;X')-\eps]\} \bigg]\\
&\leq \bbP\bigg[ \sum_m \sum_{m'\neq m} \calI(m,m')\sum_{\by \in \calT(P_{Y|XX'})}\indicator \{ Z_{mm'}(\by)> (1-4 \delta_n^2) e^{-2\delta_n} \exp\{n[\alpha(R+2\eps,P_Y)+\eps] \}\}\nn\\
&\qquad \qquad \qquad  \geq \big(\exp\{n[2R-I_P(X;X')-\eps]\}-\exp\{n[2R-I_P(X;X')-3\eps/2]\}\big).|\calT(P_{Y|XX'})|,\nn\\
&\qquad \qquad \qquad   N(P_{XX'})   \geq (1-4 \delta_n^2) e^{-2\delta_n}  \exp\{n[2R-I_P(X;X')-\eps]\} \bigg]\\
&\leq \bbP\bigg[ \sum_m \sum_{m'\neq m} \calI(m,m') \sum_{\by \in \calT(P_{Y|XX'})}\indicator \{ Z_{mm'}(\by)> (1-4 \delta_n^2) e^{-2\delta_n} \exp\{n[\alpha(R+2\eps,P_{XY})+\eps] \}\}\nn\\
&\qquad \qquad \qquad  \geq \big(\exp\{n[2R-I_P(X;X')-\eps]\}-\exp\{n[2R-I_P(X;X')-3\eps/2]\}\big).|\calT(P_{Y|XX'})|\bigg]\\
&\dotleq \frac{\bbE\big[\sum_m \sum_{m'\neq m} \calI(m,m')\sum_{\by \in \calT(P_{Y|XX'})}\indicator \{ Z_{mm'}(\by)> (1-4 \delta_n^2) e^{-2\delta_n} \exp\{n[\alpha(R+2\eps,P_Y)+\eps] \}\}\big]}{\exp\{n[2R-I_P(X;X')-\eps]\} |\calT(P_{Y|XX'})|}\label{eq:markovlong}\\
&= \frac{\sum_m \sum_{m'\neq m} \sum_{\by \in \calT(P_{Y|XX'})} \zeta(m,m',\by)}{\exp\{n[2R-I_P(X;X')-\eps]\} |\calT(P_{Y|XX'})|} \label{mat4},
\end{align} 
where \eqref{eq:markovlong} follows from Markov's inequality and
\begin{align}
\zeta(m,m',\by)&\triangleq \bbP\big[(\bX_m,\bX_{m'}) \in \calT(P_{XX'}), Z_{mm'}(\by)> (1-4 \delta_n^2) e^{-2\delta_n} \exp\{n[\alpha(R+2\eps,P_Y)+\eps] \}\big]\\
&=\sum_{(\bx_m,\bx_{m'}) \in \calT(P_{XX'}):\atop d(\bx_m,\bx_{m'})>\Delta}\bbP(\bx_m,\bx_{m'})\nn\\
&\qquad \times \bbP\big[Z_{mm'}(\by)> (1-4 \delta_n^2) e^{-2\delta_n} \exp\{n[\alpha(R+2\eps,P_Y)+\eps]\}\big|\bX_m=\bx_m,\bX_{m'}=\bx_{m'} \big] \label{B1}
\end{align}
%Observe that
%\begin{align}
%&\zeta(m,m',\by)\nn\\
%&\qquad=\sum_{(\bx_m,\bx_{m'}) \in \calT(P_{XX'})}\bbP(\bx_m,\bx_{m'})\bbP\big[Z_{mm'}(\by)> (1-4 \delta_n^2) e^{-2\delta_n} \exp\{n[\alpha(R+2\eps,P_Y)+\eps]\} \big]\\
%&\qquad=\sum_{(\bx_m,\bx_{m'}) \in \calT(P_{XX'}):\atop d(\bx_m,\bx_{m'})>\Delta}\bbP(\bx_m,\bx_{m'})\bbP\big[Z_{mm'}(\by)> (1-4 \delta_n^2) e^{-2\delta_n} \exp\{n[\alpha(R+2\eps,P_Y)+\eps]\} \big] \label{B1}
%\end{align} 
where \eqref{B1} follows from the fact that $\bbP(\bx_m,\bx_{m'})=0$ if $d(\bx_m,\bx_{m'})<\Delta$ by Lemma \ref{lem2some}.

Now, given a fixed pair $(\bx_m,\bx_{m'})$ such that $d(\bx_m,\bx_{m'})>\Delta$, define
\begin{align}
P_{X'|Y}^*&\triangleq \argmax_{P_{X'|Y}} \bbP\big[ N(P_{X'|Y})> (n+1)^{-|\calX||\calY|} (1-4 \delta_n^2) e^{-2\delta_n}\nn\\
&\qquad \qquad \times \exp\{n[\alpha(R+2\eps,P_Y)+\eps-g(P_{X' Y})]\}  \big|\bX_m=\bx_m,\bX_{m'}=\bx_{m'}\big] \label{B20},
\end{align} 
where
\begin{align}
N(P_{X'|Y}):=\sum_{\tilm\neq m,m'} \indicator\big\{(\bX_{\tilm},\by) \in \calT(P_{\tilX Y})\big\}.
\end{align}
Then, we have
\begin{align}
&\bbP\big[Z_{mm'}(\by)> (1-4 \delta_n^2) e^{-2\delta_n} \exp\{n[\alpha(R+2\eps,P_Y)+\eps]\} \big|\bX_m=\bx_m,\bX_{m'}=\bx_{m'}\big]\nn\\
& =\bbP\bigg[\sum_{\tilm \neq m,m'} \exp\{ng(\hatP_{\bX_{\tilm},\by})\}\nn\\
&\qquad \qquad > (1-4 \delta_n^2) e^{-2\delta_n} \exp\{n[\alpha(R+2\eps,P_Y)+\eps]\} \big|\bX_m=\bx_m,\bX_{m'}=\bx_{m'}\bigg]\\
&\leq \bbP\bigg[\sum_{P_{X'|Y}}N(P_{X'| Y})\exp\{ng(P_{X' Y})\}\nn\\
&\qquad \qquad >  (1-4 \delta_n^2) e^{-2\delta_n} \exp\{n[\alpha(R+2\eps,P_Y)+\eps]\} \big|\bX_m=\bx_m,\bX_{m'}=\bx_{m'} \bigg]\\
&= \bbP\bigg[\sum_{P_{X'|Y}}N(P_{X' Y})\exp\{ng(P_{X'Y})\}\nn\\
&\qquad \qquad >  (1-4 \delta_n^2) e^{-2\delta_n} \exp\{n[\alpha(R+2\eps,P_Y)+\eps]\}\big|\bX_m=\bx_m,\bX_{m'}=\bx_{m'} \bigg]\\
&\doteq  \max_{P_{X'|Y}} \bbP\big[ N(P_{X'|Y})\exp\{ng(P_{X' Y})\}\nn\\
&\quad \quad > (n+1)^{-|\calX||\calY|} (1-4 \delta_n^2) e^{-2\delta_n} \exp\{n[\alpha(R+2\eps,P_Y)+\eps]\}  \big|\bX_m=\bx_m,\bX_{m'}=\bx_{m'}\big]\label{eq:proptypeenum}\\
&= \max_{P_{X'|Y}} \bbP\big[ N(P_{X'|Y})> (n+1)^{-|\calX||\calY|} (1-4 \delta_n^2) e^{-2\delta_n}\nn\\
&\qquad \qquad  \times \exp\{n[\alpha(R+2\eps,P_Y)+\eps-g(P_{X'Y})]\}  \big|\bX_m=\bx_m,\bX_{m'}=\bx_{m'}\big]\\
&= \bbP\big[ N(P^*_{X'|Y})> (n+1)^{-|\calX||\calY|} (1-4 \delta_n^2) e^{-2\delta_n}\nn\\
&\qquad \qquad  \times \exp\{n[\alpha(R+2\eps,P_Y)+\eps-g(P^*_{X'Y})]\}  \big|\bX_m=\bx_m,\bX_{m'}=\bx_{m'}\big] \label{B2},
\end{align}  
where \eqref{B2} follows from \eqref{B20}.\\
Now, for all $\tilm \in [M]$, observe that
\begin{align}
\bbP\big[(\bX_{\tilm},\by)\in \calT(P_{X'Y})\big] &=\sum_{\bx_{\tilm} \in \calT(P_{X'|Y})}\bbP(\bx_{\tilm}) \\
&= \frac{|\calT(P_{X'|Y})}{|\calT(Q_X)|}\label{amett}\\
&:=p \label{defp},
\end{align} where \eqref{amett} follows from Lemma \cite[Lemma 4]{somekh_2019}. It is easy to see that $p$ does not depend on $\tilm$.

Now, we consider two cases:

Case 1: $I_{P^*}(X';Y)\leq R+2\eps$. Then, we have
\begin{align}
\alpha(R+2\eps, P_Y)+\eps-g(P_{X'Y}^*)&=\max_{P_{X'|Y}: P_{X'}=Q_X,\atop I_P(X'; Y)\leq R+2\eps}\big(g(P_{X'Y})-I_P(X';Y)\big)+R+ 2\eps -g(P_{X'Y}^*)\\
&\geq g(P_{X'Y}^*)-I_{P^*}(X';Y)+R+ 2\eps -g(P_{X'Y}^*)\\
&= R+2\eps- I_{P^*}(X';Y) \label{X1}.
\end{align}
On the other hand, if we let 
\begin{align}
\gamma\triangleq  \frac{p}{1-e^{-n\delta}},
\end{align}
we have
\begin{align}
(M-2)\gamma&\doteq \frac{e^{n(R- I_{P^*}(X';Y))}}{1-e^{-n \delta}} \label{v0}
\end{align}
It follows that
\begin{align}
&\bbP\big[ N(P_{X'Y}^*)> (n+1)^{-|\calX||\calY|} (1-4 \delta_n^2) e^{-2\delta_n} \exp\{n[\alpha(R+2\eps,P_Y)+\eps-g(P_{X'Y}^*)]\}  \big|\bX_m=\bx_m,\bX_{m'}=\bx_{m'}\big]\nn\\
&\qquad \dotleq \bbP\big[ N(P_{X'Y}^*)> (M-2)\gamma e^{2n\eps}\big|\bX_m=\bx_m,\bX_{m'}=\bx_{m'}\big] \label{v1}
\end{align}
where the last step follows from \eqref{v0} and \eqref{X1}.
Now, let $Z_{\tilm}\triangleq \indicator \{(\bX_{\tilm},\by) \in \calT(P_{X'Y})\}$. Then, for all $\calA \subset [M]\setminus \{m,m'\}$, under the condition \eqref{keycond}, by Lemma \ref{lem:aux0}, it holds that
\begin{align}
&\bbE\bigg[\prod_{\tilm \in \calA} Z_{\tilm}\big|\bX_m=\bx_m,\bX_{m'}=\bx_{m'}\bigg]\nn\\
&\qquad =\sum_{\bx_1,\bx_2,\cdots,\bx_{|\calA|}}\bbP\bigg[\bigcap_{\tilm \in \calA} \{\bX_{\tilm}=\bx_{\tilm}\}\big|\bX_m=\bx_m,\bX_{m'}=\bx_{m'}\bigg]\prod_{\tilm \in \calA}\indicator \{(\bx_{\tilm},\by) \in \calT(P_{X'Y})\}.
\end{align}
Now, observe that
\begin{align}
&\bbP\bigg[\bigcap_{\tilm \in \calA} \{\bX_{\tilm}=\bx_{\tilm}\}\big|\bX_m=\bx_m,\bX_{m'}=\bx_{m'}\bigg]\nn\\
&\qquad = \frac{\bbP\big(\bigcap_{\tilm \in \calA\cup \{m,m'\}} \{\bX_{\tilm}=\bx_{\tilm}\}\big)}{\bbP\big(\bX_m=\bx_m,\bX_{m'}=\bx_{m'}\big)}\\
&\qquad \leq \frac{1}{(1-e^{-\delta n})^{|\calA|+2}}\bigg(\frac{1}{|\calT(Q_X)|^{|\calA|+2}}\bigg)\frac{|\calT(Q_X)|^2}{1-4\delta_n^2} e^{2\delta_n} \label{maku1}
\end{align} where \eqref{maku1} follows from Lemma \ref{lem:aux0} and Lemma \ref{lem2some} with noting that $d(\bx_m,\bx_{m'})>\Delta$.

Hence, it holds that
\begin{align}
&\bbE\bigg[\prod_{\tilm \in \calA} Z_{\tilm}\big|\bX_m=\bx_m,\bX_{m'}=\bx_{m'}\bigg]\nn\\
&\leq \bigg(\frac{e^{2\delta_n}}{1-4\delta_n^2}\bigg)\bigg(\frac{1}{(1-e^{-\delta n})^{|\calA|+2}}\bigg)\sum_{\bx_1,\bx_2,\cdots,\bx_{|\calA|}}\prod_{\tilm \in \calA} \bbP[\bX_{\tilm}=\bx_{\tilm}]\prod_{\tilm \in \calA}\indicator \{(\bx_{\tilm},\by) \in \calT(P_{X'Y})\} \label{D0} \\
&=\bigg(\frac{e^{2\delta_n}}{1-4\delta_n^2}\bigg)\bigg(\frac{1}{(1-e^{-\delta n})^{|\calA|+2}}\bigg)\prod_{\tilm \in \calA}\sum_{\bx_{\tilm}} \bbP[\bX_{\tilm}=\bx_{\tilm}]\indicator \{(\bx_{\tilm},\by) \in \calT(P_{X'Y})\} \label{D01} \\
&=\bigg(\frac{e^{2\delta_n}}{1-4\delta_n^2}\bigg)\bigg(\frac{1}{(1-e^{-\delta n})^{|\calA|+2}}\bigg) \prod_{\tilm \in \calA} \bbP\big[(\bX_{\tilm},\by)\in \calT(P_{X'Y})\big]\\
&\doteq \bigg(\frac{p}{1-e^{-\delta n}}\bigg)^{|\calA|}  \label{D2},
\end{align} where \eqref{D0} follows from Lemma \ref{lem:aux0} (under the condition \eqref{keycond}).
Hence, by applying Lemma \ref{lemkey2}, we have
\begin{align}
\bbP\big[ N(P_{X'Y}^*)> (M-2)\gamma e^{2n\eps}\big|\bX_m=\bx_m,\bX_{m'}=\bx_{m'}\big]\dotleq \exp\bigg\{-e^{nR}D(e^{-na}\| e^{-nb}) \bigg\} \label{g1}
\end{align} 
where $D(p\|q)$ is the relative entropy between two Bernouilli distributions, with success probability $p, q$, respectively, and  $a\triangleq I_{P^*}(X';Y)-2\eps+(1/n)\log (1-e^{-n\delta})$ and $b\triangleq I_{P^*}(X';Y)+(1/n)\log(1-e^{-n\delta})$. Since $b-a= 2\eps$, by using the following fact \cite[Sec.~ 6.3]{CIT-052}:
\begin{align}
D(a\|b)\geq a \log \frac{a}{b}+b-a,
\end{align} we have
\begin{align}
D\big(e^{-an}\|e^{-bn}\big)&\geq  e^{-bn}\big[1+e^{(b-a)n} ((b-a)n-1)\big]\\
&\doteq e^{-n I_{P^*}(X';Y)} e^{2n\eps} 2n\eps  \label{g2}.
\end{align}
From \eqref{g1} and \eqref{g2}, for any pair $(\bx_m,\bx_{m'})$ such that $d(\bx_m,\bx_{m'})>\Delta$, we obtain
\begin{align}
\bbP\big[ N(P_{X'Y}^*)> (M-2)\gamma e^{2n\eps}\big|\bX_m=\bx_m,\bX_{m'}=\bx_{m'}\big] &\dotleq \exp\bigg\{-e^{n(R- I_{P^*}(X';Y))} e^{2n\eps} 2n\eps\bigg\}\\
&\leq \exp\bigg\{-e^{-2 n\eps} e^{2n\eps} 2n\eps\bigg\}\label{lan1}\\
&= \exp\big\{-2n\eps\big\} \label{eqv},
\end{align} where \eqref{lan1} follows from the condition $I_{P^*}(X';Y)\leq R+2\eps$.

Case 2: $I_{P^*}(X';Y)>R+2\eps$. For this case, for any pair $(\bx_m,\bx_{m'})$ such that $d(\bx_m,\bx_{m'})>\Delta$, we have
\begin{align}
&\bbP\big[ N(P_{X'Y}^*)> (n+1)^{-|\calX||\calY|} (1-4 \delta_n^2) e^{-2\delta_n} \exp\{n[\alpha(R+2\eps,P_Y)+\eps-g(P_{X'Y}^*)]\}  \big|\bX_m=\bx_m,\bX_{m'}=\bx_{m'}\big]\nn\\
&\qquad \leq \bbP\big[ N(P_{X'Y}^*)\geq 1 \big|\bX_m=\bx_m,\bX_{m'}=\bx_{m'}\big]\label{messy}\\
&\qquad \leq \bbE[ N(P_{X'Y}^*)\big|\bX_m=\bx_m,\bX_{m'}=\bx_{m'}]\label{messy1}\\
&\qquad= \sum_{\tilm \neq m,m'} \bbP\big((\bX_{\tilm},\by) \in \calT(P_{X'Y})\big|\bX_m=\bx_m,\bX_{m'}=\bx_{m'}\big) \label{amu},
\end{align} where \eqref{messy} follows from the fact that $N(P_{X'Y}^*) \in \bbZ_{+}$, and \eqref{messy1} follows from the Markov's inequality.

Now, by using \eqref{maku1} with $\calA=\{\tilm\}$, we have
\begin{align}
\bbP\bigg[ \{\bX_{\tilm}=\bx_{\tilm}\}\big|\bX_m=\bx_m,\bX_{m'}=\bx_{m'}\bigg]&\leq \frac{1}{(1-e^{-\delta n})^3}\bigg(\frac{1}{|\calT(Q_X)|^3}\bigg)\frac{|\calT(Q_X)|^2}{1-4\delta_n^2} e^{2\delta_n} \\
& \doteq \frac{1}{|\calT(Q_X)|} \label{maku1b}.
\end{align}
From \eqref{amu} and \eqref{maku1b}, for any pair $(\bx_m,\bx_{m'})$ such that $d(\bx_m,\bx_{m'})>\Delta$, we obtain
\begin{align}
&\bbP\big[ N(P_{X'Y}^*)> (n+1)^{-|\calX||\calY|} (1-4 \delta_n^2) e^{-2\delta_n} \exp\{n[\alpha(R+2\eps,P_Y)+\eps-g(P_{X'Y}^*)]\}  \big|\bX_m=\bx_m,\bX_{m'}=\bx_{m'}\big]\nn\\
&\qquad \dotleq (M-2)p\\
&\qquad \doteq e^{n (R-I_{P^*}(X';Y))} \label{messy2}\\
&\qquad \leq e^{-2n\eps} \label{v2},
\end{align} where  \eqref{messy2} follows from \eqref{defp}, and \eqref{v2} follows from condition $I_{P^*}(X';Y)>R+2\eps$. 

From \eqref{eqv} and \eqref{v2}, for any pair $(\bx_m,\bx_{m'})$ such that $d(\bx_m,\bx_{m'})>\Delta$, we have
\begin{align}
&\bbP\big[ N(P_{X'Y}^*)> (n+1)^{-|\calX|} (1-4 \delta_n^2) e^{-2\delta_n} \exp\{n[\alpha(R+2\eps,P_Y)+\eps-g(P_{X'Y}^*)]\}  \big|\bX_m=\bx_m,\bX_{m'}=\bx_{m'}\big]\nn\\
&\qquad \leq e^{-2n\eps}  \label{v}.
\end{align}
From \eqref{B2} and \eqref{v}, we obtain
\begin{align}
\bbP\big[Z_{mm'}(\by)> (1-4 \delta_n^2) e^{-2\delta_n} \exp\{n[\alpha(R+2\eps,P_Y)+\eps]\} \big|\bX_m=\bx_m, \bX_{m'}=\bx_{m'}\big]\dotleq  e^{-2n\eps}\label{v10}
\end{align} where the constant in $\dotleq$ does not depends on $\bx_m,\bx_{m'}$.

It follows from \eqref{B1} and \eqref{v10} that
\begin{align}
&\bbP\big[(\bX_m,\bX_{m'}) \in \calT(P_{XX'}), Z_{mm'}(\by)> (1-4 \delta_n^2) e^{-2\delta_n} \exp\{n[\alpha(R+2\eps,P_Y)+\eps] \}\big]\nn\\
&\qquad \dotleq \sum_{(\bx_m,\bx_{m'}) \in \calT(P_{XX'})\atop d(\bx_m,\bx_{m'})>\Delta}\bbP(\bx_m,\bx_{m'})e^{-2n\eps}\\
&\qquad \leq \sum_{(\bx_m,\bx_{m'}) \in \calT(P_{XX'})\atop d(\bx_m,\bx_{m'})>\Delta} \bigg(\frac{1}{1-e^{-n\delta}}\bigg)^2 \frac{1}{|\calT(Q_X)|^2}e^{-2n\eps}\label{x10}\\
&\qquad \dotleq e^{-n I_P(X;X')} e^{-2n\eps} \label{x11},
\end{align} where \eqref{x10} follows from Lemma \ref{lem:aux0}.
By combining \eqref{mat4} and \eqref{x11}, we obtain
\begin{align}
\bbP\big[\calG_n^c(P_{Y|XX'})\cap \calE^c(P_{XX'})\big]\dotleq e^{-2n\eps} \label{key1}.
\end{align}
On the other hand, by Lemma \ref{lem:aux1}, we also have
\begin{align}
\Pr[\calE^c(P_{XX'})]\to 1 \label{key2}.
\end{align} 
Now, for any fixed joint-type $P_{XX'} \in \calQ(Q_X)$ such that  $I_P(X;X')<2R-\eps$, define
\begin{align}
\calF_n(P_{XX'})\triangleq \bigcap_{P_{Y|XX'}}\big\{\calG_n(P_{Y|XX'}) \cap \calE^c(P_{XX'})\big\} \label{key0}.
\end{align}
Then, from \eqref{key1} and \eqref{key2}, for any fixed joint-type $P_{XX'} \in \calQ(Q_X)$ such that  $I_P(X;X')<2R-\eps$, we have
\begin{align}
\bbP[\calF_n^c(P_{XX'})]&=  \bbP\bigg[ \bigcup_{P_{Y|XX'}}\{\calG_n^c(P_{Y|XX'})\cap \calE^c(P_{XX'})\} \cup \calE(P_{XX'}) \bigg]\\
&\leq \bbP\bigg[ \bigcup_{P_{Y|XX'}}\{\calG_n^c(P_{Y|XX'})\cap \calE^c(P_{XX'})\} \bigg]+\bbP[\calE(P_{XX'})]\\
&\leq \sum_{P_{Y|XX'}}\bbP\big[\calG_n^c(P_{Y|XX'})\cap \calE^c(P_{XX'})\big]+ \bbP[\calE(P_{XX'})]\\
&\dotleq |\calT(P_{Y|XX'})| e^{-2n\eps} +o(1)\\
&\to 0 \label{key4},
\end{align} 
which leads to $\bbP[\calF_n(P_{XX'})]\to 1$ as $n\to \infty$. 

Now, for a given code $C_n \in \calF_n(P_{XX'})$, define
\begin{align}
\calV(C_n,P_{Y|XX'})=\{(m,m',\by): Z_{mm'}(\by)\leq \exp[n(\alpha(R+2\eps,P_Y)+\eps)]\},
\end{align}
and
\begin{align}
\calV_{m,m'}(C_n,P_{Y|XX'})=\{\by: (m,m',\by) \in \calV(C_n,P_{Y|XX'}) \}.
\end{align}
Then, by definition of $\calG_n(P_{Y|XX'})$ in \eqref{defGPYX}, for any fixed joint type $P_{XX'} \in \calQ(Q_X)$ such that $I_P(X;X')<2R-\eps$ and $d(P_{XX'})>\Delta$, and for any $C_n \in \calF_n(P_{XX'})$, it holds that
\begin{align}
&\sum_{m,m'} \indicator \{(\bx_m,\bx_{m'}) \in \calT(P_{XX'})\} \frac{|\calT(P_{Y|XX'}) \cap \calV_{m,m'}(C_n,P_{Y|XX'})|}{|\calT(P_{Y|XX'})|}\nn\\
&\qquad \geq (1-4\delta_n^2)e^{-2\delta_n} \exp\big[ n(2R-I_P(X;X')-3\eps/2)\big] \label{av1}. 
\end{align}
Now, let
\begin{align}
P_{XX'}^*:=\argmin_{P_{XX'}:P_{X'}=P_X ,\atop I_P(X;X')\leq 2R, d(P_{XX'})>\Delta} \big\{\Gamma(P_{XX'},R)+I_P(X;X')-R\big\}.
\end{align}
Then, for any $\rho>1$, we have
\begin{align}
&\bbE[(P_{\rme}(\calC_n))^{1/\rho}]\nn\\
&= \bbE\bigg[\bigg(\frac{1}{M}\sum_m \sum_{m'\neq m} \sum_{\by} W(\by|\bX_m) \frac{\exp\{ng(\hatP_{\bX_{m'}\by})\}}{\exp\{ng(\hatP_{\bX_m \by})\} + \exp\{ng(\hatP_{\bX_{m'}\by})\}+Z_{mm'}(\by)}  \bigg)^{1/\rho}  \bigg]\\
&=\sum_{\calC_n}  \PP[\calC_n]\bigg(\frac{1}{M}\sum_m \sum_{m'\neq m} \sum_{\by} W(\by|\bx_m) \frac{\exp\{ng(\hatP_{\bx_{m'}\by})\}}{\exp\{ng(\hatP_{\bx_m \by})\} + \exp\{ng(\hatP_{\bx_{m'}\by})\}+Z_{mm'}(\by)}  \bigg)^{1/\rho}\\
&=  \sum_{\calC_n}  \PP[\calC_n]\bigg(\frac{1}{M} \sum_{P_{XX'} \in \calQ(Q_X)} \sum_m \sum_{m'\neq m}\indicator \{(\bx_m,\bx_{m'}) \in \calT(P_{XX'})\}\nn\\
&\qquad \times  \sum_{P_{Y|XX'}} \sum_{\by \in \calT(P_{Y|XX'})} W(\by|\bx_m) \frac{\exp\{ng(\hatP_{\bx_{m'}\by})\}}{\exp\{ng(\hatP_{\bx_m \by})\} + \exp\{ng(\hatP_{\bx_{m'}\by})\}+Z_{mm'}(\by)}  \bigg)^{1/\rho}\label{m2t2}\\
&= \sum_{\calC_n} \PP[\calC_n]\bigg(\frac{1}{M}\sum_{P_{XX'} \in \calQ(Q_X)}  \sum_m \sum_{m'\neq m}\indicator \{(\bx_m,\bx_{m'}) \in \calT(P_{XX'})\}\nn\\
&\qquad \times  \sum_{P_{Y|XX'}} \sum_{\by \in \calT(P_{Y|XX'})} W(\by|\bx_m) \frac{\exp\{ng(\hatP_{\bx_{m'}\by})\}}{\exp\{ng(\hatP_{\bx_m \by})\} + \exp\{ng(\hatP_{\bx_{m'}\by})\}+Z_{mm'}(\by)}  \bigg)^{1/\rho}\label{m2t3}\\
&\geq \sum_{\calC_n \in \calF_n(P_{XX'}^*)} \PP[\calC_n] \bigg(\frac{1}{M} \sum_m \sum_{m'\neq m}\indicator \{(\bx_m,\bx_{m'}) \in \calT(P_{XX'}^*)\}\nn\\
&\qquad \times  \sum_{P_{Y|XX'}} \sum_{\by \in \calT(P_{Y|XX'}) \cap \calV_{m,m'}(\calC_n,P_{Y|XX'})} W(\by|\bx_m) \frac{\exp\{ng(\hatP_{\bx_{m'}\by})\}}{\exp\{ng(\hatP_{\bx_m \by})\} + \exp\{ng(\hatP_{\bx_{m'}\by})\}+Z_{mm'}(\by)}  \bigg)^{1/\rho}\label{m2t4}\\
&\doteq \sum_{\calC_n \in \calF_n(P_{XX'}^*)} \PP[\calC_n]
\bigg(\frac{1}{M} \sum_{P_{Y|XX'}}  \sum_m \sum_{m'\neq m}\indicator \{(\bx_m,\bx_{m'}) \in \calT(P_{XX'}^*)\}\nn\\
&\qquad \times  \frac{|\calT(P_{Y|XX'})\cap \calV_{m,m'}(\calC_n,P_{Y|XX'})|}{|\calT(P_{Y|XX'})|} 
\exp\bigg\{-n [D(P_{Y|X}\|W|Q_X)]+ I_P(X';Y|X)\nn\\
&\qquad \qquad + [\max\{g(P_{XY}),\alpha(R+2\eps,P_Y)+\eps\}-g(P_{X'Y})]_+\bigg\} \bigg)^{1/\rho}\label{m2t6}\\
&\geq \sum_{\calC_n \in \calF_n(P_{XX'}^*)} \PP[\calC_n]  \bigg(\frac{1}{M}  \sum_{P_{Y|XX'}} (1-4\delta_n^2)e^{-2\delta_n} \exp\big[ n(2R-I_{P^*}(X;X')-3\eps/2)\big]\nn\\
&\quad \times \exp\bigg\{-n [D(P_{Y|X}\|W|Q_X)]+ I_P(X';Y|X)+[\max\{g(P_{XY}),\alpha(R+2\eps,P_Y)+\eps\}-g(P_{X'Y})]_+\bigg\} \bigg)^{1/\rho}\label{m2t7}\\
&\doteq  \bbP[\calF_n^c(P_{XX'}^*)] \bigg(\sum_{P_{Y|XX'}}  (1-4\delta_n^2)e^{-2\delta_n} \exp\big[ n(R-I_{P^*}(X;X')-3\eps/2)\big]\nn\\
&\quad \times \exp\bigg\{-n [D(P_{Y|X}\|W|Q_X)]+ I_P(X';Y|X)+[\max\{g(P_{XY}),\alpha(R+2\eps,P_Y)+\eps\}-g(P_{X'Y})]_+\bigg\} \bigg)^{1/\rho}\label{m2t8}\\
&\doteq  \bbP[\calF_n^c(P_{XX'}^*)] \bigg(  \exp\big[ n(R-I_{P^*}(X;X')-3\eps/2)\big]\exp[-n \Gamma(P_{XX'},R+2\eps)]
\bigg)^{1/\rho}\label{m2t9b},
\end{align} where \eqref{m2t3} follows from Tonelli's theorem \cite{Royden}, \eqref{m2t6} follows from \eqref{v10}, and \eqref{m2t7} follows from \eqref{av1}, \eqref{m2t9b} follows from $\delta_n \to 0$ and the definition of $\Gamma(P_{XX'},R)$. 

From \eqref{m2t9b}, it follows that
\begin{align}
E_{\rm{trc}}^{\rm{rgv}}(R,Q_X,d,\Delta)&=-\frac{1}{n}\lim_{\rho \to \infty} \rho \log \Big(\bbE[P_{\rme}(\calC_n)^{1/\rho}]\Big)\nn\\
&\leq \Gamma(P_{XX'}^*,R)+I_{P^*}(X;X')-R+O(\eps)\\
&\qquad  = \min_{P_{XX'}:P_{X'}=P_X ,\atop I_P(X;X')\leq 2R, d(P_{XX'})>\Delta} \bigg\{\Gamma(P_{XX'},R)+I_P(X;X')-R\bigg\}+O(\eps)
\end{align} 
for any $\eps>0$. By taking $\eps \to 0$, we obtain \eqref{akab}. This concludes the proof of Proposition \ref{lem:upTRC}.
\end{IEEEproof}

%%%%%%%%%%%%%%%%%%%%%%%%%%%%%%%%
%%%%%%%%%%%%%%%%%%%%%%%%%%%%%%%%
\section{Concentration Properties} \label{sec:concen}

In this section, we study the concentration properties of the RGV ensemble with GLD. In particular, we study the lower tail $\bbP\big[-\frac{1}{n} \log P_{\rme}(\calC_n) \leq E_0\big]$ and derive both upper and lower bounds. We show that both bounds exhibit an exponential decay. We also derive upper and lower bounds to the upper tail $\bbP\big[-\frac{1}{n} \log P_{\rme}(\calC_n) \geq E_0\big]$. We show that the lower tail exhibits a doubly-exponential behavior.

\subsection{Lower Tail}
\label{sec:lowtail}

In this section, we derive exponential upper and lower bounds to the lower tail probability. Before proceeding, we define the following sets
\begin{align}
	\calL(R,E_0)&\triangleq \{P_{XX'}\in \calQ(Q_X): d(P_{XX'})>\Delta, [2R-I_P(X;X')]_+\geq \Gamma(P_{XX'},R)+R-E_0\},\\
	\calM(R,E_0)&\triangleq \big\{P_{XX'}\in \calQ(Q_X): d(P_{XX'})>\Delta, [2R-I_P(X;X')]_+\geq \Lambda(P_{XX'},R)+R-E_0\big\}
\end{align}
where
\begin{align}
\Lambda(P_{XX'},R)&=\min_{P_{Y|XX'}}\big\{D(P_{Y|X}\|W|Q_X) + I_P(X';Y|X)+\beta(R,P_Y)-g(P_{X'Y})\big\},\\
\beta(R,P_Y)&=\max_{P_{\tilX|Y}:P_{\tilX}=Q_X} \big\{ g(P_{\tilX Y})+[R-I_P(\tilX;Y)]_+\big\}.
 \label{defbeta}
\end{align}

We have the following result.

\begin{theorem} \label{thm:concen1} Consider the ensemble of RGV codes $\calC_n$ of rate $R$ and composition $Q_X$ satisfying condition \eqref{keycond}. Then, it holds that
	\begin{align}
	\bbP\bigg[-\frac{1}{n} \log P_{\rme}(\calC_n) \leq E_0\bigg] &\dotleq \exp\big\{-n E_{\rm{lt}}^{\rm{ub}}(R,E_0)\big\}, \label{lama1}\\
	\bbP\bigg[-\frac{1}{n} \log P_{\rme}(\calC_n) \leq E_0\bigg] &\dotgeq \exp\big\{-n E_{\rm{lt}}^{\rm{lb}}(R,E_0)\big\} \label{lama1}.
	\end{align}
where
	\begin{align}
E_{\rm{lt}}^{\rm{ub}}(R,E_0)&\triangleq \min_{P_{XX'}\in \calL(R,E_0)}[I_P(X;X')-2R]_+ \label{defEltub},\\
E_{\rm{lt}}^{\rm{lb}}(R,E_0)&\triangleq \min_{P_{XX'}\in \calM(R,E_0)}[I_P(X;X')-2R]_+ \label{defEltub},
	\end{align}
	respectively.
\end{theorem}

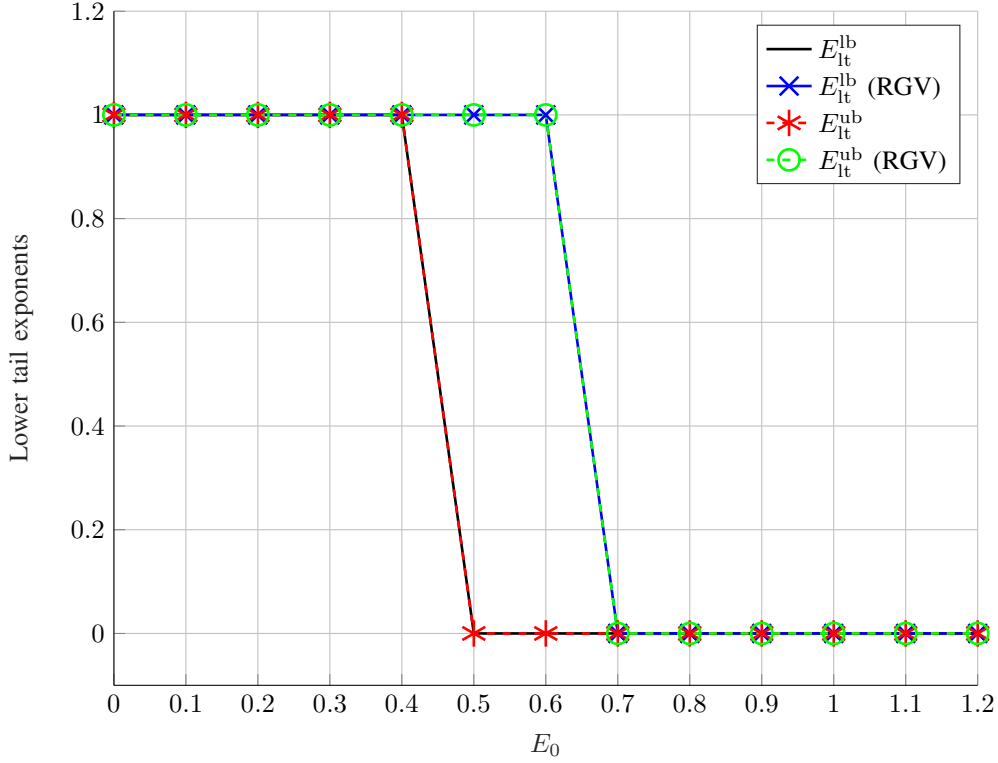
\begin{figure}[htp]
	\centering
\begin{tikzpicture}

\begin{axis}[%
width=4.521in,
height=3.531in,
at={(0.758in,0.516in)},
scale only axis,
xmin=0,
xmax=1.2,
xlabel style={font=\color{white!15!black}},
xlabel={$E_0$},
ymin=-0.1,
ymax=1.2,
ylabel style={font=\color{white!15!black}},
ylabel={Lower tail exponents},
axis background/.style={fill=white},
axis x line*=bottom,
axis y line*=left,
xmajorgrids,
ymajorgrids,
legend style={legend cell align=left, align=left, draw=white!15!black}
]
\addplot [color=black, line width=1.0pt]
  table[row sep=crcr]{%
0	1\\
0.1	1\\
0.2	1\\
0.3	1\\
0.4	1\\
0.5	0\\
0.6	0\\
0.7	0\\
0.8	0\\
0.9	0\\
1	0\\
1.1	0\\
1.2	0\\
1.3	0\\
1.4	0\\
};
\addlegendentry{$E_{\rm lt}^{\rm lb}$}

\addplot [color=blue, line width=1.0pt, mark size=5.0pt, mark=x, mark options={solid, blue}]
  table[row sep=crcr]{%
0	1\\
0.1	1\\
0.2	1\\
0.3	1\\
0.4	1\\
0.5	1\\
0.6	1\\
0.7	0\\
0.8	0\\
0.9	0\\
1	0\\
1.1	0\\
1.2	0\\
1.3	0\\
1.4	0\\
};
\addlegendentry{$E_{\rm lt}^{\rm lb}$ (RGV)}

\addplot [color=red, dashed, line width=1.0pt, mark size=5.0pt, mark=asterisk, mark options={solid, red}]
  table[row sep=crcr]{%
0	1\\
0.1	1\\
0.2	1\\
0.3	1\\
0.4	1\\
0.5	0\\
0.6	0\\
0.7	0\\
0.8	0\\
0.9	0\\
1	0\\
1.1	0\\
1.2	0\\
1.3	0\\
1.4	0\\
};
\addlegendentry{$E_{\rm lt}^{\rm ub}$}

\addplot [color=green, dashed, line width=1.0pt, mark size=4.0pt, mark=o, mark options={solid, green}]
  table[row sep=crcr]{%
0	1\\
0.1	1\\
0.2	1\\
0.3	1\\
0.4	1\\
0.5	1\\
0.6	1\\
0.7	0\\
0.8	0\\
0.9	0\\
1	0\\
1.1	0\\
1.2	0\\
1.3	0\\
1.4	0\\
};
\addlegendentry{$E_{\rm lt}^{\rm ub}$ (RGV)}

\end{axis}

\end{tikzpicture}%
	\caption{Lower tail exponents for constant composition and RGV codes for the $Z$-channel.}
	\label{fig:fig12}
\end{figure}

Before proceeding with the proof, we discuss an example in Figure \ref{fig:fig12} where the lower tail bounds are shown for the $Z$-channel with crossover probability $w=0.001$ and $R=0.2$. In particular, we show the lower tail upper and lower bounds on the tail exponent for constant composition and for the RGV ensemble with $d(P_{XX'})=-I_P(X;X')$ and $\Delta=-R$. The numerical results show that $E_{\rm{lt}}^{\rm{ub}}=E_{\rm{lt}}^{\rm{lb}}$ for the both constant composition and RGV ensembles. This can be explained by the fact that there is only one empirical channel $P_{X'Y}$ for each output type $P_Y$ for this case \cite[p.~5046]{Merhav2017a}. Hence, $[\max\{g(P_{XY}),\alpha(R,P_Y)\}-g(P_{X'Y})]=[R-I(q)]_+= \beta(R,P_Y)-g(P_{X'Y})$, which leads to $\Lambda=\Gamma$ for any $R$ and crossover probability. Fig. \ref{fig:fig12} illustrates that the lower tail for the RGV code ensemble decays faster than that for the constant composition ensemble. This can be explained by the the fact that at $R=0.2$ the typical error exponent of the RGV ensemble is higher than that for constant composition (see Figure \ref{fig:fig11}).
%\AGF{Add a comment on constant composition and explain the comparison between RGV snd constant composition.}

 %The main  %\AGF{the two bounds coincide? this means that $\Lambda$ and $\Gamma$ are equal for this case? How can this be if they are different for constant composition?}

%%%%%%%%%%%%%%%%%%%%%%%%%%%%%
\subsubsection{Proof of the Lower Tail Upper Bound} 
Let
\begin{align}
\calB_{\eps}(m,\by)=\bigg\{\calC_n: Z_m(\by)\leq \exp\{n\alpha(R-\eps,\hatP_{\by})\}\bigg\},
\end{align}
and
\begin{align}
\calB_{\eps}\triangleq \bigcup_{m=1}^M \bigcup_{\by} \calB_{\eps}(m,\by). 
\end{align}
Then, under the condition \eqref{keycond}, by Lemma \ref{lem:aut1}, we have
\begin{align}
\bbP\big\{\calB_{\eps}(m,\by)\big\}\leq \exp\bigg\{- e^{n\eps}\bigg[1- \frac{e^{-n (\eps+\delta)}}{1-e^{-n\delta}}-e^{-n\eps}(1+n\eps)\bigg]  \bigg\} \label{pus2}.
\end{align}
Hence, by the union bound, we have
\begin{align}
\bbP\{\calB_{\eps}\}&\leq \sum_{m=1}^M \sum_{\by} \bbP\big\{\calB_{\eps}(m,\by)\big\}\\
&\leq \sum_{m=1}^M \sum_{\by} \exp\bigg\{- e^{n\eps}\bigg[1- \frac{e^{-n (\eps+\delta)}}{1-e^{-n\delta}}-e^{-n\eps}(1+n\eps)\bigg]  \bigg\} \label{pus3}\\
&\leq e^{nR} |\calY|^n \exp\bigg\{- e^{n\eps}\bigg[1- \frac{e^{-n (\eps+\delta)}}{1-e^{-n\delta}}-e^{-n\eps}(1+n\eps)\bigg]  \bigg\} \label{ap}
\end{align} where \eqref{pus3} follows from \eqref{pus2}, which decays double-exponentially fast. 

Now, by using the same arguments as \cite[Proof of Theorem 1]{Tamir2020a}, we have
\begin{align}
\bbP\bigg[-\frac{1}{n}\log P_{\rme}(\calC_n)\leq E_0\bigg] & \leq \bbP\bigg[\calC_n\in \calB_{\eps}^c, \frac{1}{M}\sum_{m=1}^M \sum_{m'\neq m} e^{-n \Gamma(\hatP_{\bX_m,\bX_{m'}},R-\eps)}\geq e^{-n E_0}\bigg] + \bbP\{\calB_{\eps}\} \label{ast}\\
& \dotleq \bbP\bigg[ \frac{1}{M}\sum_{m=1}^M \sum_{m'\neq m} e^{-n \Gamma(\hatP_{\bX_m,\bX_{m'}},R-\eps)}\geq e^{-n E_0}\bigg]\\
& = \bbP\bigg[ \frac{1}{M}\sum_{m=1}^M \sum_{m'\neq m} e^{-n \Gamma(\hatP_{\bX_m,\bX_{m'}},R-\eps)}\indicator \big\{ d(\bX_m,\bX_{m'})>\Delta\big\}\geq e^{-n E_0}\bigg] \label{ast2}\\
&= \bbP\bigg[\sum_{P_{XX'} \in \calQ(Q_X): d(P_{XX'})>\Delta} N(P_{XX'})\exp\big\{-n \Gamma(P_{XX'},R-\eps)\big\}\geq e^{n(R-E_0)}\bigg]\\
&\doteq \max_{P_{XX'} \in \calQ(Q_X): d(P_{XX'})>\Delta}\bbP\bigg[N(P_{XX'})\geq \exp\{ n(\Gamma(P_{XX'},R-\eps)+R-E_0)\}\bigg] \label{mass}
\end{align}  where \eqref{ast} follows from \cite[Eq.~(60)]{Tamir2020a}, and \eqref{ast2} follows from the fact that all codes $\calC_n$ in the RGV ensemble satisfy $d(\bx_m,\bx_{m'})>\Delta$ for all $m\neq m'$. 

Now, define
\begin{align}
\calS_{\eps}(R,E_0)\triangleq \bigg\{P_{XX'} \in \calQ(Q_X): [2R-I_P(X;X')]_+\geq \Gamma(P_{XX'},R-\eps)+R-E_0\bigg\}.
\end{align}
Then, from \eqref{mass} and Lemma \ref{prop3}, under the condition \ref{keycond}, we obtain
\begin{align}
\bbP\bigg[-\frac{1}{n}\log P_{\rme}(\calC_n)\leq E_0\bigg] \dotleq e^{-n E_{\rm{lt}}^{\rm{ub}}(R,E_0,\eps)} \label{eq177},
\end{align}
where
\begin{align}
E_{\rm{lt}}^{\rm{ub}}(R,E_0,\eps)
&\triangleq \min_{P_{XX'} \in \calQ(Q_X): d(P_{XX'})> \Delta} \begin{cases}[I_P(X;X')-2R]_+,&\quad  P_{XX'}\in \calS_{\eps}(R,E_0)\\ +\infty,&\quad \mbox{otherwise} \end{cases}\\
& =\min_{P_{XX'}\in \calS_{\eps}(R,E_0):  d(P_{XX'})>\Delta} [I_P(X;X')-2R]_+ \label{eq179},
\end{align} with the convention that the minimum over an empty set is defined as infinity. Since $\eps$ can take any positive value, from \eqref{eq177} and \eqref{eq179}, we obtain \eqref{lama1}. This concludes our proof of the upper bound in Theorem \ref{thm:concen1}.

%%%%%%%%%%%%%%%%%%%%%%%%%%%%%%%%%%
\subsubsection{Proof of the Lower Tail Lower Bound} 

The proof follows similar arguments as \cite[Section B]{Tamir2020a}. 
For the RGV ensemble, however, existing techniques to lower bound on the probability of the lower tail for the constant composition codes cannot be applied. For example, due to the dependence among codewords, key proposition \cite[Prep.~4]{Tamir2020a} can no longer be applied. We develop new techniques to deal with the dependence among codewords.	

For a given $(m,m')\in [M]_*^2$, and $\by \in \calY^n$, define
\begin{align}
Z_{m,m'}(\by)=\sum_{\tilm \in \{1,2,\cdots,M\}\setminus \{m,m'\}} \exp\big\{ng(\hatP_{\bX_{\tilm},\by}) \big\} \label{mu1}.
\end{align}
Let $\sigma>0$ and define the set
\begin{align}
\hat{\calB}_n(\sigma,m,m',\by)=\Big\{C_n: Z_{mm'}(\by)\geq \exp\{n(\beta(R,\hatP_{\by})+\sigma)\}\Big\},
\end{align}
and its complement $\hat{\calG}_n(\sigma,m,m',\by)=\hat{\calB}_n^c(\sigma,m,m',\by)$, where $\beta(R,P_Y)$ is defined in \eqref{defbeta}. Let
\begin{align}
\hat{\calB}_n(\sigma)=\bigcup_{m=1}^M \bigcup_{m'\neq m} \bigcup_{\by} \hat{\calB}_n(\sigma,m,m',\by) \label{mu2},
\end{align} and
\begin{align}
\hat{\calG}_n(\sigma)=\hat{\calB}_n^c(\sigma) \label{mu3}.
\end{align}
Let $\eps>0$ and define
\begin{align}
\tilde{\Lambda}(P_{XX'},R,\eps)=\min_{P_{Y|XX'}}\big\{D(P_{Y|X}\|W|Q_X)+I_P(X';Y|X)+[\max\{g(P_{XY}),\beta(R,P_Y)+\eps\}-g(P_{X'Y})]_+\big\} \label{deftilLambda}.
\end{align}
Then, we have
\begin{align}
\bbP\bigg[-\frac{1}{n}\log P_{\rme}(\calC_n)\leq E_0\bigg]
& \dotgeq \bbP\bigg[\calC_n \in \hat{\calG}_n(\eps), \frac{1}{M}\sum_{m=1}^M \sum_{m'\neq m} e^{-n \tilde{\Lambda}(\hatP_{\bX_{m'},\bX_m},R,\eps)}\geq e^{-n E_0} \bigg] \label{gt1}\\
& = \bbP\bigg[\calC_n \in \hat{\calG}_n(\eps), \frac{1}{M}\sum_{m=1}^M \sum_{m'\neq m \atop d(\bX_m,\bX_{m'})>\Delta} e^{-n \tilde{\Lambda}(\hatP_{\bX_{m'},\bX_m},R,\eps)}\geq e^{-n E_0} \bigg] \label{matta1},
\end{align} where \eqref{gt1} follows from \cite[Eq.~(83)]{Tamir2020a}, and \eqref{matta1} follows from the fact that $d(\bx_m,\bx_{m'})>\Delta$ for any RGV code. 

On the other hand, we also have
\begin{align}
\tilde{\Lambda}(P_{XX'},R,\eps)=\Lambda(P_{XX'},R)+\eps \label{matta2}.
\end{align}
Hence, from \eqref{matta1} and \eqref{matta2}, we obtain
\begin{align}
\bbP\bigg[-\frac{1}{n}\log P_{\rme}(\calC_n)\leq E_0\bigg] \dotgeq \bbP\big[\hat{\calG}_n(\eps)\cap \calG_0\big] \label{sma1},
\end{align} where
\begin{align}
\calG_0=\bigg\{\calC_n: \sum_{m=1}^M \sum_{m'\neq m \atop d(\bX_m,\bX_{m'})>\Delta} e^{-n \tilde{\Lambda}(\hatP_{\bX_{m'},\bX_m},R,\eps)}\geq e^{n (R-E_0)} \bigg\}.
\end{align}
It then follows that
\begin{align}
\bbP\bigg[-\frac{1}{n}\log P_{\rme}(\calC_n)\leq E_0\bigg] &\dotgeq \bbP\big[\hat{\calG}_n(\eps)\cap \calG_0\big]\\
&=\bbP[\calG_0]-\bbP\big[\calG_0 \cap \hat{\calB}_n(\eps) \big] \label{ame0}\\
&\geq \bbP[\calG_0]-\sum_{m=1}^M \sum_{m'\neq m} \sum_{\by} \bbP\big[\hat{\calB}_n(\eps,m,m',\by) \cap \calG_0 \big] \label{matget}.
\end{align}
Now, observe that 
\begin{align}
\bbP[\calG_0]&=\bbP\bigg[\sum_{P_{XX'} \in \calQ(Q_X):\atop d(P_{XX'})>\Delta} N(P_{XX'}) e^{-n(\Lambda(P_{XX'},R)+\eps)} \geq e^{n(R-E_0)}\bigg]\\
&\doteq \sum_{P_{XX'}\in \calQ(Q_X): \atop d(P_{XX'})>\Delta}\bbP\bigg[N(P_{XX'})\geq e^{n(\Lambda(P_{XX'},R)+R-E_0+\eps)} \bigg] \label{GH}.
\end{align}
Define the set $\calS_{\eps}'(R,E_0)=\{P_{XX'}:[2R-I_P(X;X')]_+\geq \Lambda(P_{XX'},R)+R-E_0+\eps \}$.

Then, under condition \eqref{keycond}, by Proposition \ref{prop3}, it holds that
\begin{align}
\bbP[\calG_0]\doteq \exp\{-n E_{\rm{lt}}^{\rm{lb}}(R,E_0,\eps)\} \label{matget1},
\end{align}
where
\begin{align}
E_{\rm{lt}}^{\rm{lb}}(R,E_0,\eps)&=\min_{P_{XX'} \in \calQ(Q_X):\atop d(P_{XX'})>\Delta}\begin{cases} [I_P(X;X')-2R]_+ &\qquad P_{XX'} \in \calS_{\eps}'(R,E_0)\\ +\infty &\qquad P_{XX'} \notin S_{\eps}'(R,E_0)  \end{cases}\\
&=\min_{P_{XX'}\in \{P_{XX'}\in \calQ(Q_X): d(P_{XX'})>\Delta\}\cap \calS_{\eps}'(R,E_0)}[I_P(X;X')-2R]_+ \label{asay1}. 
\end{align}

Now, we study the second term in \eqref{matget}. For any joint type $P_{XY} \in \calP_n(\calX \times \calY)$, define
\begin{align}
N_{\by}(P_{XY})\triangleq \sum_{\tilm \in [M]\setminus \{\hatm,\ddot{m}\}}\indicator \big\{(\bX_{\tilm},\by) \in \calT(P_{XY}) \big\}.
\end{align}
Then, we have
\begin{align}
&\bbP\big[\hat{\calB}_n(\eps,\hatm,\ddot{m},\by)\cap \calG_0\big]\nn\\
&\qquad=\bbP\bigg[\sum_{\tilm \in [M]\setminus \{\hatm,\ddot{m}\}} e^{ng(\hatP_{\bX_{\tilm},\by})}\geq e^{n(\beta(R,\hatP_{\by})+\eps)},\sum_{m=1}^M \sum_{m'\neq m} e^{-n(\Lambda(\hatP_{\bx_{\tilm},\bx_m},R)+\eps)}\geq e^{n(R-E_0)}\bigg]\\
&\qquad\leq \bbP\bigg[\sum_{\tilm \in [M]\setminus \{\hatm,\ddot{m}\}} e^{ng(\hatP_{\bX_{\tilm},\by})}\geq e^{n(\beta(R,\hatP_{\by})+\eps)}\bigg]\\
&\qquad \leq \bbP\bigg[\sum_{P_{XY}:P_X=Q_X}N_{\by}(P_{XY}) e^{ng(P_{XY})}\geq e^{n(\beta(R,\hatP_{\by})+\eps)}\bigg] \label{motif1}\\
& \qquad \dotleq\sum_{P_{XY}:P_X=Q_X} \bbP\bigg[N_{\by}(P_{XY}) \geq e^{n(\beta(R,\hatP_{\by})-g(P_{XY})+\eps)}\bigg]\\ 
& \qquad \leq\sum_{P_{XY}:P_X=Q_X} \bbP\bigg[N_{\by}(P_{XY}) \geq e^{n([R-I_P(X;Y)]_++\eps)}\bigg] \label{motif2}\\
&\qquad= \sum_{P_{XY}:P_X=Q_X} \bbP\bigg[\sum_{\tilm \in [M]\setminus \{\hatm,\ddot{m}\}}\indicator \big\{(\bX_{\tilm},\by) \in \calT(P_{XY})\big\} \geq e^{n([R-I_P(X;Y)]_++\eps)}\bigg] \label{motif3a}\\
&\qquad= \sum_{P_{XY}: P_X=Q_X, I_P(X;Y)>0} \bbP\bigg[\sum_{\tilm \in [M]\setminus \{\hatm,\ddot{m}\}}\indicator \big\{(\bX_{\tilm},\by) \in \calT(P_{XY})\big\} \geq e^{n([R-I_P(X;Y)]_++\eps)}\bigg] \label{motif3},
\end{align}
where \eqref{motif2} follows from \eqref{defbeta}, and \eqref{motif3} follows from 
\begin{align}
\bbP\bigg[\sum_{\tilm \in [M]\setminus \{\hatm,\ddot{m}\}}\indicator \big\{(\bX_{\tilm},\by) \in \calT(P_{XY})\big\} \geq e^{n([R-I_P(X;Y)]_++\eps)}\bigg]=0
\end{align} if $I_P(X;Y)=0$. 

Now, in order to bound $\bbP\big[\sum_{\tilm \in [M]\setminus \{\hatm,\ddot{m}\}}\indicator \big\{(\bX_{\tilm},\by) \in \calT(P_{XY})\big\} \geq e^{n([R-I_P(X;Y)]_++\eps)}\big]$ for each joint type $P_{XY}$ such that $P_X=Q_X$, we will use the following lemma.
\begin{lemma} \cite[Lemma 1.8]{Pelekis2015} \label{aule} Suppose that $X_1,X_2,\cdots,X_n$	are random variables such that $0\leq X_i\leq 1$, for $i=1,2,\cdots,n$. Set $p=\frac{1}{n}\sum_i\bbE[X_i]$ and fix a real number $t$ such that $np+1<t<n$. If $\eps_0>0$ is such that $t-1=np(1+\eps_0)$, then
	\begin{align}
	\bbP\bigg[\sum_{i=1}^n X_i \geq t \bigg]\leq 2 e^{-nD(p(1+\eps_0)\|p)} \label{matt}.
	\end{align}
\end{lemma}
More specifically, for any $\tilm \in  [M]\setminus \{\hatm,\ddot{m}\}$, observe that
\begin{align}
\bbE\big[\indicator \big\{(\bX_{\tilm},\by) \in \calT(P_{XY})\big\}\big]&=\bbP\big[(\bX_{\tilm},\by) \in \calT(P_{XY}) \big]\\
&\qquad=\sum_{\bx_{\tilm} \in \calT(Q_X): (\bx_{\tilm},\by)\in \calT(P_{XY})} \bbP(\bx_{\tilm})\\
&\qquad= \sum_{\bx_{\tilm} \in \calT(Q_X): (\bx_{\tilm},\by)\in \calT(P_{XY})} \frac{1}{|\calT(Q_X)|} \label{motif4}\\
&\qquad\doteq e^{-n I_P(X;Y)} \label{motif5},
\end{align} where \eqref{motif4} follows from \ref{lem4some}, and \eqref{motif5} follows from \cite{Csis00}.   

It follows from \eqref{motif5} that
\begin{align}
p&\triangleq \frac{1}{M-2}\sum_{\tilm \in [M]\setminus \{\hatm,\ddot{m}\}} \bbE\big[\indicator \big\{(\bX_{\tilm},\by) \in \calT(P_{XY})\big\}\big]\\
&\doteq e^{-n I_P(X;Y)} \label{bale}.
\end{align} 

Now, there exists a $\delta(\eps)<\eps$ such that $\min\{I_P(X;Y):I_P(X;Y)>0\}>\delta(\eps)$. Then, we have
\begin{align}
&\sum_{P_{XY}: P_X=Q_X, I_P(X;Y)>0} \bbP\bigg[\sum_{\tilm \in [M]\setminus \{\hatm,\ddot{m}\}}\indicator \big\{(\bX_{\tilm},\by) \in \calT(P_{XY})\big\} \geq e^{n([R-I_P(X;Y)]_++\eps)}\bigg]\nn\\
&\qquad \leq \sum_{P_{XY}: P_X=Q_X, I_P(X;Y)>0}  \bbP\bigg[\sum_{\tilm \in [M]\setminus \{\hatm,\ddot{m}\}}\indicator \big\{(\bX_{\tilm},\by) \in \calT(P_{XY})\big\} \geq e^{n([R-I_P(X;Y)]_++\delta(\eps))}\bigg] \label{abay}.
\end{align}

By applying Lemma \ref{aule} for the sequence of Bernoulli random variables $\{\big\{(\bX_{\tilm},\by) \in \calT(P_{XY})\big\}_{\tilm \in [M]\setminus \{\hatm,\ddot{m}\} }$ with $t=e^{n([R-I_P(X;Y)]_+ +\delta(\eps))}$, we obtain
\begin{align}
&\bbP\bigg[\sum_{\tilm \in [M]\setminus \{\hatm,\ddot{m}\}}\indicator \big\{(\bX_{\tilm},\by) \in \calT(P_{XY})\big\} \geq e^{n([R-I_P(X;Y)]_++\delta(\eps))}\bigg]\nn\\
&\ddleq \exp\bigg\{-(M-2)D\big(e^{n([R-I_P(X;Y)]_+-R+\delta(\eps))}\|e^{-n I_P(X;Y)}\big) \bigg\} \label{magep1}\\
&\ddleq \exp\bigg\{-e^{n([R-I_P(X;Y)]_++\delta(\eps))} \bigg\}  \label{abay2}
\end{align}  for any joint type $P_{XY} \in \calP_n(\calX \times \calY)$ such that $P_X=Q_X$, where \eqref{abay2} follows from the fact that $D(a\|b)\geq a\big(\log  \frac{a}{b}-1\big)$ \cite{merhav_FnT2}.

It follows from  that
\begin{align}
\bbP\big[\hat{\calB}_n(\eps,\hatm,\ddot{m},\by)\cap \calG_0\big] &\ddleq \max_{P_{XY}}\exp\bigg\{-e^{n([R-I_P(X;Y)]_+ +\delta(\eps))}\bigg\}\\
&\ddleq \exp\big\{-e^{n\delta(\eps)}\big\} \label{mabe}.
\end{align}

From \eqref{matget}, \eqref{matget1}, and \eqref{mabe}, we finally obtain
\begin{align}
&\bbP\bigg[-\frac{1}{n}\log P_{\rme}(\calC_n) \leq E_0\bigg] \nn\\
&\qquad \dotgeq \exp\big\{-n E_{\rm{lt}}^{\rm{lb}}(R,E_0,\eps)\big\}-\sum_{m=1}^M \sum_{m'\neq m} \sum_{\by}  \exp\big\{-e^{n\delta(\eps)}\big\} \\
&\qquad \doteq \exp\big\{-n E_{\rm{lt}}^{\rm{lb}}(R,E_0,\eps)\big\}- e^{2nR} |\calY|^n  \exp\big\{-e^{n\delta(\eps)}\big\}\\
&\qquad \doteq \exp\big\{-n E_{\rm{lt}}^{\rm{lb}}(E,E_0,\eps)\big\} \label{magot}.
\end{align}
Due to the arbitrariness of $\eps>0$, it follows that
\begin{align}
\bbP\bigg[-\frac{1}{n}\log P_{\rme}(\calC_n)\leq E_0\bigg] \dotgeq \exp\{-n E_{\rm{lt}}^{\rm{lb}}(R,E_0)\},
\end{align} which proves the lower bound of Theorem \ref{thm:concen1}.

%%%%%%%%%%%%%%%%%%%%%%%%%%%%%
\subsection{Upper Tail}
\label{sec:uptail}

In this section, we derive double-exponential upper and lower bounds to the upper tail probability. 
First, we introduce some new notation which will be used throughout this section. Recall the definitions of $\calA_1$ and $\calA_2$ in \eqref{defcalA1} and \eqref{defcalA2}, respectively. Let
\begin{align}
\calV(R,E_0)&=\big\{P_{XX'}\in \calQ(Q_X): d(P_{XX'})>\Delta, I_P(X;X')\leq 2R, \Lambda(P_{XX'},R)+I_P(X;X')-R\leq E_0 \big\} \label{defV},\\
\calU(R_,E_0)&=\big\{P_{XX'} \in \calQ(Q_X): d(P_{XX'})>\Delta, I_P(X;X')\leq 2R, \Gamma(P_{XX'},R)+I_P(X;X')-R\leq E_0 \big\}.\label{defU}
\end{align}
and
\begin{align}
\calA_3&=\big\{P_{XX'} \in \calQ(Q_X): d(P_{XX'})>\Delta, I_P(X;X')\leq 2R, \Gamma(P_{XX'},R-\varepsilon)+I_P(X;X')-R> E_0+\varepsilon \big\}.\label{defA3}
\end{align}

\begin{theorem} \label{mainthm3} Consider the RGV ensemble $\calC_n$ of rate $R$ and composition $Q_X$ satisfying condition \eqref{keycond}.
%Recall the definition of $\tilde{\Lambda}(P_{XX'},R,\eps)$ in \eqref{deftilLambda}. 
	Assume that the conditions in Lemma \ref{lemSuen} hold for $\calD=\tilde{\calV}(R,E_0,\sigma)$. Then, the upper tail can be bounded as
\begin{align}
\bbP\bigg[-\frac{1}{n}\log P_{\rme}(\calC_n)\geq E_0\bigg] &\ddleq \exp\big\{-\exp\big\{n E_{\rm{ut}}^{\rm{ub}}(R,E_0)\big\}\big\} \label{amad1}
\end{align}
where
\begin{align}
E_{\rm{ut}}^{\rm{ub}}(R,E_0)&=\max_{P_{XX'} \in \calV(R,E_0)} \min\big\{2R-I_P(X;X'), E_0-\Lambda(P_{XX'},R)-I_P(X;X')+R,R \big\}.
\end{align}
In addition, under the conditions 	\begin{align}
	\max_{P_{XX'} \in \calA_3} I_P(X;X')&\leq \min_{P_{XX'}\in \calA_2} I_P(X;X') \label{cond0}\\
	\min_{P_{XX'}: d(P_{XX'})\leq \Delta} I_P(X;X') &\geq  \max_{P_{XX'}: d(P_{XX'})> \Delta} I_P(X;X') \label{ek1condmod},
	\end{align}
	we have that
\begin{align}
\bbP\bigg[-\frac{1}{n}\log P_{\rme}(\calC_n)\geq E_0 \bigg] &\ddgeq \exp\big\{-\exp\big\{ n E_{\rm{ut}}^{\rm{lb}}(R,E_0)\big\}\big\}, \qquad \forall E_0<E_{\rm{ex}}(R,Q_X). 
\end{align}
where
\begin{align}
E_{\rm{ut}}^{\rm{lt}}(R,E_0)&=\max_{P_{XX'} \in \calU(R,E_0)}\big\{2R-I_P(X;X')\}.
\end{align}

\end{theorem}

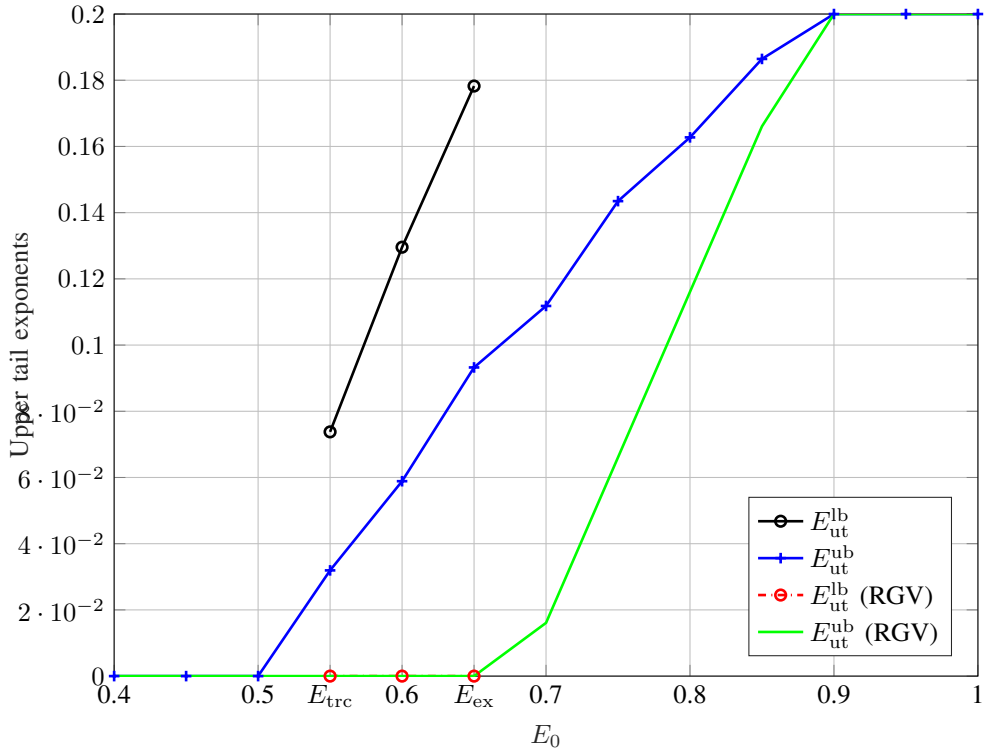
\begin{figure}[htbp]
	\centering
\begin{tikzpicture}

\begin{axis}[%
width=4.521in,
height=3.468in,
at={(0.758in,0.578in)},
scale only axis,
xmin=0.4,
xmax=1,
xtick={0.4,0.5,0.55,0.6,0.65,0.7,0.8,0.9,1},
xticklabels={{0.4},{0.5},{$E_{\rm trc}$},{0.6},{$E_{\rm ex}$},{0.7},{0.8},{0.9},{1}},
xlabel style={font=\color{white!15!black}},
xlabel={$E_0$},
ymin=0,
ymax=0.2,
ylabel style={font=\color{white!15!black}},
ylabel={Upper tail exponents},
axis background/.style={fill=white},
xmajorgrids,
ymajorgrids,
legend style={at={(0.97,0.03)}, anchor=south east, legend cell align=left, align=left, draw=white!15!black}
]
\addplot [color=black, line width=1.0pt, mark=o, mark options={solid, black}]
  table[row sep=crcr]{%
0.55	0.0737778107127643\\
0.6	0.129561907246046\\
0.65	0.178246306250149\\
};
\addlegendentry{$E_{\rm ut}^{\rm lb}$}

\addplot [color=blue, line width=1.0pt, mark=+, mark options={solid, blue}]
  table[row sep=crcr]{%
0	0\\
0.05	0\\
0.1	0\\
0.15	0\\
0.2	0\\
0.25	0\\
0.3	0\\
0.35	0\\
0.4	0\\
0.45	0\\
0.5	0\\
0.55	0.0319357928315029\\
0.6	0.0588722738270596\\
0.65	0.0932395257286585\\
0.7	0.111816304503993\\
0.75	0.143482066382722\\
0.8	0.162739032467413\\
0.85	0.186433829963443\\
0.9	0.2\\
0.95	0.2\\
1	0.2\\
1.05	0.2\\
1.1	0.2\\
1.15	0.2\\
1.2	0.2\\
1.25	0.2\\
1.3	0.2\\
1.35	0.2\\
1.4	0.2\\
1.45	0.2\\
};
\addlegendentry{$E_{\rm ut}^{\rm ub}$}

\addplot [color=red, dashdotted, line width=1.0pt, mark=o, mark options={solid, red}]
  table[row sep=crcr]{%
0.55	0\\
0.6	0\\
0.65	0\\
};
\addlegendentry{$E_{\rm ut}^{\rm lb}$ (RGV)}

\addplot [color=green, line width=1.0pt]
  table[row sep=crcr]{%
0	0\\
0.05	0\\
0.1	0\\
0.15	0\\
0.2	0\\
0.25	0\\
0.3	0\\
0.35	0\\
0.4	0\\
0.45	0\\
0.5	0\\
0.55	0\\
0.6	0\\
0.65	0\\
0.7	0.0160795149465958\\
0.75	0.0660795149465958\\
0.8	0.116079514946596\\
0.85	0.166079514946596\\
0.9	0.2\\
0.95	0.2\\
1	0.2\\
1.05	0.2\\
1.1	0.2\\
1.15	0.2\\
1.2	0.2\\
1.25	0.2\\
1.3	0.2\\
1.35	0.2\\
1.4	0.2\\
1.45	0.2\\
};
\addlegendentry{$E_{\rm ut}^{\rm ub}$ (RGV)}

\end{axis}

\end{tikzpicture}%
	\caption{Upper Tails}
	\label{fig:fig2b}
\end{figure}

In Figure \ref{fig:fig2b} we show the double-exponential bounds for the upper tail for constant composition and the RGV ensemble with $d(P_{XX'})=-I_P(X;X')$ and $\Delta=-R$ for $R=0.2$. We observe that for constant composition the decay is indeed double-exponential even if the bounds only coincide for high values of $E_0$ (above the TRC exponent). Instead, for the RGV ensemble, the bound $E_{\rm{ut}}^{\rm{lt}}(R,E_0)=0$ for values of $E_0$ of interest. This implies that the decay of the upper tail for $E_{\rm trc}\leq E_0\leq E_{\rm ex}$ is sub-double-exponential; for $E_0>E_{\rm ex}$ the behavior of the upper tail is double-exponential as suggested by $E_{\rm{ut}}^{\rm{ut}}$ for the RGV ensemble. Figure.~3 also shows that the decay rate of RGV code is slower than the constant composition code. This can be explained by the the fact that the error probability in RGV code is expected to be smaller than the constant composition codes since the later is more structured as in the Fig.~2. %\AGF{Can we compute exponents for RGV with $d(P_{XX'})=-I_P(X;X')$ and $\Delta=-R+2\delta$, with $\delta=0.001$, for example. Check what happens to the $E_{\rm{ut}}^{\rm{lt}}(R,E_0)$ bound when $E_0 = E_{\rm ex}(R,Q_X) -\epsilon$. I simulated this, but it does not change the shape.}

%%%%%%%%%%%%%%%%%%%%%%%%%%%%%%%

%%%%%%%%%%%%%%%%%%%%%%%%%%%%%%%%%
\subsubsection{Proof of the Upper Tail Upper Bound} 

The proof is based on \cite[Proof of Theorem 2]{Tamir2020a} with important changes to account for the dependency among codewords in the RGV codebook ensemble. See also the proofs of Lemma \ref{lemSuen} and Lemma \ref{smolem} below for specific changes. 

\begin{lemma} \label{smolem} For every $\sigma>0$, under condition \eqref{keycond}
the following holds
	\begin{align}
	\bbP\big\{ \hat{\calB}_n(\sigma)\big\}\ddleq \exp\{-e^{n\sigma}\}
	\end{align}
	where $\hat{\calB}_n(\sigma)$ has been defined in \eqref{mu2}.
\end{lemma}
\begin{IEEEproof}
See Appendix \ref{smolemproof}.
\end{IEEEproof}

We start by defining the following set
\begin{align}
\tilde{\calV}(R,E_0,\sigma)&\triangleq \big\{P_{XX'} \in \calQ(Q_X): d(P_{XX'})>\Delta, I_P(X;X')\leq 2R, \tilde{\Lambda}(P_{XX'},R,\sigma)+I_P(X;X')-R\leq E_0-\varepsilon \big\} \label{Ag}
%\tilde\calU(R_,E_0,\varepsilon)&\triangleq\big\{P_{XX'} \in \calQ(Q_X): d(P_{XX'})>\Delta, I_P(X;X')\leq 2R,  \Gamma(P_{XX'},R-\eps) +I_P(X;X')-R > E_0+\eps  \big\},
\end{align}
for $\sigma>0, \varepsilon>0$, where $\tilde{\Lambda}(P_{XX'},R,\eps)$ was defined in \eqref{deftilLambda}.

Under condition \eqref{keycond}, we have that
\begin{align}
\bbE[N(P_{XX'})]&=\bbE\bigg[\sum_{(m,m') \in [M]_*^2}\indicator\{(\bX_m,\bX_{m'})\in \calT(P_{XX'})\}\bigg]\\
&=\sum_{(m,m') \in [M]_*^2} \bbP\big[(\bX_m,\bX_{m'})\in \calT(P_{XX'})\big]\\
&=\sum_{(m,m') \in [M]_*^2}\sum_{\bx_m,\bx_{m'} \in \calT(P_{XX'})} \bbP\big(\bx_m,\bx_{m'}\big)\\
&\doteq e^{n(2R-I_P(X;X'))} \label{kasu},
\end{align} where \eqref{kasu} follows from Lemma \ref{lem2some}.

For a given message pair $m,m'\in [M]_*^2$, and $\by \in \calY^n$, recall the definitions of $Z_{m,m'}(\by)$, $\hat{\calB}_n(\sigma)$, and $\hat{\calG}_n(\sigma)$ in \eqref{mu1}, \eqref{mu2}, and \eqref{mu3}, respectively. 
Then, we have
	\begin{align}
	&\bbP\big[\calC_n \in \hat{\calG}_n(\sigma), -\frac{1}{n}\log P_{\rme}(\calC_n)\geq E_0 \bigg]\\
	&\quad\leq \bbP\bigg[\calC_n \in \hat{\calG}_n(\sigma), \frac{1}{M}\sum_{m=1}^M \sum_{m'\neq m} \sum_{\by} W(\by|\bX_m) \frac{e^{ng(\hatP_{\bX_{m'}\by})}}{e^{ng(\hatP_{\bX_m \by})}+e^{ng(\hatP_{\bX_{m'}\by})}+Z_{mm'}(\by) }\leq e^{-n E_0}\bigg]\label{eq230}\\
	&\quad = \bbP\bigg[\calC_n \in \hat{\calG}_n(\sigma), \frac{1}{M}\sum_{m=1}^M \sum_{m'\neq m: d(\bX_m,\bX_{m'})>\Delta} \sum_{\by} W(\by|\bX_m) \frac{e^{ng(\hatP_{\bX_{m'}\by})}}{e^{ng(\hatP_{\bX_m \by})}+e^{ng(\hatP_{\bX_{m'}\by})}+Z_{mm'}(\by) }\leq e^{-n E_0}\bigg]\label{facso1} \\
	&\quad \ddleq \min_{P_{XX'}\in \tilde{\calV}(R,E_0,\sigma)} \bbP\bigg[N(P_{XX'}) \leq e^{n(\tilde{\Lambda}(P_{XX'},R,\sigma)+ R-E_0)} \bigg]\label{eq232}\\
	&\quad \ddleq \min_{P_{XX'}\in \tilde{\calV}(R,E_0,\sigma)} \bbP\bigg[N(P_{XX'}) \leq e^{n\big(2R-I_P(X;X')-\eps\big)} \bigg]\label{eq232b}\\
	&\quad \ddleq \min_{p_{XX'}\in \tilde{\calV}(R,E_0,\sigma)}\exp\big\{-\min\big(e^{n(2R-I_P(X;X'))}, e^{nR}\big)\big\} \label{mat} 
	\end{align} where \eqref{eq230} follows from \eqref{defPeCn}, \eqref{facso1} follows from the fact that $d(\bX_m,\bX_{m'})>\Delta$ with probability $1$ by the RGV random codebook generation, \eqref{eq232} follows the same arguments to achieve \cite[Eq.~(146)]{Tamir2020a}, \eqref{eq232b} follows from \eqref{Ag}, and \eqref{mat} follows from \eqref{kasu} and Lemma \ref{lemSuen}.

It follows from \eqref{mat} that for $\sigma>0$,
\begin{align}
\bbP\bigg[\calC_n \in \hat{\calG}_n(\sigma), -\frac{1}{n}\log P_{\rme}(\calC_n)\geq E_0 \bigg\}\ddleq \exp\big\{-\exp\big\{n E_1(R,E_0,\sigma)\big\} \big\}\label{mas},
\end{align}
where
\begin{align}
E_1(R,E_0,\sigma)=\max_{P_{XX'} \in \tilde{\calV}(R,E_0,\sigma)} \min\{2R-I_P(X;X'),R\}.
\end{align}
Therefore, we have
\begin{align}
\bbP\bigg[-\frac{1}{n}\log P_{\rme}(\calC_n) \geq E_0 \bigg]
&=\bbP\bigg[\calC_n \in \hat{\calG}_n(\sigma),-\frac{1}{n}\log P_{\rme}(\calC_n) \geq E_0 \bigg]+ \bbP\bigg[\calC_n \in \hat{\calG}_n^c(\sigma),-\frac{1}{n}\log P_{\rme}(\calC_n) \geq E_0 \bigg]\\
&=\bbP\bigg[\calC_n \in \hat{\calG}_n(\sigma),-\frac{1}{n}\log P_{\rme}(\calC_n) \geq E_0 \bigg]+ \bbP\bigg[\calC_n \in \hat{\calB}_n(\sigma),-\frac{1}{n}\log P_{\rme}(\calC_n) \geq E_0 \bigg] \label{lie1}\\
&\leq \bbP\bigg[\calC_n \in \hat{\calG}_n(\sigma),-\frac{1}{n}\log P_{\rme}(\calC_n) \geq E_0 \bigg]+ \bbP\big[\calC_n \in \hat{\calB}_n(\sigma)\big]\\
&\ddleq \exp\big\{-\exp\big\{n E_1(R,E_0,\sigma)\big\} \big\}+ \exp\{-e^{n\sigma}\}\label{mut1}
\end{align} where \eqref{lie1} follows from $\hat{\calB}_n(\sigma)=\hat{\calG}_n^c(\sigma)$, \eqref{mut1} follows from Lemma \ref{smolem} and \eqref{mas}. 

Finally, by using the same arguments as to obtain \cite[Eq.~(175)]{Tamir2020a} from \cite[Eq.~(153)]{Tamir2020a}, from \eqref{mut1}, we obtain
\begin{align}
\bbP\bigg[-\frac{1}{n}\log P_{\rme}(\calC_n) \geq E_0 \bigg]\ddleq \exp\Big\{-e^{n E_{\rm{ut}}^{\rm{ub}}(R,E_0)}\Big\},
\end{align} which concludes our proof of the upper bound on the upper tail.

%%%%%%%%%%%%%%%%%%%%%%%%%%

\subsubsection{Proof of the Upper Tail Lower Bound} 

%\begin{theorem} Consider the ensemble of RGV $\calC_n$ of rate $R$ and composition $Q_X$. Then, under the conditions that
%	\begin{align}
%	\max_{P_{XX'} \in \calA_3} I_P(X;X')&\leq \min_{P_{XX'}\in \calA_2} I_P(X;X') \label{cond0}\\
%	\min_{P_{XX'}: d(P_{XX'})\leq \Delta} I_P(X;X') &\geq  \max_{P_{XX'}: d(P_{XX'})> \Delta} I_P(X;X') \label{ek1condmod},
%	\end{align}
%	and
%	\begin{align}
%	R\leq \min_{P_{XX'}: d(P_{XX'})\leq \Delta} I_P(X;X')-2 \delta \label{condkeymod}
%	\end{align} for some $\delta>0$, it holds that
%	\begin{align}
%	\bbP\bigg\{-\frac{1}{n}\log P_{\rme}(\calC_n)\geq E_0 \bigg\} \ddgeq \exp\bigg\{-\exp\big\{ n E_{\rm{ut}}^{\rm{lb}}(R,E_0)\big\}\bigg\}, \qquad \forall E_0<E_{\rm{ex}}(R,Q_X,d,\Delta). 
%	\end{align}
%\end{theorem}
%\begin{IEEEproof}
Let
\begin{align}
\calB_{\eps}(m,\by)=\bigg\{\calC_n: Z_m(\by)\leq \exp\{n\alpha(R-\eps,\hatP_{\by})\}\bigg\},
\end{align}
and
\begin{align}
\calB_{\eps}\triangleq \bigcup_{m=1}^M \bigcup_{\by} \calB_{\eps}(m,\by). 
\end{align}
Then, under condition \eqref{keycond}, by Lemma \ref{lem:aut1} and the union bound, we have
\begin{align}
\bbP\{\calB_{\eps}\} \leq e^{nR} |\calY|^n \exp\bigg\{- e^{n\eps}\bigg[1- \frac{e^{-n (\eps+\delta)}}{1-e^{-n\delta}}-e^{-n\eps}(1+n\eps)\bigg]  \bigg\} \label{apmod}.
\end{align}
Now, define $\calG_{\eps}(m,\by)=\calB_{\eps}^c(m,\by)$ and $\calG_{\eps}=\calB_{\eps}^c$.

Recall the definition of $Z_m(\by)$ in \eqref{defZmy}.
%\begin{align}
%Z_m(\by)\triangleq \sum_{\tilm \neq m} e^{n g(\hatP_{\bX_{\tilm},\by})}.
%\end{align}
We have that
\begin{align}
&\bbP\bigg[-\frac{1}{n}\log P_{\rme}(\calC_n)\geq E_0 \bigg]\nn\\
&\qquad =\bbP\bigg[\frac{1}{M}\sum_{m=1}^M \sum_{m'\neq m} \sum_{\by} W(\by|\bX_m)  \frac{ \exp\{ng(\hatP_{\bX_{m'}\by})\}}{\exp\{ng(\hatP_{\bX_{m}\by})\}+ Z_m(\by)}\leq e^{-n E_0}\bigg]\\
&\qquad =\bbP\bigg[\frac{1}{M}\sum_{m=1}^M \sum_{m'\neq m: d(\bX_m,\bX_{m'})>\Delta} \sum_{\by} W(\by|\bX_m)  \frac{ \exp\{ng(\hatP_{\bX_{m'}\by})\}}{\exp\{ng(\hatP_{\bX_{m}\by})\}+ Z_m(\by)}\leq e^{-n E_0}\bigg] \label{ahm1}\\
&\qquad \geq \bbP\bigg[\frac{1}{M}\sum_{m=1}^M \sum_{m'\neq m: d(\bX_m,\bX_{m'})>\Delta} \sum_{\by} W(\by|\bX_m)  \frac{ \exp\{ng(\hatP_{\bX_{m'}\by})\}}{\exp\{ng(\hatP_{\bX_{m}\by})\}+ Z_m(\by)}\leq e^{-n E_0}, \calC_n \in \calG_{\eps} \bigg] \label{ahm2}\\
&\qquad \ddgeq \bbP\bigg[\frac{1}{M} \sum_{m=1}^M \sum_{m'\neq m: d(\hatP_{\bX_{m},\bX_{m'}})>\Delta} \exp\big\{-n \Gamma(\hatP_{\bX_{m},\bX_{m'}},R-\eps)\big\}\leq e^{-n E_0}, \calC_n \in \calG_{\eps}\bigg] \label{eq261},
\end{align} where \eqref{ahm1} follows from the fact that $\min_{i\neq j} d(\bx_i,\bx_j)>\Delta$ for all RGV code $(\bx_1,\bx_2,\cdots,\bx_M)$, and \eqref{eq261} follows from the same arguments to obtain \cite[Eq.~(178)]{Tamir2020a}.

Now, define
\begin{align}
\calE_0\triangleq \bigg\{\frac{1}{M} \sum_{m=1}^M \sum_{m'\neq m: d(\hatP_{\bX_{m},\bX_{m'}})>\Delta} \exp\big\{-n \Gamma(\hatP_{\bX_{m},\bX_{m'}},R-\eps)\big\} \leq e^{-n E_0}\bigg\}.
\end{align}
Then, we have 
\begin{align}
&\bbP\bigg[\frac{1}{M} \sum_{m=1}^M \sum_{m'\neq m: d(\hatP_{\bX_{m},\bX_{m'}})>\Delta} \exp\big\{-n \Gamma(\hatP_{\bX_{m},\bX_{m'}},R-\eps)\big\} \leq e^{-n E_0}, \calC_n \in \calG_{\eps} \bigg]\nn\\
&\qquad =\bbP\bigg[\calC_n \in \calE_0, \calC_n \in \calG_{\eps} \bigg]\\
&\qquad=\bbP\bigg[\bigcap_{\tilm=1}^M \bigcap_{\by} \calG_{\eps}(\tilm,\by)\big|\calE_0 \bigg] \bbP(\calE_0)\\
&\qquad=\bigg(1-\bbP\bigg[\bigcup_{\tilm=1}^M \bigcup_{\by} \calG_{\eps}^c(\tilm,\by)\big|\calE_0 \bigg]\bigg) \bbP[\calE_0]\\
&\qquad \geq \bigg(1-\sum_{m=1}^M \sum_{\by} \bbP\big[\calG_{\eps}^c(\tilm,\by)\big|\calE_0 \big]\bigg) \bbP[\calE_0]\\
&\qquad=\bbP[\calE_0]-\sum_{m=1}^M \sum_{\by} \bbP\big[\calB_{\eps}(\tilm, \by) \cap \calE_0 \big] \label{es1}.
\end{align}
Now, observe that
\begin{align}
\bbP[\calE_0]&=\bbP\bigg[\frac{1}{M} \sum_{m=1}^M \sum_{m'\neq m: d(\hatP_{\bX_{m},\bX_{m'}})>\Delta} \exp\big\{-n \Gamma(\hatP_{\bX_{m},\bX_{m'}},R-\eps)\big\} \leq e^{-n E_0}\bigg]\\
&\ddeq \bbP\bigg[\bigcap_{P_{XX'}\in \calQ(Q_X):d(P_{XX'})>\Delta} \bigg\{N(P_{XX'})\leq e^{n(\Gamma(P_{XX'},R-\eps)+R-E_0)} \bigg\} \bigg] \label{eq269},
\end{align} where \eqref{eq269} follows by using the same arguments to achieve \cite[Eq.~(187)]{Tamir2020a}.

Recall the definition of $\calF_0$ in \eqref{defF0} in Lemma \ref{uplem1}, i.e., 
\begin{align}
\calF_0=\bigcap_{P_{XX'} \in \calA_1 \cup \calA_2} \big\{ N(P_{XX'})=0 \big\} \label{defF0modc}.
\end{align}
Define
\begin{align}
\calF(P_{XX'})\triangleq \bigg\{ N(P_{XX'}) \leq e^{n(\Gamma(P_{XX'},R-\eps)+R-E_0)} \bigg\} \label{defcalFP}. 
\end{align}
Then, from \eqref{eq269} and \eqref{defcalFP}, we obtain
\begin{align}
\bbP[\calE_0] &\ddeq  \bbP\bigg[\bigcap_{P_{XX'}\in \calQ(Q_X):d(P_{XX'})>\Delta} \calF(P_{XX'})\bigg]\\
&=\bbP\bigg[ \bigcap_{P_{XX'} \in \calA_1 \cup \calA_2 \cup \calA_3} \calF(P_{XX'}) \bigg]\\
&= \bbP\bigg[ \bigcap_{P_{XX'} \in \calA_3} \calF(P_{XX'})\cap  \bigcap_{P_{XX'} \in \calA_1\cup \calA_2} \calF(P_{XX'}) \bigg]\\
&\geq \bbP\bigg[\bigcap_{P_{XX'} \in \calA_3} \calF(P_{XX'})\cap \calF_0 \bigg] \label{matte1} \\
&=\bbP\bigg[ \bigcap_{P_{XX'} \in \calA_3} \calF(P_{XX'})\big|\calF_0 \bigg]\bbP[\calF_0]\\
&=\bigg(1- \bbP\bigg[ \bigcup_{P_{XX'} \in  \calA_3} \calF^c(P_{XX'})\big|\calF_0 \bigg]\bigg)\bbP[\calF_0]\\
&\geq \bbP[\calF_0]- \sum_{P_{XX'} \in \calA_3} \bbP\big[\calF^c(P_{XX'})\big|\calF_0 \big] \bbP[\calF_0] \label{amotta}\\
&\geq \bbP[\calF_0]- \sum_{P_{XX'} \in \calA_3} \bbP\big[\calF^c(P_{XX'})\cap \calF_0 \big]\\
&\geq \bbP[\calF_0]- \sum_{P_{XX'} \in \calA_3} \bbP\big[\calF^c(P_{XX'})\big] \label{amotta2},
\end{align} where \eqref{matte1} follows from the fact that for each joint type $P_{XX'} \in \calA_1 \cup \calA_2$, it holds that $\{N(Q_{XX'})=0\} \subset \{N(Q_{XX'})\leq e^{n(\Gamma(P_{XX'},R-\eps)+R-E_0)}\}$. 

Equation \eqref{amotta2} resembles \cite[Eq.~(205)]{Tamir2020a} with subtle differences in the definition of sets $\calA_1,\calA_2$ and $\calA_3$. However, since all the codewords in RGV are dependent, \cite[Eq.~(218)]{Tamir2020a} does not hold. We proceed with different arguments.  
For any $P_{XX'}\in \calA_3$, we have
\begin{align}
\bbP\big[\calF^c(P_{XX'}) \big\}&=\bbP\bigg\{N(P_{XX'})\geq e^{n(\Gamma(P_{XX'},R-\eps)+R-E_0)} \bigg]\\
&\leq \bbP\bigg[N(P_{XX'})\geq e^{n\eps}e^{n(2R-I_P(X;X'))} \bigg] \label{by1},
\end{align} where \eqref{by1} follows from the definition of the set $\calA_3$, which implies that
\begin{align}
\Gamma(P_{XX'},R-\eps)+R-E_0> 2R-I_P(X;X')+\eps.
\end{align}
On the other hand, by Lemma \ref{lem:ba1}, we have
\begin{align}
\sum_{P_{XX'} \in \calA_3} \bbP\bigg[N(P_{XX'})\geq e^{n\eps}e^{n(2R-I_P(X;X'))} \bigg] &\ddleq \max_{P_{XX'}\in \calA_3} \exp\bigg\{-e^{nR(2R-I_P(X;X')+\eps)} \bigg\}\\
&\qquad=  \exp\bigg\{-e^{n(2R-\max_{P_{XX'}\in \calA_3} I_P(X;X')+\eps)} \bigg\}\\
&\qquad\leq  \exp\bigg\{-e^{n(2R-\min_{P_{XX'}\in \calA_2} I_P(X;X')+\eps)} \bigg\} \label{kacond},
\end{align} where \eqref{kacond} follows from the condition \eqref{cond0}. 

Now, under the condition \eqref{ek1condmod}, by Lemma \ref{uplem1}, we have
\begin{align}
\bbP\{F_0\} \ddgeq \exp\big\{-e^{n \max_{P_{XX'} \in \calA_2} (2R-I_P(X;X'))}\big\} \label{exmod}.
\end{align}
From \eqref{amotta2}, \eqref{kacond}, and \eqref{exmod}, we obtain
\begin{align}
\bbP[\calE_0]&\ddgeq \exp\big\{-e^{n \max_{P_{XX'} \in \calA_2} (2R-I_P(X;X'))}\big\} -\exp\bigg\{-e^{n(2R-\min_{P_{XX'}\in A_2} I_P(X;X')+\eps)} \bigg\}\\
&\ddeq \exp\big\{-e^{n \max_{P_{XX'} \in \calA_2} (2R-I_P(X;X'))}\big\} \label{defE0}.
\end{align}

To bound $\bbP[\calB_{\eps}(\tilm,\by) \cap \calE_0]$, we use the following arguments. As \cite{Tamir2020a}, let
\begin{align}
\calN^2:=\bigg\{(m,m'): m\neq m', m,m' \in \{1,2,\cdots, \lfloor M/2 \rfloor-1\}\bigg\}.
\end{align}
Define
\begin{align}
\calS:= \bigg\{(\bx_1,\bx_2, \cdots, \bx_{\floor M/2\rfloor})\in \underbrace{\bbR^n \times \bbR^n \cdots \times \bbR^n}_{\lfloor M/2 \rfloor \enspace \mbox{times}}:\min_{i,j \in \{1, 2,\cdots,\lfloor M/2 \rfloor \}, i\neq j}\{d(\bx_i,\bx_j)\}> \Delta \bigg\}.
\end{align}
%In addition, for any vector $\bx \in \bbR^n$, define
%\begin{align}
%d(\bx,\calS):=\min_{\bx' \in \calS} d(\bx,\bx'). 
%\end{align}
Since the distance between two codewords in a RGV ensemble is at least $\Delta$, we have
\begin{align}
&\bbP[\calB_{\eps}(\tilm,\by) \cap \calE_0]\nn\\
&\qquad \leq \bbP\bigg[ \sum_{(m,m')\in \calN^2} e^{-n \Gamma(\hatP_{\bX_m,\bX_{m'}},R-\eps)} \leq e^{n(R-E_0)}\bigg]\nn\\
&\qquad \times \bbP\bigg[\sum_{m' \in \{\lfloor M/2 \rfloor,\cdots,M\}\setminus \{\tilm\}}e^{ng(\hatP_{\bX_{m'}\by})}\leq e^{n\alpha(R-\eps,\hatP_{\by})}\nn\\
&\qquad \qquad \qquad\bigg|\bigg\{\sum_{(m,m')\in \calN^2} e^{-n \Gamma(\hatP_{\bX_m,\bX_{m'}},R-\eps)} \leq e^{n(R-E_0)}\bigg\}\cap \bigg\{(\bX_1,\bX_2, \cdots, \bX_{\floor M/2\rfloor})\in \calS \bigg\} \bigg] \label{amo1}.
\end{align}
Now, for any tuple $(\bx_1,\bx_2, \cdots, \bx_{\lfloor M \rfloor})$ such that $\min_{i,j \in \{1,2,\cdots,\lfloor M/2 \rfloor\}, i\neq j}\{d(\bx_i,\bx_j)\}> \Delta$, it holds that
\begin{align}
&\bbP\bigg(\bX_{\lfloor M/2\rfloor+1}=\bx_{\lfloor M/2\rfloor+1}, \bX_{\lfloor M/2 \rfloor+2}=\bx_{\lfloor M/2\rfloor+2},\cdots, \bX_M=\bx_M \big|\bX_1=\bx_1, \cdots, \bX_{\lfloor M/2 \rfloor}=\bx_{\lfloor M/2\rfloor}\bigg)\nn\\
&\qquad =\frac{\bbP(\bX_1=\bx_1,\bX_2=\bx_2,\cdots,\bX_M=\bx_M)}{\bbP(\bX_{\lfloor M/2 \rfloor }=\bx_{\lfloor M/2 \rfloor}, \cdots, \bX_1=\bx_1)}\\
&\qquad \dotleq \frac{1}{|\calT(Q_X)|^{\lceil M/2 \rceil}} \label{mege},
\end{align} where \eqref{mege} follows from Lemma \ref{lem:aux0}. Hence, by using the same arguments as the proof of Lemma \ref{lem:aut1}, we obtain
\begin{align}
\bbP[\calB_{\eps}(\tilm,\by) \cap \calE_0] \leq \exp\bigg\{- e^{n\eps}\bigg[1- \frac{e^{-n (\eps+\delta)}}{1-e^{-n\delta}}-e^{-n\eps}(1+n\eps)\bigg]  \bigg\}\bbP(\calE_0) \label{megu}. 
\end{align}  
From \eqref{es1}, \eqref{defE0}, and \eqref{megu}, we have
\begin{align}
&\bbP\bigg[-\frac{1}{n}\log P_{\rme}(\calC_n)\geq E_0 \bigg]\nn\\
&\qquad \ddgeq \bigg(1-e^{nR} |\calY|^n \exp\bigg\{- e^{n\eps}\bigg[1- \frac{e^{-n (\eps+\delta)}}{1-e^{-n\delta}}-e^{-n\eps}(1+n\eps)\bigg]  \bigg\}\bigg)\exp\big\{-e^{n \max_{P_{XX'} \in \calA_2} (2R-I_P(X;X'))}\big\}  \label{hence}\\
&\qquad \ddeq \exp\big\{-e^{n \max_{P_{XX'} \in \calA_2} (2R-I_P(X;X'))}\big\}
\end{align}
which concludes the proof.

%%%%%%%%%%%%%%%%%%%%%%%%%%%%%
\subsection{Convergence in Probability}

This section enumerates properties of the tail exponents derived in Sections \ref{sec:lowtail} and \ref{sec:uptail}, respectively,
 and establishes the convergence in probability to the TRC exponent of the RGV.
In particular, the following results can be obtained by using the same arguments as the proofs of \cite[Prop.~1]{Tamir2020a}, \cite[Prop.~3]{Tamir2020a}, \cite[Prop.~2]{Tamir2020a}, respectively, and are therefore stated without proof. Define
\begin{align}
\tilE(R)\triangleq\min_{P_{XX'} \in \calQ(Q_X): I_P(X;X')\leq 2R, d(P_{XX'})>\Delta} \big\{\Lambda(P_{XX'},R)+I_P(X;X')-R \big\}.
\end{align}
\begin{proposition}[Lower tail] \label{prep3} $E_{\rm{lt}}^{\rm{ub}}(R,E_0)$ and $E_{\rm{lt}}^{\rm{lb}}(R,E_0)$ have the following properties
\begin{enumerate}
	\item  For fixed $R$, $E_{\rm{lt}}^{\rm{ub}}(R,E_0)$ and $E_{\rm{lt}}^{\rm{lb}}(R,E_0)$ are decreasing in $E_0$.
	\item $E_{\rm{lt}}^{\rm{ub}}(R,E_0)>0$ if and only if $E_0<E_{\rm{trc}}^{\rm{rgv}}(R,Q_X,\Delta,d)$.
	\item $E_{\rm{lt}}^{\rm{lb}}(R,E_0)>0$ if $E_0<\tilE(R)$.
	\item $E_{\rm{lt}}^{\rm{lb}}(R,E_0)=\infty$ for any $E_0<E_0^{\min}(R)$, where
	\begin{align}
	E_0^{\min}(R)\triangleq \min_{P_{XX'}\in \calQ(Q_X):d(P_{XX'})>\Delta}\big\{\Gamma(P_{XX'},R)-[2R-I_P(X;X')]_+ + R\big\}
	\end{align}
\end{enumerate}
\end{proposition} 

\begin{proposition}[Upper tail]  \label{prep4} $E_{\rm{ut}}^{\rm{ub}}(R,E_0)$ and $E_{\rm{ut}}^{\rm{lb}}(R,E_0)$ have the following properties
	\begin{enumerate}
		\item  For fixed $R$, $E_{\rm{ut}}^{\rm{ub}}(R,E_0)$ and $E_{\rm{ut}}^{\rm{lb}}(R,E_0)$ are increasing in $E_0$.
		\item $E_{\rm{ut}}^{\rm{ub}}(R,E_0)>0$ if and only if $E_0>E_{\rm{trc}}^{\rm{rgv}}(R,Q_X,\Delta,d)$.
		\item $E_{\rm{ut}}^{\rm{lb}}(R,E_0)>0$ if $E_0>\tilE(R)$.
	\end{enumerate}
\end{proposition} 

From Propositions \ref{prep3} and \ref{prep4}, the following result states the convergence in probability to the TRC of the RGV ensemble. 
\begin{corollary} 
\label{thm:bac} 
For any RGV ensemble with GLD, under the conditions in Lemma \ref{lemSuen} and Lemma \ref{uplem1}, we have
that	\begin{align}
	-\frac{1}{n}\log P_{\rme}(\calC_n) \pto E_{\rm{trc}}^{\rm rgv}(R,Q_X,d,\Delta) \label{labaan}.
	\end{align}
\end{corollary}

Recall that for $d(P_{XX'})=-I_P(X;X')$ and $\Delta=-(R+2\delta)$, the conditions in Lemma \ref{lemSuen} and Lemma \ref{uplem1} hold. Hence, Corollary \ref{thm:bac} holds for this important case for which $E_{\rm{rce}}^{\rm rgv}(R,Q_X,d,\Delta)=E_{\rm{trc}}^{\rm rgv}(R,Q_X,d,\Delta)=E_{\rm ex}(R,Q_X)$.
%In the following, we provide a direct proof of \ref{thm:bac} without using the conditions in Lemma \ref{lemSuen} and Lemma \ref{uplem1}.
%\begin{theorem}
%For any RGV ensemble with GLD, we have that 
%\begin{align}
%E_{\rm{rce}}^{\rm rgv}(R,Q_X,d,\Delta)=E_{\rm{trc}}^{\rm rgv}(R,Q_X,d,\Delta)
%\end{align}
% and that 
% \begin{align}
%-\frac{1}{n}\log P_{\rme}(\calC_n) \pto E_{\rm{trc}}^{\rm rgv}(R,Q_X,d,\Delta) \label{labaab}.
%\end{align}
%\end{theorem}

%\AGF{Is Corollary \ref{thm:bac} redundant given this theorem?}
%\begin{IEEEproof}
%The proof follows the same steps as \cite{somekh_2019} for the  RGV ensemble with ML decoding and \cite{Merhav2017a}. Specifically, we can show that
%
%By using the same proofs as \cite{somekh_2019} and \cite{Merhav2017a} for the RGV ensemble with ML decoder, we can show that \textcolor{blue}{The proof was not provided in that paper. This is a conjecture. Hence, we can remove this.
%\begin{align}
%E_{\rm{rce}}^{\rm rgv}(R,Q_X,d,\Delta)= E_{\rm{trc}}^{\rm rgv}(R,Q_X,d,\Delta), 
%\end{align}	where $E_{\rm{trc}}^{\rm rgv}(R,Q_X,d,\Delta)$ is determined in Theorem \ref{mainthm1}. 

%Then, by using the same proof as \cite[Proof of Theorem 1]{Truong2022PO} (case 1), we obtain
%\begin{align}
%-\frac{1}{n}\log P_{\rme}(\calC_n) \pto E_{\rm{trc}}^{\rm rgv}(R,Q_X,d,\Delta) \label{labamo}.
%\end{align}
%\AGF{I have not been able to find case 1 in the proof of Theorem 1 in \cite{Truong2022PO}. Perhaps we could give sections, or equation ranges, or something else that can be easily verified. } }
%\end{IEEEproof}

%%%%%%%%%%%%%%%%%%%%%%%%%%%%%%
\section{Conclusions}

We have studied the RGV code ensemble and have studied the typical error exponent and upper and lower error exponent tails. We have shown that the lower tail decays exponentially while the upper tail exhibits a decay that is between exponential and double-exponential; it is sub-double-exponential below the expurgated exponent and double-exponential above the expurgated exponent. In addition, we have shown that the error exponent of a sufficiently long RGV code concentrates in probability around the typical error exponent; this is also shown to coincide with the random coding exponent of the RGV ensemble, known to coincide with the maximum of the expurgated and the random-coding exponent. This suggests that every code in the ensemble asymptotically attains as high an error exponent as it is known from random codes.

%%%%%%%%%%%%%%%%%%%%%%%%%%%%%%

\newpage
\appendices
\section{Proof of Lemma \ref{lem:aux0}} \label{lem:aux0proof}
Assume that $\calA=\{i_1,i_2,\cdots,i_l\}$ where $1\leq i_1<i_2<\cdots <i_l\leq M$ for some $l \in [M]$. 
First, if $\min_{j,k \in [l], j\neq k} d(\bx_{i_j},\bx_{i_k})\leq \Delta$, then by the RGV generation, we have
\begin{align}
\bbP(\bx_{i_1},\bx_{i_2},\cdots,\bx_{i_l})=0.
\end{align}
Hence, \eqref{amote} trivially holds.

Now, under the condition $\min_{j,k \in [l], j\neq k} d(\bx_{i_j},\bx_{i_k})>\Delta$, we have
\begin{align}
&\bbP\bigg[\bigcap_{k\in A} \{\bX_k=\bx_k\}\bigg]=\bbP(\bx_{i_1},\bx_{i_2},\cdots,\bx_{i_l})\\
&\qquad =\sum_{x_1^{i_1-1}, x_{i_1+1}^{i_2-1},\cdots, x_{i_{l-1}+1}^{i_l-1}: d(\bx_k,\bx_l)>\Delta \forall k,l \in [i_l], k\neq l}\bbP(\bx_1^{i_1-1})\bbP(\bx_{i_1}|\bx_1^{i_1-1})\bbP(\bx_{i_1+1}^{i_2-1}|\bx_1^{i_1})\bbP(\bx_{i_2}|\bx_1^{i_2-1})\nn\\
&\qquad \qquad \qquad \times \bbP(\bx_{i_2+1}^{i_3-1}|\bx_1^{i_2})\bbP(\bx_{i_3}|\bx_1^{i_3-1})\times \cdots \times   \bbP(\bx_{i_{l-1}+1}^{i_l-1}|\bx_1^{i_{l-1}}) \bbP(\bx_{l_l}|\bx_1^{i_l-1}) \label{ah1}\\
&\qquad= \sum_{x_1^{i_1-1}, x_{i_1+1}^{i_2-1},\cdots, x_{i_{l-1}+1}^{i_l-1}: d(\bx_k,\bx_l)>\Delta \forall k,l \in [i_l], k\neq l}\bbP(\bx_1^{i_1-1})\bbP(\bx_{i_1+1}^{i_2-1}|\bx_1^{i_1})\bbP(\bx_{i_2+1}^{i_3-1}|\bx_1^{i_2}) \cdots \bbP(\bx_{i_{l-1}+1}^{i_l-1}|\bx_1^{i_{l-1}}) \nn\\
&\qquad \qquad \qquad \times  \prod_{j=1}^l \bbP(\bx_{i_j}|\bx_1^{i_j-1}) \\
&\qquad= \sum_{x_1^{i_1-1}, x_{i_1+1}^{i_2-1},\cdots, x_{i_{l-1}+1}^{i_l-1}:d(\bx_k,\bx_l)>\Delta \forall k,l \in [i_l], k\neq l }\bbP(\bx_1^{i_1-1})\bbP(\bx_{i_1+1}^{i_2-1}|\bx_1^{i_1})\bbP(\bx_{i_2+1}^{i_3-1}|\bx_1^{i_2}) \cdots\bbP(\bx_{i_{l-1}+1}^{i_l-1}|\bx_1^{i_{l-1}})\nn\\
&\qquad \qquad \qquad \times   \prod_{j=1}^l \frac{1}{|\calT(Q_X,\bx_1^{i_j-1})|}
\label{ah2}.
\end{align}
On the other hand, under the condition \ref{keycond},
by Lemma \ref{lem1some}, we have
\begin{align}
|\calT(Q_X)|\geq \calT(Q_X,\bx_1^{i-1})|\geq (1-e^{-n\delta}) |\calT(Q_X)|, \qquad \forall i \in[M] \label{ah3}
\end{align} for all $\bx_1^{i-1}$ occurring with non-zero probability.

From \eqref{ah2} and \eqref{ah3}, if $\min_{j,k \in [l], j\neq k} d(\bx_{i_j},\bx_{i_k})> \Delta$, we obtain
\begin{align}
&\bbP\bigg[\bigcap_{k\in A} \{\bX_k=\bx_k\}\bigg]\nn\\
&\qquad \leq \sum_{x_1^{i_1-1}, x_{i_1+1}^{i_2-1},\cdots, x_{i_{l-1}+1}^{i_l-1}:d(\bx_k,\bx_l)>\Delta \forall k,l \in [i_l], k\neq l}\bbP(\bx_1^{i_1-1})\bbP(\bx_{i_1+1}^{i_2-1}|\bx_1^{i_1})\bbP(\bx_{i_2+1}^{i_3-1}|\bx_1^{i_2}) \nn\\
&\qquad \qquad \times \cdots \times \bbP(\bx_{i_{l-1}+1}^{i_l-1}|\bx_1^{i_{l-1}}) \frac{1}{(1-e^{-n\delta})^l |\calT(Q_X)|^l}\\
&\qquad \leq \sum_{x_1^{i_1-1}, x_{i_1+1}^{i_2-1},\cdots, x_{i_{l-1}+1}^{i_l-1}}\bbP(\bx_1^{i_1-1})\bbP(\bx_{i_1+1}^{i_2-1}|\bx_1^{i_1})\bbP(\bx_{i_2+1}^{i_3-1}|\bx_1^{i_2}) \nn\\
&\qquad \qquad \times \cdots \times \bbP(\bx_{i_{l-1}+1}^{i_l-1}|\bx_1^{i_{l-1}}) \frac{1}{(1-e^{-n\delta})^l |\calT(Q_X)|^l}\\
&\qquad= \frac{1}{(1-e^{-n\delta})^l |\calT(Q_X)|^l}\\
&\qquad= \frac{1}{(1-e^{-n\delta})^{|\calA|} |\calT(Q_X)|^{|\calA|}}.
\end{align}
In addition, for any $M'\leq M$, from \eqref{ah2} and \eqref{ah3}, if $\min_{k,l \in [M']: k\neq l} d(\bx_k,\bx_l)> \Delta$, we also have
\begin{align}
&\bbP\bigg[\bigcap_{k\in [M']} \{\bX_k=\bx_k\}\bigg]\nn\\
&\qquad \geq \sum_{x_1^{i_1-1}, x_{i_1+1}^{i_2-1},\cdots, x_{i_{l-1}+1}^{i_l-1}}\bbP(\bx_1^{i_1-1})\bbP(\bx_{i_1+1}^{i_2-1}|\bx_1^{i_1})\bbP(\bx_{i_2+1}^{i_3-1}|\bx_1^{i_2}) \nn\\
&\qquad \qquad \times \cdots \times \bbP(\bx_{i_{l-1}+1}^{i_l-1}|\bx_1^{i_{l-1}}) \frac{1}{|\calT(Q_X)|^{M'}}\\
&\qquad= \frac{1}{|\calT(Q_X)|^{M'}}.
\end{align}
This concludes our proof of Lemma~\ref{lem:aux0}.
\section{Proof of Lemma \ref{aux1:lem}} \label{aux1lemproof}
First, we prove \eqref{Ap0}. Observe that
	\begin{align}
	\bbE[\calI(i,j)]&=\bbP\big[(\bX_i,\bX_j) \in \calT(P_{XX'}) \big]\\
	&=\sum_{(\bx_i,\bx_j) \in \calT(P_{XX'})}\bbP(\bx_i,\bx_j)\label{CD1}.
	\end{align}	
	Now, let
	\begin{align}
	\delta_n\triangleq \frac{e^{-n\delta}}{1-e^{-n\delta}}.
	\end{align}
	Then, under the condition \eqref{keycond} and $d(P_{XX'})>\Delta$, by Lemma \ref{lem2some}, we have
	\begin{align}
	\frac{(1-4 \delta_n^2)}{|\calT(Q_X)|^2}e^{-2\delta_n}\leq \bbP(\bx_i,\bx_j)\leq \frac{1}{(1-e^{-n\delta})^2|\calT(Q_X)|^2} , \qquad \forall (\bx_i,\bx_j) \in \calT(P_{XX'}) \label{CD2}
	\end{align} since $d(\bx_i,\bx_j)=d(P_{XX'})>\Delta$.
	From \eqref{CD1} and \eqref{CD2}, we have
	\begin{align}
	(1-4\delta_n^2) e^{-2\delta_n} \frac{|\calT(P_{XX'})|}{|\calT(Q_X)|^2}\leq \bbE[\calI(i,j)] \leq \frac{1}{(1-e^{-n\delta})^2} \frac{|\calT(P_{XX'})|}{|\calT(Q_X)|^2} \label{amattp}.
	\end{align}	
	Recall the definition of $L(P_{XX'})$ in \eqref{ta1}. From \eqref{amattp}, we have
	\begin{align}
	(1-4\delta_n^2) e^{-2\delta_n} L(P_{XX'}) \leq \bbE[\calI(i,j)]\leq \frac{1}{(1-e^{-n\delta})^2}  L(P_{XX'}).
	\end{align}
	Now, we prove \eqref{Ap1}. We consider three cases:
	\begin{itemize}
		\item Case 1: $i=k, j\neq l$.
		Observe that
		\begin{align}
		&\bbE[\calI(i,j)\calI(i,l)]\nn\\
		&\qquad =\bbP\big[(\bX_i,\bX_j) \in \calT(P_{XX'}), (\bX_i,\bX_l) \in \calT(P_{XX'})  \big]\\
		&\qquad =\sum_{(\bx_i,\bx_j,\bx_l)  \in \calT^3(Q_X)} \bbP(\bx_i,\bx_j,\bx_l)\indicator \{\{(\bx_i,\bx_j) \in \calT(P_{XX'})\}\cap\{(\bx_i,\bx_l) \in \calT(P_{XX'})\}\}\\
		&\qquad \leq \frac{1}{(1-e^{-n\delta})^3}\sum_{(\bx_i,\bx_j,\bx_l)  \in \calT^3(Q_X)} \bbP(\bx_i)\bbP(\bx_j) \bbP(\bx_l)\indicator \{\{(\bx_i,\bx_j) \in \calT(P_{XX'})\}\cap\{(\bx_i,\bx_l) \in \calT(P_{XX'})\}\} \label{ap2}\\
		&\qquad=\frac{1}{(1-e^{-n\delta})^3}\sum_{\bx_i \in \calT(Q_X)} \bbP(\bx_i) \bbP\big[(\bx_i, \bX_j) \in \calT(P_{XX'})\big] \bbP\big[(\bx_i, \bX_l) \in \calT(P_{XX'})\big]\\
		&\qquad= \frac{1}{(1-e^{-n\delta})^3}\sum_{\bx_i \in \calT(Q_X)} \bbP(\bx_i)  L^2(P_{XX'})\\
		&\qquad=\frac{1}{(1-e^{-n\delta})^3}  L^2(P_{XX'}) \label{mmu1}, 
		\end{align} where \eqref{ap2} follows from Lemma \ref{lem:aux0} and Lemma \ref{lem4some}. 
		\item $i\neq k, j=l$. The proof is similar to Case 1.
		\item $i\neq k, j\neq l$. Then, we have
		\begin{align}
		&\bbE[\calI(i,j)\calI(k,l)]\nn\\
		&\qquad =\bbP\big[(\bX_i,\bX_j) \in \calT(P_{XX'}), (\bX_k,\bX_l) \in \calT(P_{XX'})  \big]\\
		&\qquad =\sum_{(\bx_i,\bx_j,\bx_k,\bx_l)  \in \calT^4(Q_X)} \bbP(\bx_i,\bx_j,\bx_k,\bx_l)\indicator \{\{(\bx_i,\bx_j) \in \calT(P_{XX'})\}\cap\{(\bx_k,\bx_l) \in \calT(P_{XX'})\}\}\\
		&\qquad \leq \frac{1}{(1-e^{-n\delta})^4}\sum_{(\bx_i,\bx_j,\bx_k,\bx_l)  \in \calT^4(Q_X)} \bbP(\bx_i)\bbP(\bx_j) \bbP(\bx_k)\bbP(\bx_l)\nn\\
		&\qquad \qquad \times \indicator \{\{(\bx_i,\bx_j) \in \calT(P_{XX'})\}\cap\{(\bx_k,\bx_l) \in \calT(P_{XX'})\}\} \label{ap2b}\\
		&\qquad=\frac{1}{(1-e^{-n\delta})^4} \bbP\big[(\bX_i, \bX_j) \in \calT(P_{XX'})\big] \bbP\big[(\bX_k, \bX_l) \in \calT(P_{XX'})\big]\\
		&\qquad=\frac{1}{(1-e^{-n\delta})^4}  L^2(P_{XX'}) \label{mmu2},
		\end{align} where \eqref{ap2b} follows from Lemma \ref{lem:aux0} and Lemma \ref{lem4some}.
	\end{itemize}
From \eqref{mmu1} and \eqref{mmu2}, for any pairs $(i,j) \in [M]_*^2$ and $(k,l)\in [M]_*^2$ such that $(i,j)\neq (k,l)$, we have
\begin{align}
\bbE[\calI(i,j)\calI(k,l)]\leq \frac{1}{(1-e^{-n\delta})^4}  L^2(P_{XX'}),
\end{align} and we obtain \eqref{aAp1}.

Finally, by \cite{Csis00}, it is easy to see that
\begin{align}
L(P_{XX'})\doteq e^{-n I_P(X;X')} \label{ta1b}.
\end{align}
Hence, we obtain \eqref{Ap0} and \eqref{Ap1} from \eqref{aAp0} and \eqref{aAp1}, respectively. 

This concludes our proof of Lemma \ref{aux1:lem}.
\section{Proof of Lemma \ref{lem:aux1}}\label{lem:aux1proof}
Observe that
\begin{align}
\bbE[N(P_{XX'})]&=\bbE\bigg[\sum_m \sum_{m'\neq m} \indicator \big\{(\bX_m,\bX_{m'}) \in \calT(P_{XX'})\big\}\bigg]\\
&=\sum_m \sum_{m'\neq m}\bigg\{ \sum_{(\bx_m,\bx_{m'}) \in \calT(P_{XX'}):\atop d(\bx_m,\bx_{m'})>\Delta}\bbP(\bx_m,\bx_{m'}) +\sum_{(\bx_m,\bx_{m'}) \in \calT(P_{XX'}):\atop d(\bx_m,\bx_{m'})\leq \Delta} \bbP(\bx_m,\bx_{m'})  \bigg\} \label{mat1}.
\end{align}
On the other hand, by Lemma \ref{lem2some}, under the condition \ref{keycond}, it holds that 
\begin{align}
\bbP(\bx_m,\bx_{m'})=0 \label{mat2a}
\end{align}
if $d(\bx_m,\bx_{m'})\leq \Delta$, and
\begin{align}
\frac{1-4 \delta_n^2}{|\calT(Q_X)|^2}e^{-2\delta_n}\leq \bbP(\bx_m,\bx_{m'})\leq \frac{1}{(1-e^{-n\delta})^2|\calT(Q_X)|^2} \label{mat2}
\end{align} if $d(\bx_m,\bx_{m'})>\Delta$.

From \eqref{mat1}, \eqref{mat2a} and \eqref{mat2}, for any joint type $P_{XX'}$ such that $d(P_{XX'})>\Delta$, we obtain
\begin{align}
\bbE[N(P_{XX'})]\geq 
&\sum_m \sum_{m'\neq m} \sum_{(\bx_m,\bx_{m'}) \in \calT(P_{XX'}):\atop d(\bx_m,\bx_{m'})>\Delta} \bbP(\bx_m,\bx_{m'})\\
&\geq \sum_m \sum_{m'\neq m} \sum_{(\bx_m,\bx_{m'}) \in \calT(P_{XX'}):\atop d(\bx_m,\bx_{m'})>\Delta} \frac{1-4 \delta_n^2}{|\calT(Q_X)|^2}e^{-2\delta_n}\\
&=M(M-1)\sum_{(\bx_m,\bx_{m'}) \in \calT(P_{XX'})\atop d(P_{XX'})>\Delta} \frac{1-4 \delta_n^2}{|\calT(Q_X)|^2}e^{-2\delta_n}\\
&= M(M-1)|\calT(P_{XX'})|\frac{1-4 \delta_n^2}{|\calT(Q_X)|^2}e^{-2\delta_n} \label{mat3a}\\
&\geq (n+1)^{-3|\calX|^2} (1-4 \delta_n^2) e^{-2\delta_n} e^{n(2R- I_P(X;X'))} \label{mat3},
\end{align} where \eqref{mat3} follows from \cite{Csis00}.

Then, as $n$ sufficiently large, we have
\begin{align}
\bbP\big[\calE(P_{XX'})\big]&= 
\bbP\big[N(P_{XX'})<(1-4 \delta_n^2) e^{-2\delta_n}  \exp\{n[2R-I_P(X;X')-\eps]\}\big]\\
&\leq \bbP\big[N(P_{XX'})< e^{-n\eps/2}\bbE[N(P_{XX'})]\big] \label{mat10}\\
&=\bbP\bigg[\frac{N(P_{XX'})}{\bbE[N(P_{XX'})]}-1<-(1-e^{-n\eps/2})\bigg] \\
&\leq \frac{\var(N(P_{XX'}))}{(1-e^{-n\eps/2})^2 \big(\bbE[N(P_{XX'})]\big)^2} \label{mat11},
\end{align} where \eqref{mat10} follows from \eqref{mat3}, and \eqref{mat11} follows from Cauchy-Schwarz inequality.

Now, let
\begin{align}
\calI(m,m')\triangleq \indicator \big\{(\bX_m,\bX_{m'}) \in \calT(P_{XX'})\big\},
\end{align}
and
\begin{align}
L(P_{XX'})\triangleq \frac{|\calT(P_{XX'})|}{|\calT(Q_X)|^2}.
\end{align}
Then, it holds that \cite{Csis00}, 
\begin{align}
L(P_{XX'})\geq (n+1)^{-3|\calX|} e^{- nI_P(X;X')}
 \label{matsu}.
\end{align}
Hence, as $n$ sufficiently large, we have
\begin{align}
M(M-1) L(P_{XX'})&\geq (n+1)^{-3|\calX|} e^{n(2R-\frac{1}{n}-I_P(X;X'))}\\
&\geq (n+1)^{-3|\calX|} e^{n(\eps-\frac{1}{n})}\\
&\geq e^{n\eps/2} \label{pathai},
\end{align} where \eqref{pathai} follows from $\eps>>(\log n)/\sqrt{n}$. 

In addition, for any two fixed pairs $(m,m')$ and $(\tilm,\hatm)$ in $[M]_*^2$ such that $(m,m')\neq (\tilm,\hatm)$, by Lemma \ref{aux1:lem}, we have
\begin{align}
(1-4\delta_n^2) e^{-2\delta_n} L(P_{XX'}) \leq \bbE[\calI(m,m')]\leq \frac{1}{(1-e^{-n\delta})^2}  L(P_{XX'}) \label{ama3},
\end{align} and
\begin{align}
\bbE[\calI(m,m')\calI(\tilm,\hatm)]\leq \frac{1}{(1-e^{-n\delta})^4}  L^2(P_{XX'}) \label{ama2}.
\end{align} 
It follows that
\begin{align}
&\var(N(P_{XX'}))=\bbE[N^2(P_{XX'})]-\big(\bbE[N(P_{XX'})]\big)^2\\
&\qquad =\sum_{m,m',\tilm,\hatm}\bbE[\calI(m,m')\calI(\tilm,\hatm)] - \big(\bbE[N(P_{XX'})]\big)^2\\
&\qquad =\sum_{m,m'} \bbE[\calI(m,m')] + \sum_{(m,m')\neq (\tilm,\hatm)} \bbE\big[\calI(m,m') \calI(\tilm,\hatm)\big]-\big(\bbE[N(P_{XX'})]\big)^2\\
&\qquad \leq M(M-1) \frac{1}{(1-e^{-n\delta})^2} L(P_{XX'}) \nn\\
&\qquad \qquad + M(M-1)[M(M-1)-1] \frac{1}{(1-e^{-n\delta})^4} L^2(P_{XX'})-\big(\bbE[N(P_{XX'})]\big)^2\label{akay2}.
\end{align}
From \eqref{mat3a}, \eqref{mat11}, and \eqref{akay2}, as $n$ sufficiently large, we have
\begin{align}
&\bbP\big[\calE(P_{XX'})\big]\nn\\
&\leq  \frac{\var(N(P_{XX'}))}{(1-e^{-n\eps/2})^2 \big(\bbE[N(P_{XX'})]\big)^2}\\
&\leq \frac{1}{(1-e^{-n\eps/2})^2}\bigg[\frac{M(M-1) \frac{1}{(1-e^{-n\delta})^2} L(P_{XX'}) 
	+ M(M-1)[M(M-1)-1] \frac{1}{(1-e^{-n\delta})^4} L^2(P_{XX'})}{\big((1-4 \delta_n^2) e^{-2\delta_n} M(M-1) L(P_{XX'})\big)^2}-1 \bigg]\\
&\leq  \frac{1}{(1-e^{-n\eps/2})^2}\bigg[\frac{e^{4\delta_n}}{\big(1-4\delta_n^2\big)^2\big(1-e^{-n\delta}\big)^2}\bigg(\frac{1}{M(M-1) L(P_{XX'})}\bigg) + \frac{e^{4\delta_n}}{(1-4\delta_n^2)^2(1-e^{-n\delta})^4}-1\bigg] \label{bmat}\\
&\leq \frac{1}{(1-e^{-n\eps/2})^2}\bigg[\frac{e^{4\delta_n}}{\big(1-4\delta_n^2\big)^2\big(1-e^{-n\delta}\big)^2}e^{-n\eps/2} + \frac{e^{4\delta_n}}{(1-4\delta_n^2)^2(1-e^{-n\delta})^4}-1\bigg] \label{cmat},
\end{align} where \eqref{cmat} follows from \eqref{pathai}.  
\section{Proof of Lemma \ref{lem:ba1}}\label{lem:ba1proof}
It is clear that \eqref{afact} holds if $I_P(X;X')=0$ since the LHS of this inequality is equal to $0$ for this case. Now, we consider the case $I_P(X;X')>0$. Then, we can choose $\delta(\eps)$ such that $0<\delta(\eps)<<$ such that $I_P(X;X')>\delta(\eps)$. With an abuse of notation, we assume that $\delta(\eps)=\eps$. 
	
Now, observe that
	\begin{align}
	N(P_{XX'})=\sum_{m=1}^M \sum_{m'\neq m} \indicator \{(\bX_m,\bX_{m'}) \in \calT(P_{XX'})\}. 
	\end{align}
By Lemma \ref{aux1:lem}, we have
	\begin{align}
	\bbE[\indicator \{(\bX_m,\bX_{m'}) \in \calT(P_{XX'})\}]\doteq e^{-n I_P(X;X')}, \qquad \forall (m,m') \in [M]_*^2,
	\end{align} which leads to
	\begin{align}
	p&\triangleq  \frac{1}{M(M-1)}\bbE[N(P_{XX'})]\\
	&\doteq e^{-nI_P(X;X')} \label{mag1}.
	\end{align}
By choosing $t=e^{n(2R-I_P(X;X')+\eps)}+1$, then it is clear that
	\begin{align}
	M(M-1)p \leq t-1<M(M-1)-1
	\end{align} as $n$ sufficiently large if $I_P(X;X')>0$ and choose $\eps$ such that $0<\eps<<$. Then, by applying Lemma \ref{aule}, we obtain
	\begin{align}
	\bbP\big[N(P_{XX'})\geq e^{n(2R-I_P(X;X')+\eps)}\big]&\ddleq \exp\big\{-M(M-1)D\big(e^{-n(I_P(X;X')-\eps)}\|e^{-n I_P(X;X')}\big) \big\} \label{masu1}.
	\end{align} 
	Now, by using the fact that $D(a\|b)\geq a \big(\log  \frac{a}{b}-1\big)$ \cite{merhav_FnT2}, we have
	\begin{align}
	D\big(e^{-n(I_P(X;X')-\eps)}\|e^{-n I_P(X;X')}\big) \geq e^{-n(I_P(X;X')-\eps)} \big(n\eps-1\big) \label{masu2}.
	\end{align}
	From \eqref{masu1} and \eqref{masu2}, we obtain \eqref{afact}. Finally, \eqref{bfact} is a straightforward consequence of \eqref{afact}. 
	This concludes our proof of Lemma \ref{lem:ba1}.
\section{Proof of Lemma \ref{lem:ba2}}\label{lem:ba2proof}
Similar to the proof of Lemma \ref{lem:ba2}, by applying Lemma \ref{aule} with $t=e^{n\eps}$, we finally have
	\begin{align}
	\bbP\big[N(P_{XX'})\geq e^{n\eps}\big]& \ddleq \exp\big\{-M(M-1) D(e^{n(\eps-2R)}\|e^{-n I_P(X;X')})\big\} \label{matt1}.
	\end{align}
	On the other hand, we have
	\begin{align}
	D(e^{n(\eps-2R)}\|e^{-n I_P(X;X')})\geq e^{n(\eps-2R)} \big(n(\eps-2R+I_P(X;X')-1) \label{matt2b}.
	\end{align}	
	From \eqref{matt1} and \eqref{matt2b}, we obtain \eqref{uv} and \eqref{uv2}.
\section{Proof of Lemma \ref{lem:ab2}} \label{lem:ab2proof}
Observe that
	\begin{align}
	\bbE[N(P_{XX'})]&=\sum_{m=1}^M \sum_{m'\neq m} \bbE[\calI(m,m')]\\
	&\doteq e^{n(2R-I_P(X;X'))} \label{gh1} 
	\end{align} where \eqref{gh1} follows from Lemma \ref{aux1:lem}.
	An upper bound in \eqref{eq190} simply follows from Markov's inequality and \eqref{gh1}.
	
To show the lower bound, we use Suen's correlation inequality \cite[Appendix A]{Tamir2020a}. However, the dependency graph is now different from the one in \cite[Proof of Lemma 6]{Tamir2020a}. In this new dependency graph, each vertex $(i,j)$ is connected to all other vertices or $M(M-1)-1$ vertices. Using the results of Lemma \ref{aux1:lem}, we have
	\begin{align}
	\Theta:&=\frac{1}{2}\sum_{(i,j) \in [M]_*^2} \sum_{(k,l) \in [M]_*^2, (k,l)\neq (i,j)} \bbE[\calI(i,j)\calI(k,l)]\\
	&\dotleq \frac{1}{2} e^{2nR} e^{2nR} e^{-2n I_P(X;X')}\\
	&\doteq e^{n(4R-2I_P(X;X'))} \label{K1},
	\end{align}
	and
	\begin{align}
	\Omega&=\max_{(i,j) \in [M]_*^2} \sum_{(k,l) \in [M]_*^2, (k,l)\neq (i,j)} \bbE[\calI(k,l)]\\
	&\doteq e^{2nR} e^{-nI_P(X;X')} \label{as1}\\
	&\doteq e^{n(2R-I_P(X;X'))} \label{K2}.
	\end{align}
	In addition, we have
	\begin{align}
	\Delta&=\bbE[N(P_{XX'})]\\
	&\qquad \doteq e^{n(2R-I_P(X;X'))} \label{K3}.
	\end{align} 
	From \eqref{K1}, \eqref{K2}, and \eqref{K3}, we obtain
	\begin{align}
	\frac{\Delta^2}{8\Theta}\dotgeq 1,
	\end{align}
	and
	\begin{align}
	\frac{\Delta}{6\Omega}\doteq 1.
	\end{align}
	Now, by \cite[Eq.~(A.6)]{Tamir2020a}, we have
	\begin{align}
	\bbP[N(P_{XX'})=0]&\leq \exp\bigg\{-\min\bigg(\frac{\Delta^2}{8\Theta},\frac{\Delta}{6\Omega}, \frac{\Delta}{2}\bigg)\bigg\}\\
	&\dotleq \exp\bigg\{-\min\bigg(1,1,\frac{1}{2}e^{n(2R-I_P(X;X'))}\bigg)\bigg\} \\
	&= \exp\bigg\{-\frac{1}{2}e^{n(2R-I_P(X;X'))}\bigg\} \label{amo},
	\end{align} where \eqref{amo} follows from the assumption $I_P(X;X')\geq 2R$.
	
	From \eqref{amo}, by using the same arguments as \cite[Proof of Lemma 6]{Tamir2020a}, we obtain
	\begin{align}
	\bbP\big[N(P_{XX'})\geq 1\big]\dotgeq \exp\{n(2R-I_P(X;X'))\},
	\end{align} which is compatible with the upper bound, proving Lemma \ref{lem:ab2}.
\section{Proof of Lemma \ref{prop3}} \label{proofprop3}
From Lemma \ref{lem:ba1} and the fact that $0=e^{-n\infty}$, it holds that
	\begin{align}
	\bbP[N(P_{XX'})\geq e^{ns}]\doteq \exp(-n \infty) \label{eq265}
	\end{align} if $s> [2R-I_P(X;X')]_+$.  
	
	Now, for $s < [2R-I_P(X;X')]_+$ and $2R\leq I_P(X;X')$, then $s\leq 0$. It follows that
	\begin{align}
	\bbP[N(P_{XX'})\geq e^{ns}]&= \bbP[N(P_{XX'})\geq 1]\\
	&\doteq \exp\big\{n(2R-I_P(X;X'))\big\} \label{amet}\\
	&=\exp\big\{-n[I_P(X;X')-2R]_+\big\} \label{ametb}, 
	\end{align} where \eqref{amet} follows from Lemma \ref{lem:ab2}. 
	
	On the other hand, for $s< [2R-I_P(X;X')]_+$ and $2R>I_P(X;X')$, then we have
	\begin{align}
	\bbP[N(P_{XX'})\geq e^{ns}]&\leq 1\\
	&=\exp\big\{-n[I_P(X;X')-2R]_+ \big\} \label{QM1}.
	\end{align}
	In addition, for this case, there exists $\eps>0$ such that $2\eps\leq  \min\{2R-I_P(X;X'),[2R-I_P(X;X')]_+ - s\}$. Hence, by applying Lemma \ref{lem:aux1}, we have
	\begin{align}
	\bbP\big[N(P_{XX'})\geq (1-4\delta_n^2)e^{-2\delta_n} \exp\{n[2R-I_P(X;X')-\eps]\}\big]\to 1 \label{sme}. 
	\end{align}
	Furthermore, as $n$ sufficiently large, we also have
	\begin{align}
	\bbP[N(P_{XX'})\geq e^{ns}]&\geq \bbP\big[N(P_{XX'})\geq e^{n(2R-I_P(X;X')-2\eps)}\big]\\
	&\geq  \bbP\big[N(P_{XX'})\geq (1-4\delta_n^2)e^{-2\delta_n} \exp\{n[2R-I_P(X;X')-\eps]\}\big] \\
	&=1+o(1) \label{malai}\\
	&=(1+o(1)) \exp\big\{-n[I_P(X;X')-2R]_+ \big\} \label{smes2} \\
	&\doteq \exp\big\{-n[I_P(X;X')-2R]_+ \big\} \label{QM2},
	\end{align} where \eqref{malai} follows from \eqref{sme}, and \eqref{smes2} follows from $[I_P(X;X')-2R]_+=0$ for $2R>I_P(X;X')$.
	
	From \eqref{QM1} and \eqref{QM2}, we obtain
	\begin{align}
	\bbP[N(P_{XX'})\geq e^{ns}]\doteq \exp\big\{-n[I_P(X;X')-2R]_+ \big\} \label{matt2}
	\end{align} for $s< [2R-I_P(X;X')]_+$ and $2R>I_P(X;X')$.
	
	By combining \eqref{ametb} and \eqref{matt2}, we have
	\begin{align}
	\bbP[N(P_{XX'})\geq e^{ns}]\doteq \exp\big\{-n[I_P(X;X')-2R]_+ \big\}   \label{matt3}
	\end{align} for all $s < [2R-I_P(X;X')]_+$. 
	
	Finally, from \eqref{eq265} and \eqref{matt3}, we obtain
	\begin{align}
	E(R,P,s)=\begin{cases} [I_P(X;X')-2R]_+,&\qquad [2R-I_P(X;X')]+> s\\ +\infty,&\qquad [2R-I_P(X;X')]_+<s \end{cases}.
	\end{align}
	This concludes our proof of Lemma \ref{prop3}.
\section{Proof of Lemma \ref{lemSuen}}\label{lemSuenproof}
First, we prove the following auxiliary lemma.
\begin{lemma} \label{aux} For any $x \in [0,M^{-1}]$, the following holds:
\begin{align}
1-(1-x)^M < 2e^{-Mx}
\end{align} as $M$ sufficiently large.
\end{lemma}
\begin{IEEEproof}[Proof of Lemma \ref{aux}] Let $g(x)\triangleq 1-(1-x)^M-2e^{-Mx}$. This function has positive first-order derivative, hence $g(x)$ is increasing. Hence, for any $x \in [0,M^{-1}]$, we have
	\begin{align}
	g(x) &\leq g(M^{-1})\\
	     &=1-\bigg(1-\frac{1}{M}\bigg)^M-\frac{2}{e} \\
	     & \to 1-\frac{3}{e} \quad \mbox{as} \qquad  M \to \infty \label{qua}\\
	     &<0,
	\end{align} where \eqref{qua} follows from $\big(1+\frac{1}{x}\big)^{-x} \to 1/e$ as $x \to \infty$. This concludes our proof of Lemma \ref{aux}.
\end{IEEEproof}
Now, we return to prove Lemma \ref{lemSuenproof}. Observe that
\begin{align}
N(P_{XX'})=\sum_{m=1}^M \sum_{m'\neq m} \indicator \{(\bX_m,\bX_{m'}) \in \calT(P_{XX'})\}.
\end{align}
It follows that
\begin{align}
\bbE[N(P_{XX'})]&=\sum_{m=1}^M \sum_{m'\neq m}\bbP\{(\bX_m,\bX_{m'}) \in \calT(P_{XX'})\}\\
&\doteq e^{n(2R-I_P(X;X'))} \label{smai1},
\end{align} where \eqref{smai1} follows from Lemma \ref{aux1:lem}.
Then, we have
\begin{align}
&\bbP\bigg\{N(P_{XX'})\leq e^{-n\eps} \bbE[N(P_{XX'})]\bigg\}\nn\\
&\qquad \dotleq \bbP\bigg\{\sum_{m=1}^M \sum_{m'\neq m} \indicator \big\{(\bX_m,\bX_{m'}) \in \calT(P_{XX'})\big\} \leq e^{n(2R-I_P(X;X')-\eps)}\bigg\}\label{smai2}.
\end{align}
We consider two cases:
\begin{itemize}
	\item The condition \eqref{condkeyb} holds. 
\end{itemize}
On the space $\underbrace{\calX^n \times \calX^n \cdots \times \calX^n}_{M \enspace \mbox{terms}}$ define a probability measure $\bbP_{\Pi}$ such that
\begin{align}
\bbP_{\Pi}(\bx_1,\bx_2,\cdots,\bx_M)=\prod_{m=1}^M \bbP[\bX_m=\bx_m], \qquad \forall (\bx_1,\bx_2,\cdots,\bx_M) \in \underbrace{\calX^n \times \calX^n \cdots \times \calX^n}_{M \enspace \mbox{terms}}.
\end{align}
Then, for this case, for any $P_{XX'} \in \calD$, we have
\begin{align}
&\bbP\bigg\{\sum_{m=1}^M \sum_{m'\neq m} \indicator \big\{(\bX_m,\bX_{m'}) \in \calT(P_{XX'})\big\} \leq e^{n(2R-I_P(X;X')-\eps)}\bigg\}\nn\\
& =\sum_{\bx_1,\bx_2,\cdots,\bx_M}\bbP(\bx_1,\bx_2,\cdots,\bx_M)  \bigg\{\sum_{m=1}^M \sum_{m'\neq m} \indicator \big\{(\bx_m,\bx_{m'}) \in \calT(P_{XX'})\big\} \leq e^{n(2R-I_P(X;X')-\eps)}\bigg\}\\
&\leq \frac{1}{(1-e^{-n\delta})^M } \sum_{\bx_1,\bx_2,\cdots,\bx_M} \bbP_{\Pi}(\bx_1,\bx_2,\cdots,\bx_M)\bigg\{\sum_{m=1}^M \sum_{m'\neq m} \indicator \big\{(\bx_m,\bx_{m'}) \in \calT(P_{XX'})\big\} \leq e^{n(2R-I_P(X;X')-\eps)}\bigg\} \label{AB1} \\
&= e^{-e^{nR}\log(1-e^{-n\delta})} \bbP_{\Pi}\bigg\{\sum_{m=1}^M \sum_{m'\neq m} \indicator \big\{(\bX_m,\bX_{m'}) \in \calT(P_{XX'})\big\} \leq e^{n(2R-I_P(X;X')-\eps)}\bigg\} \label{AB2}\\
&\ddleq e^{-e^{nR}\log(1-e^{-n\delta})} \exp\bigg\{-\min\bigg(e^{n(2R-I_P(X;X'))},e^{nR}\bigg)\bigg\} \label{AB3},
\end{align} where \eqref{AB1} follows from Lemma \ref{lem:aux0}, and \eqref{AB3} follows from \cite[Lemma 2]{Tamir2020a}.

From \eqref{smai2} and \eqref{AB3}, we obtain
\begin{align}
&\min_{P_{XX'} \in D} \bbP\bigg\{N(P_{XX'})\leq e^{-n\eps} \bbE[N(P_{XX'})]\bigg\}\nn\\
&\qquad  \ddleq  e^{-e^{nR}\log(1-e^{-n\delta})} \exp\bigg\{-\min\bigg(e^{n(2R-\min_{P_{XX'} \in D}I_P(X;X'))},e^{nR}\bigg)\bigg\} \\
&\qquad \ddleq e^{-e^{nR}\log(1-e^{-n\delta})} \exp\big\{-e^{n(R-2\delta)}\big\} \label{AB4}\\
&\qquad \ddeq \exp\big\{-e^{n(R-2\delta)}\big\} \label{AB5},
\end{align} where \eqref{AB4} follows from $\min_{P_{XX'} \in \calD} I_P(X;X')\leq R+2\delta$ for this case, and \eqref{AB5} follows from $-\log(1-e^{-n\delta}) \sim e^{-n\delta}$.
\begin{itemize}
	\item Case 2: The condition \eqref{condkeyc} holds.
\end{itemize}
For this case, observe that
\begin{align}
&\bbP\bigg\{N(P_{XX'})> e^{-n\eps} \bbE[N(P_{XX'})]\bigg\}\nn\\
&\geq \bbP\bigg\{\bigg\{\sum_{m=1}^M \sum_{m'\neq m} \indicator \big\{(\bX_m,\bX_{m'}) \in \calT(P_{XX'})\big\} > e^{n(2R-I_P(X;X')-\eps)}\bigg\} \cap \bigg\{ \min_{(m,m') \in [M]_*^2} d(\bX_m,\bX_{m'})>\Delta \bigg\}\bigg\}\\
&=\sum_{\bx_1,\bx_2,\cdots,\bx_M}\bbP(\bx_1,\bx_2,\cdots,\bx_M) \nn\\
&\qquad \times \indicator \bigg\{\bigg\{\sum_{m=1}^M \sum_{m'\neq m} \indicator \big\{(\bx_m,\bx_{m'}) \in \calT(P_{XX'})\big\} > e^{n(2R-I_P(X;X')-\eps)}\bigg\} \cap \bigg\{ \min_{(m,m') \in [M]_*^2} d(\bx_m,\bx_{m'})>\Delta \bigg\}\bigg\}\\
& \geq \sum_{\bx_1,\bx_2,\cdots,\bx_M}\bbP_{\Pi}(\bx_1,\bx_2,\cdots,\bx_M) \nn\\
&\qquad \times \indicator \bigg\{\bigg\{\sum_{m=1}^M \sum_{m'\neq m} \indicator \big\{(\bx_m,\bx_{m'}) \in \calT(P_{XX'})\big\} > e^{-n\eps} \bbE[N(P_{XX'})]\bigg\} \cap \bigg\{ \min_{(m,m') \in [M]_*^2} d(\bx_m,\bx_{m'})>\Delta \bigg\}\bigg\} \label{ana}\\
& = \bbP_{\Pi}\bigg\{\bigg\{\sum_{m=1}^M \sum_{m'\neq m} \indicator \big\{(\bX_m,\bX_{m'}) \in \calT(P_{XX'})\big\} > e^{-n\eps} \bbE[N(P_{XX'})]\bigg\} \cap \bigg\{ \min_{(m,m') \in [M]_*^2} d(\bX_m,\bX_{m'})>\Delta \bigg\}\bigg\} \label{laha},
\end{align} where \eqref{ana} follows from Lemma \ref{lem:aux0} with $M'=M$ and Lemma \ref{lem4some}.	

From \eqref{laha}, we have
\begin{align}
&\bbP\bigg\{N(P_{XX'})\leq e^{-n\eps} \bbE[N(P_{XX'})]\bigg\}\nn\\
&\quad \leq \Pr_{\Pi}\bigg\{\bigg\{\sum_{m=1}^M \sum_{m'\neq m} \indicator \big\{(\bX_m,\bX_{m'}) \in \calT(P_{XX'})\big\} \leq e^{-n\eps} \bbE[N(P_{XX'})]\bigg\} \cup \bigg\{ \min_{(m,m') \in [M]_*^2} d(\bX_m,\bX_{m'})\leq \Delta \bigg\}\bigg\}\label{ley1}\\
&\quad= \bbP_{\Pi}\bigg\{\sum_{m=1}^M \sum_{m'\neq m} \indicator \big\{(\bX_m,\bX_{m'}) \in \calT(P_{XX'})\big\} \leq e^{-n\eps} \bbE[N(P_{XX'})]\bigg\}+ \bbP_{\Pi} \bigg\{ \min_{(m,m') \in [M]_*^2} d(\bX_m,\bX_{m'})\leq \Delta \bigg\} \label{matto}.
\end{align}
Now, observe that
\begin{align}
&\bigg\{ \min_{(m,m') \in [M]_*^2} d(\bX_m,\bX_{m'})\leq \Delta \bigg\}\nn\\
&\qquad =  \bigg\{\bigcup_{m=1}^M \bigcup_{m'\neq m} \{d(\bX_m,\bX_{m'})\leq \Delta\}  \bigg\}\\
&\qquad = \bigg\{\bigcup_{m=1}^M \bigcup_{m'\neq m} \bigcup_{\tilP_{XX'}\in \calQ(Q_X):d(\tilP_{XX'})\leq \Delta }\{(\bX_m,\bX_{m'}) \in \calT(\tilP_{XX'})\} \bigg\}\label{ley2a}.
\end{align}

Therefore, we have
\begin{align}
&\bbP_{\Pi} \bigg\{ \min_{(m,m') \in [M]_*^2} d(\bX_m,\bX_{m'})\leq \Delta \bigg\}\nn\\
&\qquad =\bbP_{\Pi}\bigg\{\bigcup_{m=1}^M \bigcup_{m'\neq m} \bigcup_{\tilP_{XX'} \in \calQ(Q_X):d(\tilP_{XX'})\leq \Delta }\{(\bX_m,\bX_{m'}) \in \calT(\tilP_{XX'})\} \bigg\}\\
&\qquad \leq \sum_{\tilP_{XX'} \in \calQ(Q_X):d(\tilP_{XX'})\leq \Delta }\sum_{m=1}^M \bbP_{\Pi}\bigg\{\bigcup_{m'\neq m} \{(\bX_m,\bX_{m'}) \in \calT(\tilP_{XX'})\} \bigg\} \label{mahay1}.
\end{align}
Now, for any joint-type $\tilP_{XX'} \in \calQ(Q_X)$ such that $d(\tilP_{XX'})\leq \Delta$, we have
\begin{align}
&\bbP_{\Pi}\bigg\{\bigcup_{m'\neq m} \{(\bX_m,\bX_{m'}) \in \calT(\tilP_{XX'})\} \bigg\}\nn\\
&\qquad=\bbE\bigg[\bbP_{\Pi}\bigg\{\bigcup_{m'\neq m} \{(\bX_m,\bX_{m'}) \in \calT(\tilP_{XX'})\}\bigg|\bX_m \bigg\}\bigg]\\
&\qquad=1-\bbE\bigg[\bbP_{\Pi}\bigg\{\bigcap_{m'\neq m} \{(\bX_m,\bX_{m'}) \notin \calT(\tilP_{XX'})\}\bigg|\bX_m \bigg\}\bigg]\\ 
&\qquad=1-\bbE\bigg[\bigg(\bbP_{\Pi}\big\{(\bX_m,\bX_{m \mod M+1})\notin \calT(\tilP_{XX'})\bigg|\bX_m\big\}\bigg)^M\bigg]\\
&\qquad\doteq 1-\big(1-e^{-nI_{\tilP}(X;X')}\big)^M\label{sta}, 
\end{align} where \eqref{sta} follows from the standard calculation (eg.~\cite{Csis00}).

Now, from the condition \eqref{condkeyc}, we have
\begin{align}
R \leq \min_{\tilP_{XX'} \in \calQ(Q_X): d(\tilP_{XX'})\leq \Delta} I_{\tilP}(X;X')-2\delta,
\end{align}
which leads to
\begin{align}
e^{-n\min_{\tilP_{XX'} \in \calQ(Q_X): d(\tilP_{XX'})\leq \Delta}I_{\tilP}(X;X')}\leq e^{-nR}=M^{-1} \label{keyp1}.
\end{align}
From \eqref{mahay1} and \eqref{sta}, we obtain
\begin{align}
&\bbP_{\Pi} \bigg\{ \min_{(m,m') \in [M]_*^2} d(\bX_m,\bX_{m'})\leq \Delta \bigg\}\nn\\
&\qquad \dotleq M\bigg[1-\big(1-e^{-n\min_{\tilP_{XX'} \in \calQ(Q_X): d(\tilP_{XX'})\leq \Delta}I_{\tilP}(X;X')}\big)^M\bigg]\label{sta1}\\
&\qquad \ddleq 2 M \exp\bigg\{ -Me^{-n\min_{\tilP_{XX'} \in \calQ(Q_X): d(\tilP_{XX'})\leq \Delta}I_{\tilP}(X;X')}  \bigg\} \label{mulao1}\\
& \qquad \ddleq \exp\bigg\{-e^{n\big(R-\min_{\tilP_{XX'} \in \calQ(Q_X): d(\tilP_{XX'})\leq \Delta}I_{\tilP}(X;X')\big)}  \bigg\}\\
& \qquad \ddleq \exp\bigg\{-e^{n\big(2R+2\delta-\min_{\tilP_{XX'}\in D}I_{\tilP}(X;X')\big)}  \bigg\} \label{sta2}
\end{align} where \eqref{mulao1} follows from Lemma \ref{aux} with \eqref{keyp1}, \eqref{sta2} follows from the condition \eqref{condkeyc}.

On the other hand, by \cite[Prep.~6]{Tamir2020a}, we have
\begin{align}
&\bbP_{\Pi}\bigg\{N(P_{XX'})\leq e^{-n\eps} \bbE[N(P_{XX'})]\bigg\}\\
&\qquad\doteq \bbP_{\Pi}\bigg\{N(P_{XX'})\leq e^{-n\eps} e^{n(2R-I_P(X;X'))} \bigg\}\\
&\qquad \ddleq \exp\bigg\{-e^{n(2R-I_P(X;X'))}\bigg\} \label{matly3}.
\end{align}

From \eqref{sta2} and \eqref{matly3}, under the condition \eqref{condkeyc}, we have
\begin{align}
\min_{P_{XX'} \in D} \bbP\bigg\{N(P_{XX'})\leq e^{-n\eps} \bbE[N(P_{XX'})]\bigg\}\ddleq \exp\bigg\{-e^{n(2R-\min_{P_{XX'}\in D} I_P(X;X'))}\bigg\} \label{sca}.
\end{align}
Finally, we obtain by combining \eqref{AB5} for the case 1 and \eqref{sca} for the case 2.

This concludes our proof of Lemma \ref{lemSuenproof}.

\section{Proof of Lemma \ref{uplem1}}
\label{uplem1proof}
Define a new probability measure $\Pi$ on $\underbrace{\calX^n \times \calX^n \cdots \times \calX^n}_{M \quad \text{times}}$:
\begin{align}
\bbP_{\Pi}(\bx_1,\bx_2,\cdots,\bx_M)=\prod_{m=1}^M \bbP[\bX_m=\bx_m], \qquad \forall (\bx_1,\bx_2,\cdots,\bx_M).
\end{align}
Observe that
\begin{align}
&\bbP\big(F_0\big)\nn\\
&\qquad =\bbP\bigg\{\sum_{P_{XX'} \in \calA_1 \cup \calA_2} N(P_{XX'})=0 \bigg\}\\
&\qquad = \bbP\bigg\{ \sum_{P_{XX'} \in \calA_1 \cup \calA_2} \sum_{m=1}^M \sum_{m'\neq m} \indicator \{(\bX_m,\bX_{m'}) \in \calT(P_{XX'})\}=0 \bigg\}\\
&\qquad=\bbP\bigg\{\bigcap_{P_{XX'} \in A_1 \cup A_2} \bigcap_{m=1}^M \bigcap_{m'\neq m} \{(\bX_m,\bX_{m'}) \in \calT(P_{XX'})\}^c \bigg\}\\
&\qquad=\bbP\bigg\{\bigcap_{P_{XX'} \in A_1 \cup A_2} \bigcap_{m=1}^M \bigcap_{m'\neq m} \{\{(\bX_m,\bX_{m'}) \in \calT(P_{XX'})\} \cap \{d(\bX_m,\bX_{m'})>\Delta\}\}^c \bigg\} \label{keym1}\\
&\qquad= \bbP\bigg\{\bigcap_{P_{XX'} \in A_1 \cup A_2} \bigcap_{m=1}^M \bigcap_{m'\neq m} \{(\bX_m,\bX_{m'}) \notin \calT(P_{XX'})\} \cup \{d(\bX_m,\bX_{m'})\leq \Delta\} \bigg\}\\
&\qquad=\sum_{\bx_1,\bx_2,\cdots,\bx_M} \bbP(\bx_1,\bx_2,\cdots,\bx_M)\nn\\
&\qquad \qquad \times \prod_{P_{XX'} \in A_1 \cup A_2} \prod_{m=1}^M \prod_{m'\neq m}\indicator \bigg\{ \{(\bx_m,\bx_{m'}) \notin \calT(P_{XX'})\} \cup \{d(\bx_m,\bx_{m'})\leq \Delta\}\bigg\} \label{eq440b}\\
&\qquad=\sum_{\bx_1,\bx_2,\cdots,\bx_M} \bbP(\bx_1,\bx_2,\cdots,\bx_M)\nn\\
&\qquad \qquad \times \prod_{P_{XX'} \in A_1 \cup A_2} \prod_{m=1}^M \prod_{m'\neq m}\bigg(1-\indicator \big\{ \{(\bx_m,\bx_{m'}) \in \calT(P_{XX'})\} \cap \{d(\bx_m,\bx_{m'})> \Delta\}\big\}\bigg) \label{eq440c}\\
&\qquad \geq \sum_{\bx_1,\bx_2,\cdots,\bx_M} \bbP(\bx_1,\bx_2,\cdots,\bx_M)\nn\\
&\qquad \qquad \times \prod_{P_{XX'} \in A_1 \cup A_2} \prod_{m=1}^M \prod_{m'\neq m}\bigg(1-\indicator \big\{(\bx_m,\bx_{m'}) \in \calT(P_{XX'})\}\bigg)\indicator  \{d(\bx_m,\bx_{m'})> \Delta\} \label{eq440d}\\
&\qquad = \sum_{\bx_1,\bx_2,\cdots,\bx_M} \bbP(\bx_1,\bx_2,\cdots,\bx_M)\nn\\
&\qquad \qquad \times \prod_{P_{XX'} \in A_1 \cup A_2} \prod_{m=1}^M \prod_{m'\neq m}\indicator \big\{(\bx_m,\bx_{m'}) \notin \calT(P_{XX'})\}\indicator  \{d(\bx_m,\bx_{m'})> \Delta\} \label{eq440e}\\
&\qquad \geq \sum_{\bx_1,\bx_2,\cdots,\bx_M} \bbP_{\Pi}(\bx_1,\bx_2,\cdots,\bx_M)\nn\\
&\qquad \qquad \times \Pi_{P_{XX'} \in A_1 \cup A_2} \Pi_{m=1}^M \Pi_{m'\neq m}\indicator \big\{(\bx_m,\bx_{m'}) \notin \calT(P_{XX'})\}\indicator  \{d(\bx_m,\bx_{m'})> \Delta\} \label{eq441}\\
&\qquad=\bbP_{\Pi}\bigg\{\bigcap_{P_{XX'} \in A_1 \cup A_2} \bigcap_{m=1}^M \bigcap_{m'\neq m} \{(\bX_m,\bX_{m'}) \notin \calT(P_{XX'})\} \cap \{d(\bX_m,\bX_{m'})>\Delta\}\bigg\} \label{eq444}\\
&\qquad=\bbP_{\Pi}\bigg\{\sum_{P_{XX'} \in A_1 \cup A_2} \sum_{m=1}^M \sum_{m'\neq m} \indicator \big\{\{(\bX_m,\bX_{m'}) \in \calT(P_{XX'})\} \cup \{d(\bX_m,\bX_{m'})\leq \Delta\}\big\}=0 \bigg\} \label{boundkey},
\end{align} where \eqref{keym1} follows from  $d(P_{XX'})>\Delta$ for all $P_{XX'} \in A_1 \cup A_2$ and $d(\bx_m,\bx_{m'})= d(\hatP_{\bx_m,\bx_{m'}})$,  \eqref{eq440d} follows from the fact that $1-\indicator \big\{\big\{(\bx_m,\bx_{m'}) \in \calT(P_{XX'})\}\cap \{d(\bx_m,\bx_{m'})>\Delta\}\big\} =\big(1-\indicator \big\{(\bx_m,\bx_{m'}) \in \calT(P_{XX'})\}\big)\indicator  \{d(\bx_m,\bx_{m'})> \Delta\}$ if $d(\bx_m,\bx_{m'})>\Delta$ and $1-\indicator \big\{\big\{(\bx_m,\bx_{m'}) \in \calT(P_{XX'})\}\cap \{d(\bx_m,\bx_{m'})>\Delta\}\big\}\geq 0 =\big(1-\indicator \big\{(\bx_m,\bx_{m'}) \in \calT(P_{XX'})\}\big)\indicator  \{d(\bx_m,\bx_{m'})> \Delta\}$ if $d(\bx_m,\bx_{m'})\leq \Delta$, 
\eqref{eq441} follows from  \cite[Lemma 4]{somekh_2019} and Lemma \ref{lem:aux0}.

To apply Lemma \eqref{jasonlem}, we form a dependency graph as follows. Define the family of Bernoulli random variables $\{\calI(m,m',P_{XX'})\}_{P_{XX'} \in \calA_1 \cup \calA_2, (m,m') \in [M]_*^2}$, where
\begin{align}
\calI(m,m',P_{XX'})\triangleq \indicator \big\{(\bX_m,\bX_{m'}) \in \calT(P_{XX'}) \cup \{d(\bX_m,\bX_{m'})\leq \Delta\}\big\}.
\end{align}
Then, we have
\begin{align}
\bbE_{\Pi}[\calI(m,m',P_{XX'})]&=\bbP_{\Pi}\big\{(\bX_m,\bX_{m'}) \in \calT(P_{XX'}) \cup \{d(\bX_m,\bX_{m'})\leq \Delta\}\big\}\\
&\leq \bbP_{\Pi}\{ (\bX_m,\bX_{m'}) \in \calT(P_{XX'})\} +\bbP_{\Pi}\big\{d(\bX_m,\bX_{m'})\leq \Delta\} \label{E1}.
\end{align}
On the other hand, we have
\begin{align}
\bbP_{\Pi}\big\{d(\bX_m,\bX_{m'})\leq \Delta\}&=\sum_{\bx_m,\bx_{m'}}\bbP_{\Pi}(\bx_m,\bx_{m'})\indicator \{d(\bx_m,\bx_{m'})\leq \Delta\} \\
&= \sum_{\bx_m,\bx_{m'}} \bbP(\bx_m) \bbP(\bx_{m'}) \indicator \{d(\bx_m,\bx_{m'})\leq \Delta\}\\
&= \sum_{P_{XX'}\in \calQ(Q_X)} \sum_{ (\bx_m,\bx_{m'})\in \calT(P_{XX'})} \bbP(\bx_m) \bbP(\bx_{m'}) \indicator \{d(\bx_m,\bx_{m'})\leq \Delta\}\\
&= \sum_{P_{XX'}\in \calQ(Q_X)} \sum_{ (\bx_m,\bx_{m'})\in \calT(P_{XX'})} \frac{1}{|T(Q_X)|^2} \indicator \{d(\bx_m,\bx_{m'})\leq \Delta\} \label{emet1}\\
&= \sum_{P_{XX'}\in \calQ(Q_X)} \sum_{ (\bx_m,\bx_{m'})\in \calT(P_{XX'})} \frac{1}{|T(Q_X)|^2}\indicator \{d(P_{XX'})\leq \Delta\}\\
&\doteq \max_{P_{XX'} \in \calQ(Q_X): d(P_{XX'})\leq \Delta} e^{-n I_P(X;X')}\\
&= e^{-n\min_{P_{XX'} \in \calQ(Q_X): d(P_{XX'})\leq \Delta} I_P(X;X')} \\
&\leq e^{-n\max_{P_{XX'} \in \calQ(Q_X): d(P_{XX'})> \Delta} I_P(X;X')}  \label{aski},
\end{align} where \eqref{emet1} follows from \ref{lem4some}, and \eqref{aski} holds by the condition \eqref{condkeymu} under \eqref{ek1cond}.

It follows from \eqref{E1} and \eqref{aski} that
\begin{align}
\bbE_{\Pi}[\calI(m,m',P_{XX'})]&\leq \bbP_{\Pi}\{ (\bX_m,\bX_{m'}) \in \calT(P_{XX'})\}+ e^{-n\max_{P_{XX'}\in \calQ(Q_X): d(P_{XX'})> \Delta} I_P(X;X')}\\
&\doteq  e^{-n I_P(X;X')}+ e^{-n\max_{P_{XX'} \in \calQ(Q_X): d(P_{XX'})> \Delta} I_P(X;X')}\\
&\leq e^{-n I_P(X;X')}+e^{-n I_P(X;X')} \label{lasa}\\
&\doteq e^{-n I_P(X;X')} \label{lasa2},
\end{align} where \eqref{lasa} follows from the fact that $d(P_{XX'})>\Delta$ for all $P_{XX'}\in \calA_1 \cup \calA_2$.

Now, we set
\begin{align}
x(m,m',P_{XX'})\triangleq 1-\exp\big\{-e^{n I_P(X;X')}\big\}.
\end{align}
Then, under the condition $\min_{P_{XX'} \in \calA_1\cup \calA_2} I_P(X;X')>R$, for all $(m,m',P_{XX'}) \in [M]_*^2 \times (\calA_1\cup \calA_2)$, it holds that
\begin{align}
&\bbE_{\Pi}[\calI(m,m',P_{XX'})]\\
&\qquad \dotleq e^{-n I_P(X;X')}\\
&\qquad \doteq 1- \exp\bigg\{-e^{-n I_P(X;X')}\bigg\} \label{mamet1}\\
&\qquad \doteq \bigg(1- \exp\bigg\{-e^{-n I_P(X;X')}\bigg\}\bigg) \bigg(\exp\bigg\{-e^{-n I_P(X;X')}\bigg\}\bigg)^{|A_1\cup A_2| e^{nR}} \label{es}\\
&\qquad = x(m,m',P_{XX'}) \prod_{(\tilm,\tilm',\tilP_{XX'})\sim (m,m',P_{XX'})}\bigg(1-x(\tilm,\tilm',\tilP_{XX'})\bigg), 
\end{align} where \eqref{mamet1} follows from the fact that $\lim_{x\to 0} \frac{e^{-x}}{1-x}=1$, \eqref{es} follows from $|\calA_1\cup \calA_2|\leq |\calQ(Q_X)|$ which is sub-exponential in $n$ and $\min_{P_{XX'} \in \calA_1\cup \calA_2} I_P(X;X')>R$. 

Then, by applying Lemma \ref{jasonlem} with $A= [M]_*^2 \times (\calA_1\cup \calA_2)$ and $B=\emptyset$, under the condition $\min_{P_{XX'} \in \calA_1\cup \calA_2} I_P(X;X')>R$ we have
\begin{align}
&\bbP_{\Pi}\bigg\{\sum_{P_{XX'} \in \calA_1 \cup \calA_2} \sum_{m=1}^M \sum_{m'\neq m} \indicator \big\{\{(\bX_m,\bX_{m'}) \in \calT(P_{XX'})\} \cup \{d(\bX_m,\bX_{m'})\leq \Delta\}\big\} =0\bigg\}\nn\\
&\qquad \geq \min_{P_{XX'} \in \calA_1 \cup \calA_2} \bigg(\exp\bigg\{-e^{n I_P(X;X')}\bigg\}\bigg)^{|\calA_1\cup \calA_2| M(M-1)} \\
&\qquad \ddeq  \exp\bigg\{-e^{n \max_{P_{XX'} \in \calA_1 \cup \calA_2}(2R-I_P(X;X'))}\bigg\} \\
&\qquad= \exp\bigg\{-e^{n \max_{P_{XX'} \in  \calA_2}(2R-I_P(X;X'))}\bigg\}\label{buga},
\end{align} where \eqref{buga} follows from the definition of $\calA_1$ and $\calA_2$.

Finally, the condition $\min_{P_{XX'} \in \calA_1\cup \calA_2} I_P(X;X')>R$ is the same as $\min_{P_{XX'} \in \calA_2} I_P(X;X')>R$, which is equivalent to the condition that
\begin{align}
E_0<E_{\rm{ex}}^{g}(R,Q_X,d,\Delta)&\triangleq  \min_{P_{XX'}\in \calQ(Q_X): d(P_{XX'})>\Delta, I_P(X;X')\leq R}  \bigg\{\Gamma(P_{XX'},R)+ I_P(X;X')-R\bigg\} \label{motsa}\\
&=E_{\rm{ex}}^{\rm{rgv}}(R,Q_X,d,\Delta),
\end{align} where \eqref{motsa} is obtained by using the same arguments to achieve \cite[I.~(30)]{Tamir2020a}. This concludes our proof of Lemma \ref{uplem1}.

%%%%%%%%%%%%%%

\section{Concentration Inequalities for Sums of Bernoulli Random Variables}
To obtain the TRC or develop concentration inequalities for the random coding exponents, we need to develop concentration inequalities for a sum of Bernoulli random variables. Since in RGV codebooks, all the codewords are correlated, standard concentration inequalities such as Suen's correlation inequality \cite{Janson1998NewVO, Tamir2020a}  cannot be applied. The main reason is that these standard inequality require a local dependency in the sum of random variables which only holds for the fixed-composition or i.i.d. random ensembles but not for RGV ones.  We develop concentration inequalities for a sum of $n$ terms where each term depends on all the $n-1$ other terms. Thanks to the structure of all these random variables, some concentration inequalities in the probability literature can be applied. In this section, we list all these inequalities. For the newly-developed inequality, the proof can be found in appendices.    
\begin{lemma}\label{spelem} \cite[Lemma 2.1]{Pelekis2015} Fix a positive number $n$ and let $\{x_1,x_2,\cdots,x_n\}$ be real numbers from the interval $[0,1]$. For every $A \subset [n]$, let $\zeta_A$ be defined as
	\begin{align}
	\zeta_A=\prod_{i\in A}x_i\prod_{i \in [n]\setminus A} (1-x_i).
	\end{align}
	Then, 
	\begin{align}
	\sum_{A \subset [n]} \zeta_A=\sum_{j=0}^n \sum_{A \in \partial_j[n]}\zeta_A=1 \label{fact1}
	\end{align}
	and
	\begin{align}
	\sum_{i=1}^n x_i=\sum_{j=0}^n j \sum_{A \in \partial_j[n]}\zeta_A \label{fact2},
	\end{align} where $\partial_j[n]$ denotes the family consisting of all subsets of $[n]$ of cardinality $j \in \{0,1,2,\cdots,n\}$.
\end{lemma}
%Based on Lemma \ref{spelem}, we develop a new lemma, which is an extension of Chernoff-Hoeffding theorem, whose proof can be found in Appendix \ref{lemkey:app}.
%\begin{lemma} \label{lemkey}Let $X_1,X_2,\cdots, X_n$ be identically distributed Bernoulli random variables with mean $p$ such that for all $A \subset [n]$,
%	\begin{align}
%	\bbE[\zeta_A]\leq (\gamma p)^{|A|} (1-\delta p)^{n-|A|} \label{asp1}
%	\end{align}	for some $\delta, \gamma\in [0,1]$, where $\zeta_A$ is defined in Lemma \ref{spelem}, i.e.,
%	\begin{align}
%	\zeta_A=\prod_{i\in A}X_i\prod_{i \in [n]\setminus A} (1-X_i).
%	\end{align}
%	Then, for any $\nu \in (0,p]$, the following holds:
%	\begin{align}
%	\bbP\bigg[\sum_{i=1}^n X_i \leq n(p-\nu) \bigg] &\leq  \bigg[1+\frac{p(1-\delta)}{1-p}+(\gamma-1)(p-\nu)\bigg]^n e^{-nD(p-\nu\|p)} \label{R1}.
%	\end{align}
%\end{lemma}
%\begin{remark}
%	The term 
%	$e^{-nD(p-\nu\|p)}$ is the Chernoff-Hoeffding upper bound for the sum of i.i.d. Bernoulli random variables with mean $p$, the term in bracket accounts for the dependence among $X_i: i \in [n]$.
%\end{remark}
The following result can be also derived from Lemma \ref{aule}. 
\begin{lemma} \label{lemkey} Suppose that $X_1,X_2,\cdots,X_n$	are random variables such that $X_i \in \{0,1\}$, for $i=1,2,\cdots,n$. Set $p=\frac{1}{n}\sum_{i=1}^n \bbE[X_i]$. Then, for any $\nu \in [0,p)$, it holds that
	\begin{align}
	\bbP\bigg[\sum_{i=1}^n X_i \leq  n(p-\nu)-1 \bigg]\leq 2 e^{-nD(p-\nu\|p)} \label{mattc}.
	\end{align}
\end{lemma}
\begin{IEEEproof}
Let $\tilX_i\triangleq 1-X_i$ for all $i\in [n]$ and set $\tilp\triangleq 1-p$. Then, we have
	\begin{align}
	\tilp=\frac{1}{n}\sum_{i=1}^n \bbE[\tilX_i]. 
	\end{align}
	Let $t-1=n(1-p)+n(1-p)\eps_0$ for some $\eps_0>0$ such that $(1-p)(1+\eps_0)<1$. Then, by applying  Lemma \ref{aule} for the Bernoulli sequence $\tilX_1,\tilX_2,\cdots,\tilX_n$, we have
	\begin{align}
	\bbP\bigg[\sum_{i=1}^n \tilX_i \geq  t \bigg] &\leq 2 e^{-nD(\tilp(1+\eps_0)\|\tilp)} \label{sumet1}\\
	&=2 e^{-nD((1-p)(1+\eps_0)\|1-p)} \label{sumet2}.
	\end{align}
	From \eqref{sumet2} and $\tilX_i=1-X_i$ for all $i\in [n]$, we obtain
	\begin{align}
	\bbP\bigg[\sum_{i=1}^n X_i \leq  n-t \bigg]\leq 2 e^{-nD((1-p)(1+\eps_0)\|1-p)} \label{sumet3}.
	\end{align}
	Now, by setting  $\eps_0\triangleq \nu/(1-p)$, we have $t= n(1-p+\nu)+1$. Then, from \eqref{sumet3}, we have
	\begin{align}
	\bbP\bigg[\sum_{i=1}^n X_i \leq n(p-\nu)-1 \bigg]&\leq 2 e^{-nD((1-p)(1+\eps_0)\|1-p)} \label{sumet4}\\
	&=2 e^{-nD(1-p+\nu\|1-p)}\\
	&= 2e^{-n D(p-\nu\|p)} \label{G2}, 
	\end{align} where \eqref{G2} follows from $D(a\|b)=D(1-a\|1-b)$. Final note is that $(1-p)(1+\eps_0)=1-p+\nu<1$ for all $\nu \in [0,p)$.

\end{IEEEproof}

Now, we recall the following result.
\begin{lemma} \cite[Theorem 1.2]{Pelekis2015} \label{lemkey2} There exists a universal constant $c\geq 1$ satisfying the following.  Suppose  $X_1,X_2,\cdots, X_n$ are random variables such that $0\leq X_i\leq 1$, for $i=1,2,\cdots,n$. Assume further that there exists constant $\gamma \in (0,1)$ such that for all $A \subset [n]$ the following condition holds true:
	\begin{align}
	\bbE\bigg[\prod_{i\in A}X_i\bigg]\leq \gamma^{|A|} \label{asp12}
	\end{align}	where $|A|$ denotes the cardinality of $A$. Fix a real number $\nu$ from the interval $\big(0,\frac{1}{\gamma}-1\big)$ and set $t=n\gamma+ n\gamma \nu$. Then,
	\begin{align}
	\bbP\bigg[\sum_{i=1}^n X_i \geq t \bigg] &\leq  ce^{-nD(\gamma(1+\nu)\|\gamma)} \label{R2},
	\end{align}
\end{lemma} 
where $D(\gamma(1+\nu)\|\gamma)$ is the Kullback-Leibler distance between $\gamma(1+\nu)$ and $\gamma$. 

%Now, we recall the following lemma.
%\begin{lemma} \cite{Impagliazzo2010} \label{impaglem} There exists a universal constant $c\geq 1$ satisfying the following. Suppose that $X_1,X_2,\cdots,X_n$ are random variables such that $0\leq X_i \leq 1$, for $i=1,2,\cdots,n$. Assume further that there exists constant $\gamma \in (0,1)$ such that for all $A \subset [n]$, the following condition holds true:
%	\begin{align}
%	\bbE\bigg[\prod_{i \in A} X_i\bigg] \leq \gamma^{|A|},
%	\end{align} where $|A|$ denotes the cardinality of $A$. Fix a real number $\eps$ from the interval $\big(0,\frac{1}{\gamma}-1\big)$ and set $t=n\gamma+n\gamma \eps$. Then 
%	\begin{align}
	%\bbP\bigg[\sum_{i=1}^n X_i \geq t\bigg]\leq c %e^{-nD(\gamma(1+\eps)\|\gamma)},
	%\end{align} where $D(\gamma(1+\eps)\|\gamma)$ is the Kullback-Leibler distance between $\gamma(1+\eps)$ and $\gamma$.
%\end{lemma}
 Now, to bound the probability in \eqref{boundkey}, we recall the following version of Suen's correlation inequality lemma in \cite{Janson1998NewVO}.
\begin{lemma} \cite[Lemma 1]{Janson1998NewVO} \label{jasonlem} Let $\{U_k\}_{k \in \calK}$, where $\calK$ is a set of multidimensional indexes, be a family of Bernoulli random variables. Let $G$ be a dependency graph for $\{U_k\}_{k \in \calK}$, i.e., a graph with vertex set $\calK$ such that if $A$ and $B$ are two disjoint subsets of $\calK$, and $G$ contains no edge between $A$ and $B$, then the families $\{U_k\}_{k \in A}$ and $\{U_k\}_{k \in B}$ are independent. Let $S_A\triangleq \sum_{k \in A} U_k$ for any $A \subset \calK$. Moreover, we write $k \sim l$ if $(k,l)$ is an edge in the dependency graph $G$. Suppose further that $x_k, k \in \calK$ are real numbers such that $0\leq x_k<1$ and
	\begin{align}
	\bbE[U_k]\leq x_k\prod_{l\sim k} \big(1-x_l), \qquad k \in \calK.
	\end{align}
	Then, for any two subsets $A,B \subset \calK$, it holds that
	\begin{align}
	\bbP\big(S_A=0|S_B=0\big)\geq \prod_{i \in A} (1-x_i).
	\end{align} 
\end{lemma}
%%%%%%%%%%%%%%

%\section{Proof of Lemma \ref{lemkey}} \label{lemkey:app}
%Let $\tilX_i\triangleq 1-X_i$ for all $i\in [n]$ and set $\tilp\triangleq 1-p$. Then, we have
%	\begin{align}
%	\tilp=\frac{1}{n}\sum_{i=1}^n \bbE[\tilX_i]. 
%	\end{align}
%	Let $t-1=n(1-p)+n(1-p)\eps_0$ for some $\eps_0>0$ such that $(1-p)(1+\eps_0)<1$. Then, by applying  Lemma \ref{aule} for the Bernoulli sequence $\tilX_1,\tilX_2,\cdots,\tilX_n$, we have
%	\begin{align}
%	\bbP\bigg[\sum_{i=1}^n \tilX_i \geq  t \bigg] &\leq 2 e^{-nD(\tilp(1+\eps_0)\|\tilp)} \label{sumet1}\\
%	&=2 e^{-nD((1-p)(1+\eps_0)\|1-p)} \label{sumet2}.
%	\end{align}
%	From \eqref{sumet2} and $\tilX_i=1-X_i$ for all $i\in [n]$, we obtain
%	\begin{align}
%	\bbP\bigg[\sum_{i=1}^n X_i \leq  n-t \bigg]\leq 2 e^{-nD((1-p)(1+\eps_0)\|1-p)} \label{sumet3}.
%	\end{align}
%	Now, by setting  $\eps_0\triangleq \nu/(1-p)$, we have $t= n(1-p+\nu)+1$. Then, from \eqref{sumet3}, we have
%	\begin{align}
%	\bbP\bigg[\sum_{i=1}^n X_i \leq n(p-\nu)-1 \bigg]&\leq 2 e^{-nD((1-p)(1+\eps_0)\|1-p)} \label{sumet4}\\
%	&=2 e^{-nD(1-p+\nu\|1-p)}\\
%	&= 2e^{-n D(p-\nu\|p)} \label{G2}, 
%	\end{align} where \eqref{G2} follows from $D(a\|b)=D(1-a\|1-b)$. Final note is that $(1-p)(1+\eps_0)=1-p+\nu<1$ for all $\nu \in [0,p)$.

\section{Proof of Lemma \ref{lem:aut1}}\label{lem:aut1proof}
Fix an $m \in [M]$. For any conditional type $P_{X'Y} \in \calP_n(\calX \times \calY)$ such that $P_{X'}=Q_X$ and $P_Y=\hatP_{\by}$, define
\begin{align}
N_{m,\by}(P_{X'Y})&\triangleq \big|\big\{\bX_{m'}:  (\bX_{m'},\by) \in \calT(P_{X'Y}), m'\neq m \big\}\big|\\
&=\sum_{m'\neq m} \indicator \big\{(\bX_{m'},\by) \in \calT(P_{X'Y})\big\} \label{laban}. 
\end{align}
Observe that
\begin{align}
\bbE\big[\indicator \big\{(\bX_{m'},\by) \in \calT(P_{X'Y})\big\}\big]&= \bbP\big[(\bX_{m'},\by) \in \calT(P_{X'Y})\big]\\
&= \sum_{\bx_m' \in \calT(P_{X'|Y})}\bbP(\bX_{m'}=\bx_{m'}) \\
&=\sum_{\bx_m' \in \calT(P_{X'|Y})}\frac{1}{|\calT(Q_X)|}  \label{ms}\\
&\doteq e^{-n I_P(X';Y)} \label{ms2},
\end{align} where \eqref{ms} follows from Lemma \ref{lem4some}, and \eqref{ms2} follows from \cite{Csis00}. Hence, $N_{m,\by}(P_{X'Y}) $ is a sum of $M-1$ binary-valued random variables, each has the expectation $e^{-n I(X';Y) }$. 

Now, from \eqref{defZmy} and \eqref{laban}, we can express $Z_m(\by)$ as 
\begin{align}
Z_m(\by)=\sum_{P_{X'|Y}: P_{X'}=Q_X} N_{m,y}(P_{X'Y})e^{n g(P_{X'Y})} \label{AQ1}.
\end{align}
Hence, by considering the randomness of $\{\bX_{m'}\}$, we have
\begin{align}
&\bbP\big[Z_m(\by)\leq \exp\big\{n\alpha(R-\eps, \hatP_{\by} ) \} \big]\nn\\
&\qquad \leq \bbP\bigg[\sum_{P_{X'|Y}: P_{X'}=Q_X} N_{m,\by}(P_{X'Y})e^{n g(P_{X'Y})} \leq  \exp\big\{n\alpha(R-\eps, \hatP_{\by} ) \}\bigg]\\
&\qquad \leq \bbP\bigg[\max_{P_{X'|Y}: P_{X'}=Q_X} N_{m,\by}(P_{X'Y})e^{n g(P_{X'Y})} \leq  \exp\big\{n\alpha(R-\eps, \hatP_{\by} ) \}\bigg]\\
&\qquad = \bbP\bigg[\bigcap_{P_{X'|Y}: P_{X'}=Q_X} \big\{N_{m,\by}(P)e^{n g(P_{X'Y})} \leq  \exp\big\{n\alpha(R-\eps, \hatP_{\by} ) \}\big\}\bigg]\\
&\qquad=\bbP\bigg[\bigcap_{P_{X'|Y}: P_{X'}=Q_X} \big\{N_{m,\by}(P_{X'Y}) \leq  \exp\big\{n\alpha(R-\eps, \hatP_{\by} ) -g(P_{X'Y})\}\bigg\}\bigg]\label{AQ2}.
\end{align}
As mentioned above, $N_{m,\by}(P_{X'Y}) $ is a sum of $M-1$ binary-valued random variables, each has the expectation $e^{-n I(X';Y) }$. However, different from i.i.d. random codebook ensembles, these random variables are correlated.

As \cite[Appendix B]{Merhav2017a}, we argue that by the definition of $\alpha(R-\eps, \hatP_{\by})$, there must exist some $P_{X'|Y}^*$ such that for $P_{X'Y}^*\triangleq \hatP_{\by} \times P_{X'|Y}^*$, $I_{P^*}(X';Y)\leq R-\eps$ and $R-\eps-I_{P^*}(X';Y)\geq \alpha(R-\eps, \hatP_{\by})-g(P_{X'Y}^*)$. To see why this is true, assume conversely, that for every $P_{X'|Y}$, which define $P_{X'Y}\triangleq \hatP_{\by} \times P_{X'|Y}$, either $I_P(X';Y)>R-\eps$ or $R-I_P(X';Y)-\eps<\alpha(R-\eps,\hatP_{\by})-g(P_{X'Y})$, which means that for every $P_{X'Y}$,
\begin{align}
R-\eps &<\max\big\{I_P(X';Y), I_P(X';Y)+\alpha(R-\eps, \hatP_{\by})-g(P_{X'Y}) \big\}\\
&=I_P(X';Y)+\big[\alpha(R-\eps, \hatP_{\by})-g(P_{X'Y})]_+,
\end{align} which implies that for every $P_{X'|Y}$, there exists $t \in [0,1]$ such that
\begin{align}
R-\eps &< \max\big\{I_P(X';Y), I_P(X';Y)+\alpha(R-\eps, \hatP_{\by})-g(P_{X'Y}) \big\}\\
&=I_P(X';Y)+t\big[\alpha(R-\eps, \hatP_{\by})-g(P_{X'Y})],
\end{align} or equivalently,
\begin{align}
\alpha(R-\eps, \hatP_{\by})&> \max_{P_{X'|Y}: P_{X'}=Q_X}\min_{0\leq t \leq 1} g(P_{X'Y})+ \frac{R-I_P(X';Y)-\eps}{t}\\
&= \max_{P_{X'|Y}: P_{X'}=Q_X}\begin{cases} g(P_{X'Y})+R-I_P(X';Y)-\eps& \quad I_P(X';Y)\leq R-\eps\\
-\infty & \quad I_P(X';Y)>R-\eps  \end{cases}\\
&=\max_{P_{X'|XY}: P_{X'}=Q_X,\atop I_P(X';Y)\leq R-\eps}\big[g(P_{X'Y})-I_P(X';Y)\big]+R-\eps\\
&=\alpha(R-\eps,\hatP_{\by}),
\end{align} which is a contradiction.

Now, from \eqref{AQ2} and the existence of $P_{X'Y}^*$ as above, it holds that
\begin{align}
&\bbP\big[Z_m(\by)\leq \exp\big\{n\alpha(R-\eps, \hatP_{\by} ) \} \big]\nn\\
&\qquad \leq \bbP\big[N_{m,\by}(P_{X'Y}^*)\leq \exp\big\{n[\alpha(R-\eps, \hatP_{\by} )-g(P_{X'Y}^*)] \} \big] \label{AQ4}.
\end{align}
Different from \cite{Merhav2018a}, $N_{\by}(P_{X'Y}^*)$ is now not the sum of i.i.d. Bernoulli random variables but these random variables are still identically distributed and weakly dependent.

Now, let
\begin{align}
Z_{m'}\triangleq  \indicator \big\{(\bX_{m'},\by) \in \calT(P_{X'Y}^*) \big\}, \qquad \forall m'\in M_{-}\triangleq [M]\setminus \{m\},
\end{align}
and
\begin{align}
p\triangleq \bbP\big[(\bX_2,\by) \in \calT(P_{X'Y}^*) \big].
\end{align}
Now, let $\nu \in (0,p)$ be chosen such that
\begin{align}
(M-1)(p-\nu)=\exp\big\{n[\alpha(R-\eps, \hatP_{\by} )-g(P_{X'Y}^*)]\big\} \label{sat}.
\end{align}
The existence of $\nu$ is guaranteed since \eqref{sat} is equivalent to
\begin{align}
\nu&=p-\frac{\exp\big\{n[\alpha(R-\eps, \hatP_{\by} )-g(P_{X'Y}^*)]\big\}}{M-1}\\
&\geq p- \frac{\exp\big\{n[R-\eps-I_{P^*}(X';Y)]\big\}}{M-1}\\
&= p- \frac{\exp\big\{n[R-\eps-I_{P^*}(X';Y)]\big\}}{\exp(nR)-1}\\
&\doteq \exp\big\{-nI_{P^*}(X';Y) \big\}-\exp\big\{-n(I_{P^*}(X';Y)+\eps) \big\}>0 \label{ato},
\end{align} so $\nu \in (0,p)$. 

By applying Lemma \ref{lemkey} with $n=M-1$, $X_i=Z_i$, $p=\bbP\big[(\bX_2,\by) \in \calT(P_{X'Y}^*) \big]$, and $\nu$ satisfying \eqref{sat}, we have
\begin{align}
&\bbP\bigg[N_{\by}(P_{X'Y}^*)\leq \exp\big\{n[\alpha(R-\eps, \hatP_{\by} )-g(P_{X'Y}^*)]\big\} \bigg]\nn\\
&\qquad \doteq \bbP\bigg[N_{\by}(P_{X'Y}^*)\leq \exp\big\{n[\alpha(R-\eps, \hatP_{\by} )-g(P_{X'Y}^*)]\big\} \bigg] \label{x1}\\
&\qquad \leq 2 \exp\big(-(M-1) D(p-\nu\|p)\big) \label{amet1}\\
&\qquad \doteq  \exp(-e^{nR} D(p-\nu\|p)) \label{abat}.
\end{align} 

Now, since $p\doteq \exp(-n I_{P^*}(X';Y))$, from \eqref{ato}, we also have
\begin{align}
&(M-1)\big[(\gamma-1)(p-\nu)\big]\nn\\
&\qquad \dotleq \exp(nR) \bigg[\bigg(\frac{1}{1-e^{-n\delta}}-1\bigg) \exp\big\{-n(I_{P^*}(X';Y)+\eps) \big\}\bigg] \\
&\qquad \dotleq  \frac{e^{-n (\delta+\eps)}}{1-e^{-n\delta}}\exp\big[n(R- I_{P^*}(X';Y))\big] \label{amen1}.
\end{align}

On the other hand, we have
\begin{align}
\exp(-e^{nR} D(p-\nu\|p))&=\exp\bigg\{-e^{nR}D(e^{-an}\|e^{-bn}) \bigg\}
\end{align} where $a\triangleq R+g(P_{X'Y}^*)-\alpha(R-\eps,\hatP_{\by})$ and $b\triangleq I_{P^*}(X';Y)$. It is easy to see that
\begin{align}
a-b&=R+g(P_{X'Y}^*)-\alpha(R-\eps,\hatP_{\by})-I_{P^*}(X';Y)\\
&\geq \eps. 
\end{align}
Hence, by using the following fact \cite[Sec.~ 6.3]{CIT-052}:
\begin{align}
D(a\|b)\geq a \log \frac{a}{b}+b-a,
\end{align} we have
\begin{align}
D(e^{-an}\|e^{-bn})&\geq e^{-bn}\big[1+e^{(b-a)n}((b-a)n-1)\big].
\end{align}
Hence, we obtain
\begin{align}
\exp(-e^{nR} D(p-\nu\|p))&\leq \exp\bigg\{- e^{n(R-I_{P^*}(X';Y))}[1-e^{-n\eps}(1+n\eps)]  \bigg\} \label{amen2}.
\end{align}
From \eqref{abat}, \eqref{amen1}, and \eqref{amen2}, we obtain
\begin{align}
&\bbP\bigg[N_{m,\by}(P_{X'Y}^*)\leq \exp\big\{n[\alpha(R-\eps, \hatP_{\by} )-g(P_{X'Y}^*)]\big\} \bigg]\nn\\
&\qquad \dotleq \exp\bigg\{\frac{e^{-n (\delta+\eps)}}{1-e^{-n\delta}}\exp\big[n(R- I_{P^*}(X';Y))\big]\bigg\} \exp\bigg\{- e^{n(R-I_{P^*}(X';Y))}[1-e^{-n\eps}(1+n\eps)]  \bigg\} \label{amot}\\
&\qquad= \exp\bigg\{- e^{n(R-I_{P^*}(X';Y))}\bigg[1- \frac{e^{-n (\delta+\eps)}}{1-e^{-n\delta}}-e^{-n\eps}(1+n\eps)\bigg]  \bigg\}\\
&\qquad \leq \exp\bigg\{- e^{n\eps}\bigg[1- \frac{e^{-n (\delta+\eps)}}{1-e^{-n\delta}}-e^{-n\eps}(1+n\eps)\bigg]  \bigg\} \label{amot3},
\end{align} where \eqref{amot3} follows from the fact that $I_{P^*}(X';Y)\leq R-\eps$. 

From \eqref{AQ4} and \eqref{amot3}, we obtain
\begin{align}
&\Pr\big[Z_m(\by)\leq \exp\big\{n\alpha(R-\eps, \hatP_{\by} ) \} \big]\nn\\
&\qquad \dotleq \exp\bigg\{- e^{n\eps}\bigg[1- \frac{e^{-n(\delta+\eps)}}{1-e^{-n \delta}}-e^{-n\eps}(1+n\eps)\bigg]  \bigg\} \label{amat}.
\end{align}
This concludes our proof of Lemma \ref{lem:aut1}.
\section{Proof of Lemma \ref{smolem}}\label{smolemproof}
The proof is based on \cite[Proof of Prep.~5]{Tamir2020a}. However, there are some changes to account for the dependency among the codewords. One such an important change is to replace the Hoeffding's inequality in \cite[Proof of Prep.~5]{Tamir2020a} by a generalized version of this inequality in \cite{Impagliazzo2010}. 

By using the union bound, we have
\begin{align}
\bbP\{\hat{\calB}_n(\sigma)\}
&= \bbP\bigg\{\bigcup_{m=1}^M \bigcup_{m'\neq m} \bigcup_{\by} \hat{\calB}_n(\sigma,m,m',\by) \bigg\}\\
& \leq \sum_{m=1}^M \sum_{m'\neq m} \sum_{\by} \bbP\bigg\{\hat{\calB}_n(\sigma,m,m',\by)\bigg\} \label{umo1}.
\end{align}
In addition, for any joint type $P_{XY} \in \calP_n(\calX \times \calY)$, let
\begin{align}
N(P_{XY})\triangleq \sum_{\tilm \in [M]\setminus \{m,m'\}} \indicator \{(\bX_{\tilm}, \by) \in \calT(P_{XY})\},
\end{align}
then we also have
\begin{align}
&\bbP\big\{\hat{\calB}_n(\sigma,m,m',\by)\big\}\nn\\
&\qquad \doteq \sum_{P_{XY}: P_X=Q_X, I_P(X;Y)\leq R} \bbP\bigg\{N(P_{XY})\geq e^{n(\beta(R,P_Y)+\sigma-g(P_{XY}))} \bigg\}\nn\\
&\qquad \qquad +  \sum_{P_{XY}: P_X=Q_X, I_P(X;Y)> R} \bbP\bigg\{N(P_{XY})\geq e^{n(\beta(R,P_Y)+\sigma-g(P_{XY}))} \bigg\} \label{H6}
\end{align} where \eqref{H6} follows from \cite[Eq.~(H.6)]{Tamir2020a}. 

Now, observe that
\begin{align}
&\bbP\bigg\{N(P_{XY})\geq e^{n(\beta(R,P_Y)+\sigma-g(P_{XY}))} \bigg\}\nn\\
&\qquad \leq \bbP\bigg\{N(P_{XY})\geq e^{n(R+\sigma-I_P(X;Y))} \bigg\} \label{H9}\\
&\qquad =\bbP\bigg\{\sum_{\tilm \in [M]\setminus \{m,m'\}} \indicator \{(\bX_{\tilm}, \by) \in \calT(P_{XY})\}\geq e^{n(R+\sigma-I_P(X;Y))} \bigg\} \label{lam1}
\end{align} where \eqref{H9} follows from \cite[Eq.~(H.9)]{Tamir2020a}.  

Define a new probability measure $\Pi$ on $\underbrace{\calX^n \times \calX^n \cdots \times \calX^n}_{M \quad \text{times}}$:
\begin{align}
\bbP_{\Pi}(\bx_1,\bx_2,\cdots,\bx_M)=\prod_{m=1}^M \bbP(\bX_m=\bx_m), \qquad \forall (\bx_1,\bx_2,\cdots,\bx_M).
\end{align}
Note that for any $A \subset [M]\setminus \{m,m'\}$, under the condition \eqref{keycond} we have
\begin{align}
\bbE\bigg[\prod_{\tilm \in A} \indicator \{(\bX_{\tilm}, \by) \in \calT(P_{XY})\} \bigg]&\leq \frac{1}{(1-e^{-n\delta})^{|A|}} \bbE_{\Pi} \bigg[\prod_{\tilm \in A} \indicator \{(\bX_{\tilm}, \by) \in \calT(P_{XY})\} \bigg] \label{amat1}\\
&=\frac{1}{(1-e^{-n\delta})^{|A|}}\prod_{\tilm \in A} \bbP\big\{(\tilbX_m,\by) \in \calT(P_{XY}) \big\} \label{ab1}
\end{align} where \eqref{amat1} follows from the change of measure and Lemma \ref{lem:aux0}.

Now, we have
\begin{align}
\bbP\big\{(\tilbX_m,\by) \in \calT(P_{XY}) \big\}&=\sum_{\tilbx_m \in \calT(P_{XY}|\by)} \bbP(\tilbx_m)\\
&= \sum_{\tilbx_m \in \calT(P_{XY}|\by)} \frac{1}{|\calT(Q_X)|} \label{mott1}\\
& \doteq e^{-nI_P(X;Y)} \label{mott2}
\end{align} where \eqref{mott1} follows from Lemma \ref{lem4some}, and \eqref{mott2} follows from \cite{Csis00}.

From \eqref{ab1} and \eqref{mott2}, we obtain
\begin{align}
\bbE\bigg[\prod_{\tilm \in A} \indicator \{(\bX_{\tilm}, \by) \in \calT(P_{XY})\} \bigg]\dotleq \gamma^{|A|} \label{but0}
\end{align} where
\begin{align}
\gamma=(1-e^{-n\delta})^{-1} e^{-n I_P(X;Y)} \label{but1}.
\end{align}
Hence, if $R \geq I_P(X;Y)$, we have
\begin{align}
&\bbP\bigg\{\sum_{\tilm \in [M]\setminus \{m,m'\}} \indicator \{(\bX_{\tilm}, \by) \in \calT(P_{XY})\}\geq e^{n(R+\sigma-I_P(X;Y))} \bigg\}\nn\\
&\qquad \dotleq  \exp\bigg\{-e^{nR} D\bigg((1-e^{-n\delta})^{-1} e^{\sigma-I_P(X;Y)} \bigg \|(1-e^{-n\delta})^{-1} e^{-n I_P(X;Y)}  \bigg)\bigg\} \label{mattr1}\\
&\qquad \leq \exp\bigg\{-e^{nR} (1-e^{-n\delta})^{-1}e^{-n(I_P(X;Y)-\sigma)}.\bigg(\log  \frac{e^{-n(I_P(X;Y)-\sigma)}}{e^{-nI_P(X;Y)}}-1  \bigg)\bigg\}\label{mattr2} \\
&\qquad = \exp\bigg\{-(1-e^{-n\delta})^{-1}e^{n(R-I_P(X;Y)+\sigma)}(n\sigma-1) \}\\
&\qquad \ddleq \exp\{-e^{n\sigma}\} \label{H14},
\end{align} where \eqref{mattr1} follows from Lemma \ref{lemkey2}, \eqref{mattr2} follows from 
the fact that $D(a\|b)\geq a \big(\log  \frac{a}{b}-1\big)$ \cite[p.~167]{merhav_FnT2}, and \eqref{H14} follows from $R\geq I_P(X;Y)$. 

From \eqref{lam1} and \eqref{H14}, we obtain
\begin{align}
\bbP\bigg\{N(P_{XY})\geq e^{n(\beta(R,P_Y)+\sigma-g(P_{XY}))} \bigg\}\ddleq \exp\{-e^{n\sigma}\}, \quad \mbox{if} \quad I_P(X;Y)\geq R \label{tfact1}.
\end{align}

Similarly, for the case $R<I_P(X;Y)$, we have
\begin{align}
&\bbP\bigg\{N(P_{XY})\geq e^{n(\beta(R,P_Y)+\sigma-g(P_{XY}))} \bigg\}\nn\\
&\qquad \leq \bbP\big\{N(P_{XY})\geq e^{n\sigma}\big\}\\
&\qquad = \bbP\bigg\{\sum_{\tilm \in [M]\setminus \{m,m'\}} \indicator \{(\bX_{\tilm}, \by) \in \calT(P_{XY})\}\geq e^{n\sigma} \bigg\}\nn\\
&\qquad \leq \exp\bigg\{-e^{nR} D\big((1-e^{-n\delta})^{-1}e^{-n(R-\sigma)}\big\| (1-e^{-n\delta})^{-1}e^{-n I_P(X;Y)}\big) \bigg\} \label{suchi1}\\
&\qquad=\exp\bigg\{-(1-e^{-n\delta})^{-1}e^{n\sigma}[n (I_P(X;Y)-R+\sigma)-1] \bigg\}\\
&\qquad \ddleq \exp\{-e^{n\sigma}\} \label{suchi2},
\end{align} where \eqref{suchi1} is obtained by applying Lemma \ref{lemkey2} and the change of measures as the arguments to achieve \eqref{H14}, and \eqref{suchi2} follows from the same arguments to achieve \eqref{mattr2}, and \eqref{suchi2} follows from $I_P(X;Y)>R$.

From \eqref{H6}, \eqref{tfact1}, and \eqref{suchi2}, we obtain
\begin{align}
\bbP\bigg\{\hat{\calB}_n(\sigma,m,m',\by)\bigg\} \ddleq \exp\{-e^{n\sigma}\} \label{suchi3}.
\end{align}
From \eqref{umo1} and \eqref{suchi3}, we finally obtain
\begin{align}
\bbP\{\hat{\calB}_n(\sigma)\} &\ddleq \sum_{m=1}^M \sum_{m'\neq m} \sum_{\by} \exp\{-e^{n\sigma}\} \label{umo2}\\
&\ddeq \exp\{-e^{n\sigma}\}.
\end{align}
This concludes our proof of Lemma \ref{smolem}.
\bibliographystyle{ieeetr}
\bibliography{IEEEabrv,isitbib}

\begin{thebibliography}{10}

\bibitem{Shannon48}
C.~E. Shannon, ``A mathematical theory of communication,'' {\em Bell System
  Technical Journal}, vol.~27, pp.~379--423, 1948.

\bibitem{Fano}
R.~M. Fano, {\em Transmission of Information}.
\newblock New York: Wiley, 1961.

\bibitem{Gallager1965a}
R.~G. Gallager, ``Simple derivation of the coding theorem and some
  applications,'' {\em {IEEE} Trans. Inf. Theory}, vol.~11, pp.~3--18, Jan
  2008.

\bibitem{sgb}
C.~E. Shannon, R.~G. Gallager, and E.~R. Berlekamp, ``Lower bounds to error
  probability for coding in discrete memoryless channels {I-II},'' {\em
  Information and Control}, vol.~10, pp.~65--103,~522--552, 1967.

\bibitem{Nakiboglu2020}
B.~Nakibo{\u{g}}lu, ``The sphere packing bound for memoryless channels,'' {\em
  Problems of Information Transmission}, vol.~56, pp.~201--244, 2020.

\bibitem{nakibouglu2019augustin}
B.~Nakibo{\u{g}}lu, ``The {A}ugustin capacity and center,'' {\em Problems of
  Information Transmission}, vol.~55, no.~4, pp.~299--342, 2019.

\bibitem{somekh_2019}
A.~{Somekh-Baruch}, J.~{Scarlett}, and A.~{Guill\'{e}n i F\`{a}bregas},
  ``Generalized random {Gilbert-Varshamov} codes,'' {\em IEEE Trans. Inf.
  Theory}, vol.~65, no.~6, pp.~3452--3469, 2019.

\bibitem{CK81}
I.~Csisz\'{a}r and J.~K\"{o}rner, ``Graph decomposition: A new key to coding
  theorems,'' {\em IEEE Trans. Inf. Th.}, vol.~27, pp.~5--11, 1981.

\bibitem{Barg2002a}
A.~{Barg} and G.~D. {Forney}, ``Random codes: minimum distances and error
  exponents,'' {\em {IEEE} Trans. Inf. Theory}, vol.~48, no.~9, pp.~2568--2573,
  2002.

\bibitem{Nazari}
A.~Nazari, A.~Anastasopoulos, and S.~S. Pradhan, ``Error exponent for
  multiple-access channels: Lower bounds,'' {\em {IEEE} Trans. Inf. Theory},
  vol.~60, no.~9, pp.~5095--5115, 2014.

\bibitem{Merhav2018a}
N.~Merhav, ``Error exponents of typical random codes,'' {\em {IEEE} Trans. Inf.
  Theory}, vol.~64, no.~9, pp.~6223--6235, 2018.

\bibitem{merhav2019error}
N.~Merhav, ``Error exponents of typical random codes for the colored {G}aussian
  channel,'' {\em IEEE Trans. Inf. Theory}, vol.~65, no.~12, pp.~8164--8179,
  2019.

\bibitem{merhav2019error2}
N.~Merhav, ``Error exponents of typical random trellis codes,'' {\em IEEE
  Trans. Inf. Theory}, vol.~66, no.~4, pp.~2067--2077, 2019.

\bibitem{merhav2019lagrange}
N.~Merhav, ``A lagrange--dual lower bound to the error exponent of the typical
  random code,'' {\em IEEE Trans. Inf. Theory}, vol.~66, no.~6, pp.~3456--3464,
  2019.

\bibitem{Tamir2020a}
R.~Tamir, N.~Merhav, N.~Weinberger, and A.~{Guill\'{e}n i F\`{a}bregas},
  ``Large deviations behavior of the logarithmic error probability of random
  codes,'' {\em {IEEE} Trans. Inf. Theory}, vol.~66, no.~11, pp.~6635--6659,
  2020.

\bibitem{Ahlswede1982}
R.~Ahlswede and G.~Dueck, ``Good codes can be produced by a few permutations,''
  {\em IEEE Trans. Inf. Theory}, vol.~28, no.~3, pp.~430--443, 1982.

\bibitem{tamir2021universal}
R.~T. (Averbuch) and N.~Merhav, ``Universal decoding for the typical random
  code and for the expurgated code,'' {\em IEEE Trans. Inf. Theory}, 2022.

\bibitem{cocco2022}
G.~Cocco, A.~{Guill\'{e}n i F\`{a}bregas}, and J.~Font-Segura, ``Typical error
  exponents: A dual domain derivation,'' {\em IEEE Trans. Inf. Theory}, to
  appear 2022.

\bibitem{Truong2022PO}
L.~V. Truong, G.~Cocco, J.~Font-Segura, and A.~{Guill{\'e}n i F{\`a}bregas},
  ``Concentration properties of random codes,'' {\em ArXiv},
  vol.~abs/2203.07853, 2022.

\bibitem{Durrett}
R.~Durrett, {\em Probability: Theory and Examples}.
\newblock Cambridge Univ. Press, 4th~ed., 2010.

\bibitem{Merhav2017a}
N.~Merhav, ``The generalized stochastic likelihood decoder: Random coding and
  expurgated bounds,'' {\em IEEE Transactions on Information Theory}, vol.~63,
  no.~8, pp.~5039--5051, 2017.

\bibitem{yassaee}
M.~H. Yassaee, M.~R. Aref, and A.~Gohari, ``A technique for deriving one-shot
  achievability results in network information theory,'' in {\em 2013 IEEE
  International Symposium on Information Theory}, pp.~1287--1291, 2013.

\bibitem{Scarlett2015e}
J.~Scarlett, A.~Martinez, and A.~Guill{\'e}n~i F{\`a}bregas, ``The likelihood
  decoder: Error exponents and mismatch,'' in {\em 2015 IEEE International
  Symposium on Information Theory (ISIT)}, pp.~86--90, 2015.

\bibitem{Csi97}
I.~Csisz\'{a}r and J.~{K\"{o}rner}, {\em Information Theory: Coding Theorems
  for Discrete Memoryless Systems}.
\newblock Cambridge University Press, 2011.

\bibitem{Tamir2022}
R.~Tamir and N.~Merhav, ``Universal decoding for the typical random code and
  for the expurgated code,'' {\em IEEE Trans. Inf. Th.}, vol.~68,
  pp.~2156--2168, Apr. 2022.

\bibitem{Gilbert1952ACO}
E.~N. Gilbert, ``A comparison of signalling alphabets,'' {\em Bell System
  Technical Journal}, vol.~31, pp.~504--522, 1952.

\bibitem{Varshamov1957a}
R.~R. Varshamov, ``Estimate of the number of signals in error correcting
  codes,'' {\em Doklady Akademii Nauk SSSR}, vol.~117, no.~5, pp.~739--741,
  1957.

\bibitem{Csis00}
I.~Csisz\'{a}r, ``The method of types,'' {\em IEEE Trans. Inf. Th.}, vol.~44,
  no.~6, pp.~2505--23, 1998.

\bibitem{CIT-052}
N.~Merhav, ``Statistical physics and information theory,'' {\em Foundations and
  Trends{\textregistered} in Communications and Information Theory}, vol.~6,
  no.~1--2, pp.~1--212, 2010.

\bibitem{Royden}
H.~Royden and P.~Fitzpatrick, {\em Real Analysis}.
\newblock Pearson, 4th~ed., 2010.

\bibitem{Pelekis2015}
C.~Pelekis and J.~Ramon, ``Hoeffding's inequality for sums of weakly dependent
  random variables,'' {\em Mediterranean Journal of Mathematics 14 (6), 1-16},
  2017.

\bibitem{merhav_FnT2}
N.~Merhav, {\em Statistical Physics and Information Theory}, vol.~6 of {\em
  Foundations and Trends in Communications and Information Theory}.
\newblock Now Publishers Inc, 2010.

\bibitem{Janson1998NewVO}
S.~Janson, ``New versions of suen's correlation inequality,'' {\em Random
  Struct. Algorithms}, vol.~13, pp.~467--483, 1998.

\bibitem{Impagliazzo2010}
R.~Impagliazzo and V.~Kabanets, ``Constructive proofs of concentration
  bounds,'' {\em Electron. Colloquium Comput. Complex.}, vol.~17, p.~72, 2010.

\end{thebibliography}
\end{document}